\newcommand\NOTYETSOUMISSIONLICS[1]{}

\newcommand\PREUVEPASENANNEXE{}

\newcommand\CORPSPREUVEDE[1]{}
\newcommand\ANNEXEPREUVEDE[2]{
	\subsection{Proof of #1}
	#2
}
\providecommand\PREUVEPASENANNEXE[1]{#1}
\PREUVEPASENANNEXE{
	\renewcommand\CORPSPREUVEDE[1]{#1}
	\renewcommand\ANNEXEPREUVEDE[2]{}
}

\newcommand\SANSCOMMENTAIRE[1]{#1
}
\newcommand\ANONYME[1]{}
\NOTYETSOUMISSIONLICS{
\renewcommand\SANSCOMMENTAIRE[1]{#1}
\renewcommand\ANONYME[1]{#1}
}

\newcommand\NOLICSSTYLE[1]{} 
\providecommand\NOLICSSTYLE[1]{}
\documentclass{article}

\usepackage{tikz}
\usetikzlibrary{arrows,arrows.meta,automata, positioning,shapes.geometric}
\usetikzlibrary{math}
\usepackage{xargs}

\usepackage{hyperref}
\usepackage{marvosym}
\usepackage{mathrsfs}

\usepackage{color}

\usepackage{todonotes}
\newcommand\olivier[1]{\SANSCOMMENTAIRE{\todo[inline,color=blue!40,caption={2do}]{Olivier: #1}}}

\usepackage{todonotes}

\usepackage{standalone}
\usepackage{multicol}

\definecolor{rouge}{RGB}{255,77,77}
\definecolor{vert}{RGB}{0,178,102}
\definecolor{jaune}{RGB}{255,255,0}
\definecolor{violet}{RGB}{208,32,144}
\definecolor{orange}{RGB}{255,140,0}
\definecolor{roujorange}{RGB}{255,108,38}
\definecolor{bleu}{RGB}{0,0,205}
\definecolor{bleuclair}{RGB}{50,255,220}
\definecolor{rose}{RGB}{255,50,220}
\definecolor{bleuvert}{RGB}{50,255,140}

\newcommand{\wang}[6]{
\draw [black,fill=#3] (#1,#2+1)  -- (#1+0.5,#2+0.5) -- (#1+1,#2+1) -- cycle;
\draw [black,fill=#4] (#1+1,#2+1)  -- (#1+0.5,#2+0.5) -- (#1+1,#2) -- cycle;
\draw [black,fill=#5] (#1,#2)  -- (#1+0.5,#2+0.5) -- (#1+1,#2) -- cycle;
\draw [black,fill=#6] (#1,#2)  -- (#1+0.5,#2+0.5) -- (#1,#2+1) -- cycle;
}

\newcommand{\wangEdge}[4]{\vbox to 17pt{\hbox{
\begin{tikzpicture}[scale=0.5,rounded corners=0pt]
\begin{scope}
\draw [black,fill=#1] (0,1)  -- (0.5,0.5) -- (1,1) -- cycle;
\draw [black,fill=#2] (1,1)  -- (0.5,0.5) -- (1,0) -- cycle;
\draw [black,fill=#3] (0,0)  -- (0.5,0.5) -- (1,0) -- cycle;
\draw [black,fill=#4] (0,0)  -- (0.5,0.5) -- (0,1) -- cycle;
\end{scope}
 \end{tikzpicture}
}}}

\newcommand{\wangEdgeH}[8]{\vbox to 17pt{\hbox{
\begin{tikzpicture}[scale=0.5,rounded corners=0pt]
\begin{scope}
\draw [black,fill=#1] (0,1)  -- (0.5,0.5) -- (1,1) -- cycle;
\draw [black,fill=#2] (1,1)  -- (0.5,0.5) -- (1,0) -- cycle;
\draw [black,fill=#3] (0,0)  -- (0.5,0.5) -- (1,0) -- cycle;
\draw [black,fill=#4] (0,0)  -- (0.5,0.5) -- (0,1) -- cycle;
\end{scope}
\begin{scope}[shift={(1,0)}]
\draw [black,fill=#5] (0,1)  -- (0.5,0.5) -- (1,1) -- cycle;
\draw [black,fill=#6] (1,1)  -- (0.5,0.5) -- (1,0) -- cycle;
\draw [black,fill=#7] (0,0)  -- (0.5,0.5) -- (1,0) -- cycle;
\draw [black,fill=#8] (0,0)  -- (0.5,0.5) -- (0,1) -- cycle;
\end{scope}
 \end{tikzpicture}
}}}

\newcommand{\wangEdgeV}[8]{\vbox to 17pt{\hbox{
\begin{tikzpicture}[scale=0.5,rounded corners=0pt]
\begin{scope}
\draw [black,fill=#1] (0,1)  -- (0.5,0.5) -- (1,1) -- cycle;
\draw [black,fill=#2] (1,1)  -- (0.5,0.5) -- (1,0) -- cycle;
\draw [black,fill=#3] (0,0)  -- (0.5,0.5) -- (1,0) -- cycle;
\draw [black,fill=#4] (0,0)  -- (0.5,0.5) -- (0,1) -- cycle;
\end{scope}
\begin{scope}[shift={(0,1)}]
\draw [black,fill=#5] (0,1)  -- (0.5,0.5) -- (1,1) -- cycle;
\draw [black,fill=#6] (1,1)  -- (0.5,0.5) -- (1,0) -- cycle;
\draw [black,fill=#7] (0,0)  -- (0.5,0.5) -- (1,0) -- cycle;
\draw [black,fill=#8] (0,0)  -- (0.5,0.5) -- (0,1) -- cycle;
\end{scope}
 \end{tikzpicture}
}}}

\newcommand{\wangEdgeVVVun}{\vbox to 17pt{\hbox{
\begin{tikzpicture}[scale=0.5,rounded corners=0pt]
\begin{scope}
\draw [black,fill=white] (0,1)  -- (0.5,0.5) -- (1,1) -- cycle;
\draw [black,fill=bleu] (1,1)  -- (0.5,0.5) -- (1,0) -- cycle;
\draw [black,fill=rouge] (0,0)  -- (0.5,0.5) -- (1,0) -- cycle;
\draw [black,fill=rouge] (0,0)  -- (0.5,0.5) -- (0,1) -- cycle;
\end{scope}
\begin{scope}[shift={(0,1)}]
\draw [black,fill=white] (0,1)  -- (0.5,0.5) -- (1,1) -- cycle;
\draw [black,fill=rouge] (1,1)  -- (0.5,0.5) -- (1,0) -- cycle;
\draw [black,fill=white] (0,0)  -- (0.5,0.5) -- (1,0) -- cycle;
\draw [black,fill=vert] (0,0)  -- (0.5,0.5) -- (0,1) -- cycle;
\end{scope}
\begin{scope}[shift={(0,2)}]
\draw [black,fill=rouge] (0,1)  -- (0.5,0.5) -- (1,1) -- cycle;
\draw [black,fill=rouge] (1,1)  -- (0.5,0.5) -- (1,0) -- cycle;
\draw [black,fill=white] (0,0)  -- (0.5,0.5) -- (1,0) -- cycle;
\draw [black,fill=vert] (0,0)  -- (0.5,0.5) -- (0,1) -- cycle;
\end{scope}
 \end{tikzpicture}
}}}

\newcommand{\wangEdgeVVVdeux}{\vbox to 17pt{\hbox{
\begin{tikzpicture}[scale=0.5,rounded corners=0pt]
\begin{scope}
\draw [black,fill=white] (0,1)  -- (0.5,0.5) -- (1,1) -- cycle;
\draw [black,fill=vert] (1,1)  -- (0.5,0.5) -- (1,0) -- cycle;
\draw [black,fill=rouge] (0,0)  -- (0.5,0.5) -- (1,0) -- cycle;
\draw [black,fill=bleu] (0,0)  -- (0.5,0.5) -- (0,1) -- cycle;
\end{scope}
\begin{scope}[shift={(0,1)}]
\draw [black,fill=white] (0,1)  -- (0.5,0.5) -- (1,1) -- cycle;
\draw [black,fill=vert] (1,1)  -- (0.5,0.5) -- (1,0) -- cycle;
\draw [black,fill=white] (0,0)  -- (0.5,0.5) -- (1,0) -- cycle;
\draw [black,fill=rouge] (0,0)  -- (0.5,0.5) -- (0,1) -- cycle;
\end{scope}
\begin{scope}[shift={(0,2)}]
\draw [black,fill=rouge] (0,1)  -- (0.5,0.5) -- (1,1) -- cycle;
\draw [black,fill=bleu] (1,1)  -- (0.5,0.5) -- (1,0) -- cycle;
\draw [black,fill=white] (0,0)  -- (0.5,0.5) -- (1,0) -- cycle;
\draw [black,fill=rouge] (0,0)  -- (0.5,0.5) -- (0,1) -- cycle;
\end{scope}
 \end{tikzpicture}
}}}

\newcommand{\wangEdgeVVVtrois}{\vbox to 41pt{\hbox{
\begin{tikzpicture}[scale=0.5,rounded corners=0pt]
\begin{scope}
\draw [black,fill=white] (0,1)  -- (0.5,0.5) -- (1,1) -- cycle;
\draw [black,fill=rouge] (1,1)  -- (0.5,0.5) -- (1,0) -- cycle;
\draw [black,fill=rouge] (0,0)  -- (0.5,0.5) -- (1,0) -- cycle;
\draw [black,fill=vert] (0,0)  -- (0.5,0.5) -- (0,1) -- cycle;
\end{scope}
\begin{scope}[shift={(0,1)}]
\draw [black,fill=white] (0,1)  -- (0.5,0.5) -- (1,1) -- cycle;
\draw [black,fill=rouge] (1,1)  -- (0.5,0.5) -- (1,0) -- cycle;
\draw [black,fill=white] (0,0)  -- (0.5,0.5) -- (1,0) -- cycle;
\draw [black,fill=vert] (0,0)  -- (0.5,0.5) -- (0,1) -- cycle;
\end{scope}
\begin{scope}[shift={(0,2)}]
\draw [black,fill=vert] (0,1)  -- (0.5,0.5) -- (1,1) -- cycle;
\draw [black,fill=vert] (1,1)  -- (0.5,0.5) -- (1,0) -- cycle;
\draw [black,fill=white] (0,0)  -- (0.5,0.5) -- (1,0) -- cycle;
\draw [black,fill=bleu] (0,0)  -- (0.5,0.5) -- (0,1) -- cycle;
\end{scope}
 \end{tikzpicture}
}}}

\newcommand{\wangEdgeVVVquatre}{\vbox to 41pt{\hbox{
\begin{tikzpicture}[scale=0.5,rounded corners=0pt]
\begin{scope}
\draw [black,fill=white] (0,1)  -- (0.5,0.5) -- (1,1) -- cycle;
\draw [black,fill=vert] (1,1)  -- (0.5,0.5) -- (1,0) -- cycle;
\draw [black,fill=rouge] (0,0)  -- (0.5,0.5) -- (1,0) -- cycle;
\draw [black,fill=rouge] (0,0)  -- (0.5,0.5) -- (0,1) -- cycle;
\end{scope}
\begin{scope}[shift={(0,1)}]
\draw [black,fill=white] (0,1)  -- (0.5,0.5) -- (1,1) -- cycle;
\draw [black,fill=bleu] (1,1)  -- (0.5,0.5) -- (1,0) -- cycle;
\draw [black,fill=white] (0,0)  -- (0.5,0.5) -- (1,0) -- cycle;
\draw [black,fill=rouge] (0,0)  -- (0.5,0.5) -- (0,1) -- cycle;
\end{scope}
\begin{scope}[shift={(0,2)}]
\draw [black,fill=rouge] (0,1)  -- (0.5,0.5) -- (1,1) -- cycle;
\draw [black,fill=rouge] (1,1)  -- (0.5,0.5) -- (1,0) -- cycle;
\draw [black,fill=white] (0,0)  -- (0.5,0.5) -- (1,0) -- cycle;
\draw [black,fill=vert] (0,0)  -- (0.5,0.5) -- (0,1) -- cycle;
\end{scope}
 \end{tikzpicture}
}}}

\newcommand{\wangEdgeVVVcinq}{\vbox to 17pt{\hbox{
\begin{tikzpicture}[scale=0.5,rounded corners=0pt]
\begin{scope}
\draw [black,fill=white] (0,1)  -- (0.5,0.5) -- (1,1) -- cycle;
\draw [black,fill=rouge] (1,1)  -- (0.5,0.5) -- (1,0) -- cycle;
\draw [black,fill=vert] (0,0)  -- (0.5,0.5) -- (1,0) -- cycle;
\draw [black,fill=vert] (0,0)  -- (0.5,0.5) -- (0,1) -- cycle;
\end{scope}
\begin{scope}[shift={(0,1)}]
\draw [black,fill=white] (0,1)  -- (0.5,0.5) -- (1,1) -- cycle;
\draw [black,fill=vert] (1,1)  -- (0.5,0.5) -- (1,0) -- cycle;
\draw [black,fill=white] (0,0)  -- (0.5,0.5) -- (1,0) -- cycle;
\draw [black,fill=bleu] (0,0)  -- (0.5,0.5) -- (0,1) -- cycle;
\end{scope}
\begin{scope}[shift={(0,2)}]
\draw [black,fill=rouge] (0,1)  -- (0.5,0.5) -- (1,1) -- cycle;
\draw [black,fill=vert] (1,1)  -- (0.5,0.5) -- (1,0) -- cycle;
\draw [black,fill=white] (0,0)  -- (0.5,0.5) -- (1,0) -- cycle;
\draw [black,fill=rouge] (0,0)  -- (0.5,0.5) -- (0,1) -- cycle;
\end{scope}
 \end{tikzpicture}
}}}

\newcommand{\cyclicPun}{\vbox to 17pt{\hbox{
\begin{tikzpicture}[scale=0.5,rounded corners=0pt]
\begin{scope}[shift={(-5,0)}]
\wang{0}{0}{white}{white}{rouge}{rouge}
\wang{1}{0}{white}{white}{rouge}{white}
\wang{2}{0}{white}{white}{rouge}{white}
\wang{3}{0}{white}{white}{vert}{white}
\wang{4}{0}{white}{rouge}{rouge}{white}
\wang{0}{1}{white}{white}{white}{vert}
\wang{1}{1}{white}{white}{white}{white}
\wang{2}{1}{white}{white}{white}{white}
\wang{3}{1}{white}{white}{white}{white}
\wang{4}{1}{white}{vert}{white}{white}
\wang{0}{2}{rouge}{white}{white}{vert}
\wang{1}{2}{rouge}{white}{white}{white}
\wang{2}{2}{vert}{white}{white}{white}
\wang{3}{2}{rouge}{white}{white}{white}
\wang{4}{2}{rouge}{vert}{white}{white}
\end{scope}
 \end{tikzpicture}
}}}

\newcommand{\cyclicPdeux}{\vbox to 17pt{\hbox{
\begin{tikzpicture}[scale=0.5,rounded corners=0pt]
\begin{scope}[shift={(-5,0)}]
\wang{0}{0}{white}{white}{rouge}{bleu}
\wang{1}{0}{white}{white}{rouge}{white}
\wang{2}{0}{white}{white}{vert}{white}
\wang{3}{0}{white}{white}{rouge}{white}
\wang{4}{0}{white}{bleu}{rouge}{white}
\wang{0}{1}{white}{white}{white}{rouge}
\wang{1}{1}{white}{white}{white}{white}
\wang{2}{1}{white}{white}{white}{white}
\wang{3}{1}{white}{white}{white}{white}
\wang{4}{1}{white}{rouge}{white}{white}
\wang{0}{2}{rouge}{white}{white}{rouge}
\wang{1}{2}{vert}{white}{white}{white}
\wang{2}{2}{rouge}{white}{white}{white}
\wang{3}{2}{rouge}{white}{white}{white}
\wang{4}{2}{rouge}{rouge}{white}{white}
\end{scope}
 \end{tikzpicture}
}}}

\newcommand{\cyclicPtrois}{\vbox to 17pt{\hbox{
\begin{tikzpicture}[scale=0.5,rounded corners=0pt]
\begin{scope}[shift={(-5,0)}]
\wang{0}{0}{white}{white}{rouge}{vert}
\wang{1}{0}{white}{white}{vert}{white}
\wang{2}{0}{white}{white}{rouge}{white}
\wang{3}{0}{white}{white}{rouge}{white}
\wang{4}{0}{white}{vert}{rouge}{white}
\wang{0}{1}{white}{white}{white}{vert}
\wang{1}{1}{white}{white}{white}{white}
\wang{2}{1}{white}{white}{white}{white}
\wang{3}{1}{white}{white}{white}{white}
\wang{4}{1}{white}{vert}{white}{white}
\wang{0}{2}{vert}{white}{white}{bleu}
\wang{1}{2}{rouge}{white}{white}{white}
\wang{2}{2}{rouge}{white}{white}{white}
\wang{3}{2}{rouge}{white}{white}{white}
\wang{4}{2}{rouge}{bleu}{white}{white}
\end{scope}
 \end{tikzpicture}
}}}

\newcommand{\cyclicPquatre}{\vbox to 17pt{\hbox{
\begin{tikzpicture}[scale=0.5,rounded corners=0pt]
\begin{scope}[shift={(-5,0)}]
\wang{0}{0}{white}{white}{vert}{rouge}
\wang{1}{0}{white}{white}{rouge}{white}
\wang{2}{0}{white}{white}{rouge}{white}
\wang{3}{0}{white}{white}{rouge}{white}
\wang{4}{0}{white}{rouge}{rouge}{white}
\wang{0}{1}{white}{white}{white}{rouge}
\wang{1}{1}{white}{white}{white}{white}
\wang{2}{1}{white}{white}{white}{white}
\wang{3}{1}{white}{white}{white}{white}
\wang{4}{1}{white}{rouge}{white}{white}
\wang{0}{2}{rouge}{white}{white}{vert}
\wang{1}{2}{rouge}{white}{white}{white}
\wang{2}{2}{rouge}{white}{white}{white}
\wang{3}{2}{rouge}{white}{white}{white}
\wang{4}{2}{vert}{vert}{white}{white}
\end{scope}
 \end{tikzpicture}
}}}

\newcommand{\cyclicPcinq}{\vbox to 17pt{\hbox{
\begin{tikzpicture}[scale=0.5,rounded corners=0pt]
\begin{scope}[shift={(-5,0)}]
\wang{0}{0}{white}{white}{rouge}{vert}
\wang{1}{0}{white}{white}{rouge}{white}
\wang{2}{0}{white}{white}{rouge}{white}
\wang{3}{0}{white}{white}{rouge}{white}
\wang{4}{0}{white}{vert}{vert}{white}
\wang{0}{1}{white}{white}{white}{bleu}
\wang{1}{1}{white}{white}{white}{white}
\wang{2}{1}{white}{white}{white}{white}
\wang{3}{1}{white}{white}{white}{white}
\wang{4}{1}{white}{bleu}{white}{white}
\wang{0}{2}{rouge}{white}{white}{rouge}
\wang{1}{2}{rouge}{white}{white}{white}
\wang{2}{2}{rouge}{white}{white}{white}
\wang{3}{2}{vert}{white}{white}{white}
\wang{4}{2}{rouge}{rouge}{white}{white}
\end{scope}
 \end{tikzpicture}
}}}

\newcommand{\wangEdgeHH}{\vbox to 17pt{\hbox{
\begin{tikzpicture}[scale=0.5,rounded corners=0pt]
\begin{scope}
\draw [black,fill=rouge] (0,1)  -- (0.5,0.5) -- (1,1) -- cycle;
\draw [black,fill=rouge] (1,1)  -- (0.5,0.5) -- (1,0) -- cycle;
\draw [black,fill=rouge] (0,0)  -- (0.5,0.5) -- (1,0) -- cycle;
\draw [black,fill=vert] (0,0)  -- (0.5,0.5) -- (0,1) -- cycle;
\end{scope}
\begin{scope}[shift={(1,0)}]
\draw [black,fill=rouge] (0,1)  -- (0.5,0.5) -- (1,1) -- cycle;
\draw [black,fill=bleu] (1,1)  -- (0.5,0.5) -- (1,0) -- cycle;
\draw [black,fill=rouge] (0,0)  -- (0.5,0.5) -- (1,0) -- cycle;
\draw [black,fill=rouge] (0,0)  -- (0.5,0.5) -- (0,1) -- cycle;
\end{scope}
\begin{scope}[shift={(2,0)}]
\draw [black,fill=vert] (0,1)  -- (0.5,0.5) -- (1,1) -- cycle;
\draw [black,fill=vert] (1,1)  -- (0.5,0.5) -- (1,0) -- cycle;
\draw [black,fill=rouge] (0,0)  -- (0.5,0.5) -- (1,0) -- cycle;
\draw [black,fill=bleu] (0,0)  -- (0.5,0.5) -- (0,1) -- cycle;
\end{scope}
 \end{tikzpicture}
}}}

\newcommand{\wangEdgeEdge}[8]{\vbox to 17pt{\hbox{
\begin{tikzpicture}[scale=0.5,rounded corners=0pt]
\draw [black,fill=#1] (0,1)  -- (0.5,0.5) -- (1,1) -- cycle;
\draw [black,fill=#2] (1,1)  -- (0.5,0.5) -- (1,0) -- cycle;
\draw [black,fill=#3] (0,0)  -- (0.5,0.5) -- (1,0) -- cycle;
\draw [black,fill=#4] (0,0)  -- (0.5,0.5) -- (0,1) -- cycle;
\begin{scope}[shift={(1,0)}]
\draw [black,fill=#5] (0,1)  -- (0.5,0.5) -- (1,1) -- cycle;
\draw [black,fill=#6] (1,1)  -- (0.5,0.5) -- (1,0) -- cycle;
\draw [black,fill=#7] (0,0)  -- (0.5,0.5) -- (1,0) -- cycle;
\draw [black,fill=#8] (0,0)  -- (0.5,0.5) -- (0,1) -- cycle;
\end{scope}
 \end{tikzpicture}
}}}

\newcommand{\robinsoncornerbumpybis}[4]{
\begin{scope}[shift={(#1,#2)},rotate=#3]
\draw[color=black,fill=#4] (0,0.1) -- (-0.1,0) -- (0,-0.1) -- (0.1,0) -- (0.4,0) -- (0.5,-0.2) -- (0.6,0) -- (0.9,0) -- (1,-0.1) -- (1.1,0) -- (1,0.1) -- (1,0.4) -- (1.2,0.4) -- (1,0.6) -- (1,0.9) -- (1.1,1) -- (1,1.1) -- (0.9,1) -- (0.6,1) -- (0.4,1.2) -- (0.4,1) -- (0.1,1) -- (0,1.1) -- (-0.1,1) -- (0,0.9) -- (0,0.6) -- (-0.2,0.5) -- (0,0.4) --  cycle ; 
\draw[color=bleu,thick,-,shorten >=0pt] (0.5,1.1) -- (0.5,0.5) -- (1.1,0.5);
\end{scope}
}

\newcommand{\tRobiCorner}[2]{\vbox to 9pt{\hbox{
\begin{tikzpicture}[scale=0.3]
\robinsoncornerbumpybis{0}{0}{#1}{#2};
\end{tikzpicture}
}}}

\newcommand{\robinsoncornerdentedbis}[4]{
\begin{scope}[shift={(#1,#2)},rotate=#3]
\draw[color=black,fill=#4] (0,0.1) -- (0.1,0) -- (0.4,0) -- (0.5,-0.2) -- (0.6,0) -- (0.9,0) -- (1,0.1) -- (1,0.4) -- (1.2,0.4) -- (1,0.6) -- (1,0.9) -- (0.9,1) -- (0.6,1) -- (0.4,1.2) -- (0.4,1) -- (0.1,1) -- (0,0.9) -- (0,0.6) -- (-0.2,0.5) -- (0,0.4) --  cycle ;
\draw[color=bleu,thick,-,shorten >=0pt] (0.5,1.1) -- (0.5,0.5) -- (1.1,0.5);
\end{scope}
}

\newcommand{\tRobiCornerDented}[2]{\vbox to 9pt{\hbox{
\begin{tikzpicture}[scale=0.3]
\robinsoncornerdentedbis{0}{0}{#1}{#2};
\end{tikzpicture}
}}}

\newcommand{\robinsoncrossbis}[4]{
\begin{scope}[shift={(#1,#2)},rotate=#3]
\draw[color=black,fill=#4] (0,0.1) -- (0.1,0) -- (0.4,0) -- (0.4,-0.2) -- (0.6,0) -- (0.9,0) -- (1,0.1) -- (1,0.4) -- (0.8,0.4) -- (1,0.6) -- (1,0.9) -- (0.9,1) -- (0.6,1) -- (0.4,0.8) -- (0.4,1) -- (0.1,1) -- (0,0.9) -- (0,0.6) -- (0.2,0.4) -- (0,0.4) --  cycle ; 
\draw[color=bleu,thick,-,shorten >=0pt] (0.1,0.5) -- (0.9,0.5);
\draw[color=bleu,thick,-,shorten >=0pt] (0.5,-0.1) -- (0.5,0.9);
\end{scope}
}

\newcommand{\tRobiCross}[2]{\vbox to 9pt{\hbox{
\begin{tikzpicture}[scale=0.3]
\robinsoncrossbis{0}{0}{#1}{#2};
\end{tikzpicture}
}}}

\newcommand{\robinsonhorizontalbis}[4]{
\begin{scope}[shift={(#1,#2)},rotate=#3]
\draw[color=black,fill=#4] (0,0.1) -- (0.1,0) -- (0.4,0) -- (0.5,-0.2) -- (0.6,0) -- (0.9,0) -- (1,0.1) -- (1,0.4) -- (0.8,0.4) -- (1,0.6) -- (1,0.9) -- (0.9,1) -- (0.6,1) -- (0.5,0.8) -- (0.4,1) -- (0.1,1) -- (0,0.9) -- (0,0.6) -- (0.2,0.4) -- (0,0.4) --  cycle ; 
\draw[color=bleu,thick,-,shorten >=0pt] (0.1,0.5) -- (0.9,0.5);
\end{scope}
}

\newcommand{\tRobiH}[2]{\vbox to 9pt{\hbox{
\begin{tikzpicture}[scale=0.3]
\robinsonhorizontalbis{0}{0}{#1}{#2};
\end{tikzpicture}
}}}

\newcommand{\robinsonverticalbis}[4]{
\begin{scope}[shift={(#1,#2)},rotate=#3]
\draw[color=black,fill=#4] (0,0.1) -- (0.1,0) -- (0.4,0) -- (0.4,-0.2) -- (0.6,0) -- (0.9,0) -- (1,0.1) -- (1,0.4) -- (0.8,0.5) -- (1,0.6) -- (1,0.9) -- (0.9,1) -- (0.6,1) -- (0.4,0.8) -- (0.4,1) -- (0.1,1) -- (0,0.9) -- (0,0.6) -- (0.2,0.5) -- (0,0.4) --  cycle ; 
\draw[color=bleu,thick,-,shorten >=0pt] (0.5,-0.1) -- (0.5,0.9);
\end{scope}
}

\newcommand{\tRobiV}[2]{\vbox to 9pt{\hbox{
\begin{tikzpicture}[scale=0.3]
\robinsonverticalbis{0}{0}{#1}{#2};
\end{tikzpicture}
}}}

\newcommand{\robinsonempty}[4]{
\begin{scope}[shift={(#1,#2)},rotate=#3]
\draw[color=black,fill=#4] (0,0.1) -- (0.1,0) -- (0.4,0) -- (0.5,-0.2) -- (0.6,0) -- (0.9,0) -- (1,0.1) -- (1,0.4) -- (0.8,0.5) -- (1,0.6) -- (1,0.9) -- (0.9,1) -- (0.6,1) -- (0.5,0.8) -- (0.4,1) -- (0.1,1) -- (0,0.9) -- (0,0.6) -- (0.2,0.5) -- (0,0.4) --  cycle ; 
\end{scope}
}

\newcommand{\tRobiEmpty}[2]{\vbox to 9pt{\hbox{
\begin{tikzpicture}[scale=0.3]
\robinsonempty{0}{0}{#1}{#2};
\end{tikzpicture}
}}}

\newcommand{\robinsonemptybis}[4]{
\begin{scope}[shift={(#1,#2)},rotate=#3]
\draw[color=black,fill=#4] (0,0.1) -- (0.1,0) -- (0.4,0) -- (0.5,-0.2) -- (0.6,0) -- (0.9,0) -- (1,0.1) -- (1,0.4) -- (0.8,0.5) -- (1,0.6) -- (1,0.9) -- (0.9,1) -- (0.6,1) -- (0.5,0.8) -- (0.4,1) -- (0.1,1) -- (0,0.9) -- (0,0.6) -- (0.2,0.5) -- (0,0.4) --  cycle ; 
\end{scope}
}

\newcommand{\robinsoncrossreflectedbis}[4]{
\begin{scope}[shift={(#1,#2)},rotate=#3]
\draw[color=black,fill=#4] (0,0.1) -- (0.1,0) -- (0.4,0) -- (0.6,-0.2) -- (0.6,0) -- (0.9,0) -- (1,0.1) -- (1,0.4) -- (0.8,0.4) -- (1,0.6) -- (1,0.9) -- (0.9,1) -- (0.6,1) -- (0.6,0.8) -- (0.4,1) -- (0.1,1) -- (0,0.9) -- (0,0.6) -- (0.2,0.4) -- (0,0.4) --  cycle ; 
\draw[color=bleu,thick,-,shorten >=0pt] (0.1,0.5) -- (0.9,0.5);
\draw[color=bleu,thick,-,shorten >=0pt] (0.5,-0.1) -- (0.5,0.9);
\end{scope}
}

\newcommand{\tRobiCrossBis}[2]{\vbox to 9pt{\hbox{
\begin{tikzpicture}[scale=0.3]
\robinsoncrossreflectedbis{0}{0}{#1}{#2};
\end{tikzpicture}
}}}

\newcommand{\robinsonverticalreflectedbis}[4]{
\begin{scope}[shift={(#1,#2)},rotate=#3]
\draw[color=black,fill=#4] (0,0.1) -- (0.1,0) -- (0.4,0) -- (0.6,-0.2) -- (0.6,0) -- (0.9,0) -- (1,0.1) -- (1,0.4) -- (0.8,0.5) -- (1,0.6) -- (1,0.9) -- (0.9,1) -- (0.6,1) -- (0.6,0.8) -- (0.4,1) -- (0.1,1) -- (0,0.9) -- (0,0.6) -- (0.2,0.5) -- (0,0.4) --  cycle ; 
\draw[color=bleu,thick,-,shorten >=0pt] (0.5,-0.1) -- (0.5,0.9);
\end{scope}
}

\newcommand{\tRobiVbis}[2]{\vbox to 9pt{\hbox{
\begin{tikzpicture}[scale=0.3]
\robinsonverticalreflectedbis{0}{0}{#1}{#2};
\end{tikzpicture}
}}}

\newcommand{\lettredT}[1]{\vbox to 11pt{\hbox{
\begin{tikzpicture}[scale=0.5]
\draw[fill=#1] (0.5,0) -- (1,0.5) -- (0.5,1) -- (0,0.5) -- cycle ;
\draw (0.5,0) -- (0.5,1) ;
 \end{tikzpicture}
}}}

\newcommand{\lettreddT}[2]{\vbox to 28pt{\hbox{
\begin{tikzpicture}[scale=0.5, rounded corners=0mm]
\draw[fill=#1] (0.5,0) -- (1,0.5) -- (0.5,1) -- (0,0.5) -- cycle ;
\draw[fill=#2] (0.5,1) -- (1,1.5) -- (0.5,2) -- (0,1.5) -- cycle ;
\draw (0.5,0) -- (0.5,2) ;
 \end{tikzpicture}
}}}

\newcommand{\lettredddT}[3]{\vbox to 41pt{\hbox{
\begin{tikzpicture}[scale=0.5, rounded corners=0mm]
\draw[fill=#1] (0.5,0) -- (1,0.5) -- (0.5,1) -- (0,0.5) -- cycle ;
\draw[fill=#2] (0.5,1) -- (1,1.5) -- (0.5,2) -- (0,1.5) -- cycle ;
\draw[fill=#3] (0.5,2) -- (1,2.5) -- (0.5,3) -- (0,2.5) -- cycle ;
\draw (0.5,0) -- (0.5,3) ;
 \end{tikzpicture}
}}}

\newcommand{\lettrebT}[1]{\vbox to 5pt{\hbox{
\begin{tikzpicture}[scale=0.5]
\draw[fill=#1] (0.5,0.5) -- (1,0) -- (0,0) -- cycle ;
 \end{tikzpicture}
}}}

\newcommand{\lettrebbT}[2]{\vbox to 5pt{\hbox{
\begin{tikzpicture}[scale=0.5]
\draw[fill=#1] (0.5,0.5) -- (1,0) -- (0,0) -- cycle ;
\draw[fill=#2] (1.5,0.5) -- (2,0) -- (1,0) -- cycle ;
 \end{tikzpicture}
}}}

\newcommand{\lettreuT}[1]{\vbox to 5pt{\hbox{
\begin{tikzpicture}[scale=0.5]
\draw[fill=#1] (0.5,0.5) -- (1,1) -- (0,1) -- cycle ;
 \end{tikzpicture}
}}}

\newcommand{\lettreuuT}[2]{\vbox to 5pt{\hbox{
\begin{tikzpicture}[scale=0.5]
\draw[fill=#1] (0.5,0.5) -- (1,1) -- (0,1) -- cycle ;
\draw[fill=#2] (1.5,0.5) -- (2,1) -- (1,1) -- cycle ;
 \end{tikzpicture}
}}}

\newcommand\g[1]{
  \ifnum#1=1	1
  \else \ifnum#1=2 3
  \else \ifnum#1=3 7
  \else \ifnum#1=4  15
  \else \ifnum#1=5 31
  \else \ifnum#1=6 63
  \fi \fi \fi \fi \fi \fi
 }

\newcommand\twopowern[1]{
  \ifnum#1=0	1
  \else  \ifnum#1=1	2
  \else \ifnum#1=2 4
  \else \ifnum#1=3 8
  \else \ifnum#1=4 16
  \else \ifnum#1=5 32
  \else \ifnum#1=6 64
  \fi \fi \fi \fi \fi \fi\fi
  }
  
\newcommand\minusone[1]{%
  \ifnum#1=0	-1%
  \else  \ifnum#1=1	0%
  \else \ifnum#1=2 1%
  \else \ifnum#1=3 2%
  \else \ifnum#1=4 3%
  \else \ifnum#1=5 4%
  \else \ifnum#1=6 5%
 \fi\fi\fi\fi\fi\fi\fi%
 }
 
\newcommand\minustwo[1]{%
  \ifnum#1=0	-2%
  \else  \ifnum#1=1	-1%
  \else \ifnum#1=2 0%
  \else \ifnum#1=3 1%
  \else \ifnum#1=4 2%
  \else \ifnum#1=5 3%
  \else \ifnum#1=6 3%
 \fi\fi\fi\fi\fi\fi\fi%
 }

\newcommand\deuxpuissancenmoinsun[1]{
  \ifnum#1=0 0
  \else  \ifnum#1=1	1
  \else \ifnum#1=2 2
  \else \ifnum#1=3 4
  \else \ifnum#1=4 8
  \else \ifnum#1=5 16
  \else \ifnum#1=6 32
  \fi \fi \fi \fi \fi \fi\fi
}

\newcommand\deuxpuissancenmoinsdeux[1]{
  \ifnum#1=0	?
  \else  \ifnum#1=1	0
  \else \ifnum#1=2 1
  \else \ifnum#1=3 2
  \else \ifnum#1=4  4
  \else \ifnum#1=5 8
  \else \ifnum#1=6 16
  \fi \fi \fi \fi \fi \fi\fi
}

\newcommand\moitmoitun[1]{
\number\numexpr\deuxpuissancenmoinsun{#1}-\deuxpuissancenmoinsdeux{#1}\relax
}

\newcommand\moitmoitdeux[1]{
\number\numexpr\deuxpuissancenmoinsun{#1}+\deuxpuissancenmoinsdeux{#1}\relax
}

\newcommand{\myrobinsoncornerbumpybis}[4]{
\begin{scope}[shift={(#1,#2)},rotate=#3]
\draw[shift={(-0.5,-0.5)},color=black,fill=#4] (0,0.1) -- (-0.1,0) -- (0,-0.1) -- (0.1,0) -- (0.4,0) -- (0.5,-0.2) -- (0.6,0) -- (0.9,0) -- (1,-0.1) -- (1.1,0) -- (1,0.1) -- (1,0.4) -- (1.2,0.4) -- (1,0.6) -- (1,0.9) -- (1.1,1) -- (1,1.1) -- (0.9,1) -- (0.6,1) -- (0.4,1.2) -- (0.4,1) -- (0.1,1) -- (0,1.1) -- (-0.1,1) -- (0,0.9) -- (0,0.6) -- (-0.2,0.5) -- (0,0.4) --  cycle ; 
\draw[shift={(-0.5,-0.5)},color=bleu,thick,-,shorten >=0pt] (0.5,1.1) -- (0.5,0.5) -- (1.1,0.5);
\end{scope}
}

\newcommand{\myrobinsoncornerdentedbis}[4]{
\begin{scope}[shift={(#1,#2)},rotate=#3]
\draw[shift={(-0.5,-0.5)},color=black,fill=#4] (0,0.1) -- (0.1,0) -- (0.4,0) -- (0.5,-0.2) -- (0.6,0) -- (0.9,0) -- (1,0.1) -- (1,0.4) -- (1.2,0.4) -- (1,0.6) -- (1,0.9) -- (0.9,1) -- (0.6,1) -- (0.4,1.2) -- (0.4,1) -- (0.1,1) -- (0,0.9) -- (0,0.6) -- (-0.2,0.5) -- (0,0.4) --  cycle ;
\draw[shift={(-0.5,-0.5)},color=bleu,thick,-,shorten >=0pt] (0.5,1.1) -- (0.5,0.5) -- (1.1,0.5);
\end{scope}
}

\newcommand{\myrobinsoncrossbis}[4]{
\begin{scope}[shift={(#1,#2)},rotate=#3]
\draw[shift={(-0.5,-0.5)},color=black,fill=#4] (0,0.1) -- (0.1,0) -- (0.4,0) -- (0.4,-0.2) -- (0.6,0) -- (0.9,0) -- (1,0.1) -- (1,0.4) -- (0.8,0.4) -- (1,0.6) -- (1,0.9) -- (0.9,1) -- (0.6,1) -- (0.4,0.8) -- (0.4,1) -- (0.1,1) -- (0,0.9) -- (0,0.6) -- (0.2,0.4) -- (0,0.4) --  cycle ; 
\draw[shift={(-0.5,-0.5)},color=bleu,thick,-,shorten >=0pt] (0.1,0.5) -- (0.9,0.5);
\draw[shift={(-0.5,-0.5)},color=bleu,thick,-,shorten >=0pt] (0.5,-0.1) -- (0.5,0.9);
\end{scope}
}

\newcommand{\myrobinsonhorizontalbis}[4]{
\begin{scope}[shift={(#1,#2)},rotate=#3]
\draw[shift={(-0.5,-0.5)},color=black,fill=#4] (0,0.1) -- (0.1,0) -- (0.4,0) -- (0.5,-0.2) -- (0.6,0) -- (0.9,0) -- (1,0.1) -- (1,0.4) -- (0.8,0.4) -- (1,0.6) -- (1,0.9) -- (0.9,1) -- (0.6,1) -- (0.5,0.8) -- (0.4,1) -- (0.1,1) -- (0,0.9) -- (0,0.6) -- (0.2,0.4) -- (0,0.4) --  cycle ; 
\draw[shift={(-0.5,-0.5)},color=bleu,thick,-,shorten >=0pt] (0.1,0.5) -- (0.9,0.5);
\end{scope}
}

\newcommand{\myrobinsonverticalbis}[4]{
\begin{scope}[shift={(#1,#2)},rotate=#3]
\draw[shift={(-0.5,-0.5)},color=black,fill=#4] (0,0.1) -- (0.1,0) -- (0.4,0) -- (0.4,-0.2) -- (0.6,0) -- (0.9,0) -- (1,0.1) -- (1,0.4) -- (0.8,0.5) -- (1,0.6) -- (1,0.9) -- (0.9,1) -- (0.6,1) -- (0.4,0.8) -- (0.4,1) -- (0.1,1) -- (0,0.9) -- (0,0.6) -- (0.2,0.5) -- (0,0.4) --  cycle ; 
\draw[shift={(-0.5,-0.5)},color=bleu,thick,-,shorten >=0pt] (0.5,-0.1) -- (0.5,0.9);
\end{scope}
}

\newcommand{\mytRobiV}[2]{\vbox to 9pt{\hbox{
\begin{tikzpicture}[scale=0.3]
\robinsonverticalbis{0}{0}{#1}{#2};
\end{tikzpicture}
}}}

\newcommand{\myrobinsonempty}[4]{
\begin{scope}[shift={(#1,#2)},rotate=#3]
\draw[shift={(-0.5,-0.5)},color=black,fill=#4] (0,0.1) -- (0.1,0) -- (0.4,0) -- (0.5,-0.2) -- (0.6,0) -- (0.9,0) -- (1,0.1) -- (1,0.4) -- (0.8,0.5) -- (1,0.6) -- (1,0.9) -- (0.9,1) -- (0.6,1) -- (0.5,0.8) -- (0.4,1) -- (0.1,1) -- (0,0.9) -- (0,0.6) -- (0.2,0.5) -- (0,0.4) --  cycle ; 
\end{scope}
}

\newcommand{\myrobinsonemptynotdotted}[4]{
\begin{scope}[shift={(#1,#2)},rotate=#3]
\draw[shift={(-0.5,-0.5)},color=black,fill=white,-] 
(0.9,0) -- (1,0.1) -- (1,0.4) -- (0.8,0.5) -- (1,0.6) -- (1,0.9) -- (0.9,1);
\draw[shift={(-0.5,-0.5)},color=black,fill=white,-] 
(0.1,1) -- (0,0.9) -- (0,0.6) -- (0.2,0.5) -- (0,0.4) -- (0,0.1) -- (0.1,0) 
; 
\end{scope}
}

\newcommand{\mytRobiEmpty}[2]{\vbox to 9pt{\hbox{
\begin{tikzpicture}[scale=0.3]
\robinsonempty{0}{0}{#1}{#2};
\end{tikzpicture}
}}}

\newcommand{\myrobinsonemptybis}[4]{
\begin{scope}[shift={(#1,#2)},rotate=#3]
\draw[shift={(-0.5,-0.5)},color=black,fill=#4] (0,0.1) -- (0.1,0) -- (0.4,0) -- (0.5,-0.2) -- (0.6,0) -- (0.9,0) -- (1,0.1) -- (1,0.4) -- (0.8,0.5) -- (1,0.6) -- (1,0.9) -- (0.9,1) -- (0.6,1) -- (0.5,0.8) -- (0.4,1) -- (0.1,1) -- (0,0.9) -- (0,0.6) -- (0.2,0.5) -- (0,0.4) --  cycle ; 
\end{scope}
}

\newcommand{\myrobinsoncrossreflectedbis}[4]{
\begin{scope}[shift={(#1,#2)},rotate=#3]
\draw[shift={(-0.5,-0.5)},color=black,fill=#4] (0,0.1) -- (0.1,0) -- (0.4,0) -- (0.6,-0.2) -- (0.6,0) -- (0.9,0) -- (1,0.1) -- (1,0.4) -- (0.8,0.4) -- (1,0.6) -- (1,0.9) -- (0.9,1) -- (0.6,1) -- (0.6,0.8) -- (0.4,1) -- (0.1,1) -- (0,0.9) -- (0,0.6) -- (0.2,0.4) -- (0,0.4) --  cycle ; 
\draw[shift={(-0.5,-0.5)},color=bleu,thick,-,shorten >=0pt] (0.1,0.5) -- (0.9,0.5);
\draw[shift={(-0.5,-0.5)},color=bleu,thick,-,shorten >=0pt] (0.5,-0.1) -- (0.5,0.9);
\end{scope}
}

\newcommand{\myrobinsonverticalreflectedbis}[4]{
\begin{scope}[shift={(#1,#2)},rotate=#3]
\draw[shift={(-0.5,-0.5)},color=black,fill=#4] (0,0.1) -- (0.1,0) -- (0.4,0) -- (0.6,-0.2) -- (0.6,0) -- (0.9,0) -- (1,0.1) -- (1,0.4) -- (0.8,0.5) -- (1,0.6) -- (1,0.9) -- (0.9,1) -- (0.6,1) -- (0.6,0.8) -- (0.4,1) -- (0.1,1) -- (0,0.9) -- (0,0.6) -- (0.2,0.5) -- (0,0.4) --  cycle ; 
\draw[shift={(-0.5,-0.5)},color=bleu,thick,-,shorten >=0pt] (0.5,-0.1) -- (0.5,0.9);
\end{scope}
}

\newcommand{\mytRobiVbis}[2]{\vbox to 9pt{\hbox{
\begin{tikzpicture}[scale=0.3]
\robinsonverticalreflectedbis{0}{0}{#1}{#2};
\end{tikzpicture}
}}}

\newcommand{\nletterRobiH}[1]{\vbox to 3pt{\hbox{
\begin{tikzpicture}[scale=0.5,rounded corners=0pt]
\begin{scope}[rotate=#1]
\draw[]  (0,0) -- (0.4,0) -- (0.5,-0.2) -- (0.6,0) -- (1,0);
\end{scope}
 \end{tikzpicture}
}}}

\newcommand{\nletterRobiHbis}[1]{\vbox to 9pt{\hbox{
\begin{tikzpicture}[scale=0.5,rounded corners=0pt]
\begin{scope}[rotate=#1]
\draw[] (1,1) -- (0.6,1) -- (0.4,1.2) -- (0.4,1) -- (0,1) ;
\end{scope}
 \end{tikzpicture}
}}}

\newcommand{\nletterRobiHter}[1]{\vbox to 9pt{\hbox{
\begin{tikzpicture}[scale=0.5,rounded corners=0pt]
\begin{scope}[rotate=#1]
\draw[] (1,1) -- (0.6,1) -- (0.6,1.2)  -- (0.4,1) -- (0,1) ;
\end{scope}
 \end{tikzpicture}
}}}

\newcommand{\nletterRobiV}[1]{\vbox to 9pt{\hbox{
\begin{tikzpicture}[scale=0.5,rounded corners=0pt]
\begin{scope}[rotate=#1]
\draw[] (0,0) -- (0.4,0) -- (0.5,-0.2) -- (0.6,0) -- (1,0);
\end{scope}
 \end{tikzpicture}
}}}

\newcommand{\nletterRobiVbis}[1]{\vbox to 9pt{\hbox{
\begin{tikzpicture}[scale=0.5,rounded corners=0pt]
\begin{scope}[rotate=#1]
\draw[] (1,1) -- (0.6,1) -- (0.4,1.2) -- (0.4,1) -- (0,1) ;
\end{scope}
 \end{tikzpicture}
}}}

\newcommand{\nletterRobiVter}[1]{\vbox to 9pt{\hbox{
\begin{tikzpicture}[scale=0.5,rounded corners=0pt]
\begin{scope}[rotate=#1]
\draw[] (1,1) -- (0.6,1) -- (0.6,1.2)  -- (0.4,1) -- (0,1) ;
\end{scope}
 \end{tikzpicture}
}}}

\newcommand{\Hun}{\vbox to 5pt{\hbox{
\begin{tikzpicture}[scale=0.5,rounded corners=0pt]
\begin{scope}[rotate=180]
\draw[]  (0,0) -- (0.4,0) -- (0.5,-0.2) -- (0.6,0) -- (1,0);
\end{scope}
 \end{tikzpicture}
}}}

\newcommand{\Hdeux}{\vbox to 5pt{\hbox{
\begin{tikzpicture}[scale=0.5,rounded corners=0pt]
\begin{scope}[rotate=0]
\draw[] (1,1) -- (0.6,1) -- (0.4,1.2) -- (0.4,1) -- (0,1) ;
\end{scope}
 \end{tikzpicture}
}}}

\newcommand{\Htrois}{\vbox to 5pt{\hbox{
\begin{tikzpicture}[scale=0.5,rounded corners=0pt]
\begin{scope}[rotate=0]
\draw[] (1,1) -- (0.6,1) -- (0.6,1.2)  -- (0.4,1) -- (0,1) ;
\end{scope}
 \end{tikzpicture}
}}}

\newcommand{\Hquatre}{\vbox to 3pt{\hbox{
\begin{tikzpicture}[scale=0.5,rounded corners=0pt]
\begin{scope}[rotate=0]
\draw[]  (0,0) -- (0.4,0) -- (0.5,-0.2) -- (0.6,0) -- (1,0);
\end{scope}
 \end{tikzpicture}
}}}

\newcommand{\Hcinq}{\vbox to 3pt{\hbox{
\begin{tikzpicture}[scale=0.5,rounded corners=0pt]
\begin{scope}[rotate=180]
\draw[] (1,1) -- (0.6,1) -- (0.6,1.2)  -- (0.4,1) -- (0,1) ;
\end{scope}
 \end{tikzpicture}
}}}

\newcommand{\Hsix}{\vbox to 3pt{\hbox{
\begin{tikzpicture}[scale=0.5,rounded corners=0pt]
\begin{scope}[rotate=180]
\draw[] (1,1) -- (0.6,1) -- (0.4,1.2) -- (0.4,1) -- (0,1) ;
\end{scope}
 \end{tikzpicture}
}}}

\newcommand{\patternLoopRobi}[1]{\vbox to 17pt{\hbox{
\begin{tikzpicture}[scale=#1,rounded corners=0pt]
\robinsoncornerbumpybis{0}{0}{0}{black!10}
\robinsoncornerbumpybis{3}{0}{90}{black!10}
\robinsonhorizontalbis{1}{0}{0}{white}
\robinsonemptybis{3}{0}{0}{white}
 \end{tikzpicture}
}}}

\newcommand\triangledroit{\nletterRobiH{90}}
\newcommand\demitriangledroithaut{\nletterRobiVter{-90}}
\newcommand\demitriangledroitbas{\nletterRobiVbis{-90}}

\newcommand\trianglegauche{\nletterRobiH{-90}}
\newcommand\demitrianglegauchehaut{\nletterRobiHbis{90}}
\newcommand\demitrianglegauchebas{\nletterRobiHter{90}}
 
\newcommand\macrotriangledroit[1]{\pattern{\triangledroit}{#1}}
\newcommand\macrotrianglegauche[1]{\pattern{\trianglegauche}{#1}}
\newcommand\macrodemitriangledroithaut[1]{\lettertriplet{\triangledroit}{\demitriangledroithaut}{#1}}
\newcommand\macrodemitriangledroitbas[1]{\lettertriplet{\triangledroit}{\demitriangledroitbas}{#1}}
\newcommand\macrodemitrianglegauchehaut[1]{\lettertriplet{\trianglegauche}{\demitrianglegauchehaut}{#1}}
\newcommand\macrodemitrianglegauchebas[1]{\lettertriplet{\trianglegauche}{\demitrianglegauchebas}{#1}}

\newcommand\trianglebas{\nletterRobiH{0}}
\newcommand\demitrianglebasdroit{\nletterRobiVter{180}}

\newcommand\trianglehaut{\nletterRobiH{180}}
\newcommand\demitrianglehautdroit{\nletterRobiHbis{0}}

\newcommand\macrodemitrianglehautdroit[1]{\lettertriplet{\trianglehaut}{\demitrianglehautdroit}{#1}}

\newcommand{\mymacrotileone}[4]{
\begin{scope}[shift={(#1,#2)},rotate=#3] 
\myrobinsoncornerbumpybis{-1}{-1}{0}{black!10}
\myrobinsoncornerbumpybis{1}{-1}{90}{black!10}
\myrobinsoncornerbumpybis{-1}{1}{-90}{black!10}
\myrobinsoncornerbumpybis{1}{1}{180}{black!10}
\myrobinsoncornerdentedbis{0}{0}{0}{#4}
\myrobinsonhorizontalbis{0}{-1}{0}{#4}
\myrobinsonhorizontalbis{-1}{0}{-90}{#4}
\myrobinsoncrossbis{1}{0}{90}{#4}
\myrobinsoncrossreflectedbis{0}{1}{180}{#4} 
\end{scope}
}

\newcommand\leftmymacrotileone[1]{$\macrotrianglegauche{#1}$}
\newcommand\rightmymacrotileone[1]{\macrodemitriangledroithaut{#1}}
\newcommand\leftmymacrotileoneqd[1]{\macrodemitrianglegauchehaut{#1}}
\newcommand\rightmymacrotileoneqd[1]{$\macrotriangledroit{n}$}
\newcommand\leftmymacrotileonemqd[1]{$\macrotrianglegauche{#1}$}
\newcommand\rightmymacrotileonemqd[1]{\macrodemitriangledroitbas{#1}}
\newcommand\leftmymacrotileonecqv[1]{\macrodemitrianglegauchebas{#1}}
\newcommand\rightmymacrotileonecqv[1]{$\macrotriangledroit{n}$}

\newcommand\repetentimesV[6]{
\begin{scope}[shift={(#2,#3)},rotate=#4] 
\ifnum#6>1
#1{0}{1}{0}{#5}
#1{0}{-1}{0}{#5}
\fi
\ifnum#6>2
#1{0}{3}{0}{#5}
#1{0}{2}{0}{#5}
#1{0}{-2}{0}{#5}
#1{0}{-3}{0}{#5}
\fi
\ifnum#6>3
#1{0}{7}{0}{#5}
#1{0}{6}{0}{#5}
#1{0}{5}{0}{#5}
#1{0}{4}{0}{#5}
#1{0}{-4}{0}{#5}
#1{0}{-5}{0}{#5}
#1{0}{-6}{0}{#5}
#1{0}{-7}{0}{#5}
\fi
\end{scope}
}

\newcommand\repetentimesVnew[6]{
\begin{scope}[shift={(#2,#3)},rotate=#4] 
\ifnum#6=2
#1{0}{1}{0}{#5}
#1{0}{-1}{0}{#5}
\myrobinsonemptynotdotted{0}{0}{0}{#5}
\fi
\ifnum#6=25
#1{0}{1.5}{0}{#5}
#1{0}{-1.5}{0}{#5}
\myrobinsonemptynotdotted{0}{0}{0}{#5}
\draw[shift={(-0.5,+0.5)},color=black,fill=white,dotted,-] 
(1,0.5) 
-- (1,0);
\draw[shift={(-0.5,+0.5)},color=black,fill=white,dotted,-] 
(0,0.5) 
-- (0,0);
\draw[shift={(-0.5,-1)},color=black,fill=white,dotted,-] 
(1,0.5) 
-- (1,0);
\draw[shift={(-0.5,-1)},color=black,fill=white,dotted,-] 
(0,0.5) 
-- (0,0);
\fi
\ifnum#6=3
#1{0}{3}{0}{#5}
\myrobinsonemptynotdotted{0}{0}{0}{#5}
\draw[color=black,fill=white,dotted,-] 
(0.5,3) 
-- (0.5,0.5);
\draw[color=black,fill=white,dotted,-] 
(0.5,0.5) 
-- (0.5,-3);
\draw[color=black,fill=white,dotted,-] 
(-0.5,3) 
-- (-0.5,0.5);
\draw[color=black,fill=white,dotted,-] 
(-0.5,0.5) 
-- (-0.5,-3);
#1{0}{-3}{0}{#5}
\fi
\ifnum#6=4
#1{0}{7}{0}{#5}
\draw[color=black,fill=white,dotted,-] 
(0.5,7) 
-- (0.5,0.5);
\draw[color=black,fill=white,dotted,-] 
(0.5,0.5) 
-- (0.5,-7);
\draw[color=black,fill=white,dotted,-] 
(-0.5,7) 
-- (-0.5,0.5);
\draw[color=black,fill=white,dotted,-] 
(-0.5,0.5) 
-- (-0.5,-7);
\myrobinsonemptynotdotted{0}{0}{0}{#5}
#1{0}{-7}{0}{#5}
\fi
\end{scope}
}

\newcommand\macrorobinsonempty[5]{
\begin{scope}[shift={(#1,#2)},rotate=#3] 
\myrobinsonempty{0}{0}{0}{#4}
\end{scope}
\repetentimesV{\myrobinsonempty}{#1}{#2}{#3}{#4}{#5}
}

\newcommand\leftmacrorobinsonempty[1]{$\pattern{\leftrobinsonempty}{#1}$}
\newcommand\rightmacrorobinsonempty[1]{$\pattern{\rightrobinsonempty}{#1}$}
\newcommand\leftmacrorobinsonemptymqd{$\leftrobinsonemptymqd$}

\newcommand\macrorobinsonverticalreflectedbis[5]{
\begin{scope}[shift={(#1,#2)},rotate=#3] 
\myrobinsonverticalreflectedbis{0}{0}{0}{#4}
\end{scope}
\repetentimesV	{\myrobinsonverticalreflectedbis}{#1}{#2}{#3}{#4}{#5}
}

\newcommand\macrorobinsonverticalbis[5]{
\begin{scope}[shift={(#1,#2)},rotate=#3] 
\myrobinsonverticalbis{0}{0}{0}{#4}
\end{scope}
\repetentimesV	{\myrobinsonverticalbis}{#1}{#2}{#3}{#4}{#5}
}

\newcommand\macrorobinsonemptygeneric[6]{
\repetentimesVnew	{\myrobinsonempty}{#1}{#2}{#3}{#4}{#5}
\node[anchor=center,scale=0.5]  at (#1,#2) {#6}; 
}

\newcommand\macrorobinsonemptygenericnew[6]{
\repetentimesVnew	{\myrobinsonempty}{#1}{#2}{#3}{#4}{#5}
\node[anchor=center,scale=0.5]  at (#1,#2) {#6}; 
}

\newcommand\macrorobinsonverticalreflectedbisgeneric[6]{
\repetentimesVnew	{\myrobinsonverticalreflectedbis}{#1}{#2}{#3}{#4}{#5}
\node[anchor=center,scale=0.5]  at (#1,#2) {#6}; 
}

\newcommand\macrorobinsonverticalbisgeneric[6]{
\repetentimesVnew	{\myrobinsonverticalbis}{#1}{#2}{#3}{#4}{#5}
\node[anchor=center,scale=0.5]  at (#1,#2) {#6}; 
}

\newcommand{\mymacrotilen}[5]{
	\ifnum#5=1 \mymacrotileone{#1}{#2}{#3}{#4}
	\else
		\begin{scope}[shift={(#1,#2)},rotate=#3] 
		\mymacrotilen{-\deuxpuissancenmoinsun{#5}}{-\deuxpuissancenmoinsun{#5}}{0}{#4}{\minusone{#5}}
		\mymacrotilen{\deuxpuissancenmoinsun{#5}}{-\deuxpuissancenmoinsun{#5}}{90}{#4}{\minusone{#5}}
		\mymacrotilen{-\deuxpuissancenmoinsun{#5}}{\deuxpuissancenmoinsun{#5}}{-90}{#4}{\minusone{#5}}
		\mymacrotilen{\deuxpuissancenmoinsun{#5}}{\deuxpuissancenmoinsun{#5}}{180}{#4}{\minusone{#5}}
		\myrobinsoncornerdentedbis{0}{0}{0}{#4}
		\myrobinsonhorizontalbis{0}{-\deuxpuissancenmoinsun{#5}}{0}{#4}
		\myrobinsonhorizontalbis{-\deuxpuissancenmoinsun{#5}}{0}{-90}{#4}
		\myrobinsoncrossbis{\deuxpuissancenmoinsun{#5}}{0}{90}{#4}
		\myrobinsoncrossreflectedbis{0}{\deuxpuissancenmoinsun{#5}}{180}{#4}
		\macrorobinsonempty{0}{-\moitmoitun{#5}}{0}{#4}{\minusone{#5}}
		\macrorobinsonempty{0}{-\moitmoitdeux{#5}}{0}{#4}{\minusone{#5}}
		\macrorobinsonempty{-\moitmoitun{#5}}{0}{-90}{#4}{\minusone{#5}}
		\macrorobinsonempty{-\moitmoitdeux{#5}}{0}{-90}{#4}{\minusone{#5}}
		\macrorobinsonverticalreflectedbis{0}{\moitmoitun{#5}}{180}{#4}{\minusone{#5}}
		\macrorobinsonverticalreflectedbis{0}{\moitmoitdeux{#5}}{180}{#4}{\minusone{#5}}
		\macrorobinsonverticalbis{\moitmoitun{#5}}{0}{90}{#4}{\minusone{#5}}
		\macrorobinsonverticalbis{\moitmoitdeux{#5}}{0}{90}{#4}{\minusone{#5}}
		\end{scope}
	\fi
}
\newcommand{\mymacrotilengeneric}[6]{
	\mymacrotilen{#1}{#2}{#3}{#4}{#5}
	\begin{scope}[shift={(#1,#2)},rotate=#3] 
	\draw[fill,white] (-\g{#5}-.1,-\g{#5}-0.1) rectangle (\g{#5}+.1,\g{#5}+.1);
	\draw[color=bleu,thick,-,shorten >=0pt] (0,\g{#5}+0.5) -- (0,0) -- (\g{#5}+0.5,0);
	\end{scope}
	\draw[fill,white] (#1-0.5,#2-0.5) rectangle (#1+0.5,#2+0.5);
	\node[anchor=center] at (#1,#2) {#6};
}

\newcommand{\mymacrotiletwo}[4]{
	\mymacrotilen{#1}{#2}{#3}{#4}{2}
}

\newcommand{\mymacrotiletwogeneric}[5]{
	\mymacrotilengeneric{#1}{#2}{#3}{#4}{2}{#5}
}

\newcommand\lettertriplet[3]{
\tikz{
\node at (0,0) {#2};
\node[anchor=west,draw,circle,scale=0.5] at (0.25,0.25) {#3};
}
}

\newcommand\patterntriplet[3]
{
\pattern{#2}{#3}
}

\newcommand\pattern[2]{
#1^{\tikz{ \node[draw,circle,scale=0.5,xshift=-0.1] {#2};}}
}

\newcommand\patternplusone[2]{
\tikz{
\node at (0,0) {
$#1^{\tikz{ \node[draw,circle,scale=0.5,xshift=-0.1] {#2};}}$};
\node at (0,1) {#1};
}}

\newcommand\rightatrightmacrotiletwo[1]{\lettertriplet{\rightrobinsonempty}{\rightrobinsonhorizontalbis}{#1}}

\newcommand\macrorobinsonhorizontalgenericnew[6]{
\begin{scope}[shift={(#1,#2)},rotate=#3] 
\macrorobinsonemptygeneric{0}{2.5}{0}{#4}{#5}{#6}
\myrobinsonhorizontalbis{0}{0}{0}{#4}
\macrorobinsonemptygeneric{0}{-2.5}{0}{#4}{#5}{#6}

\end{scope}
}

\newcommand\atrightmacrotilengeneric[6]{
\begin{scope}[shift={(#1,#2)},rotate=#3] 
\macrorobinsonemptygeneric{0}{\deuxpuissancenmoinsun{#5}}{0}{#4}{#5}{#6}
\myrobinsonhorizontalbis{0}{0}{0}{#4}
\macrorobinsonemptygeneric{0}{-\deuxpuissancenmoinsun{#5}}{0}{#4}{#5}{#6}
\end{scope}
}

\newcommand\atrightmacrotiletwogenericnew[5]{
\begin{scope}[shift={(#1,#2)},rotate=#3] 
\macrorobinsonemptygenericnew{0}{2.5}{0}{#4}{25}{#5}
\myrobinsonhorizontalbis{0}{0}{0}{#4}
\macrorobinsonemptygenericnew{0}{-2.5}{0}{#4}{25}{#5}
\end{scope}
}

\newcommand\macrorobinsoncrossbisgenericnew[6]{
\begin{scope}[shift={(#1,#2)},rotate=#3] 
\macrorobinsonverticalreflectedbisgeneric{0}{2.5}{0}{#4}{#5}{#6}
\myrobinsoncrossreflectedbis{0}{0}{0}{#4}
\macrorobinsonverticalreflectedbisgeneric{0}{-2.5}{0}{#4}{#5}{#6}
\end{scope}
}

\newcommand\rightVNEmacrotiletwo[1]{\lettertriplet{\rightrobinsonverticalreflectedbis}{\rightrobinsoncrossreflectedbiscqv}{#1}}

\newcommand\VNEmacrotiletwogenericnew[5]{
\begin{scope}[shift={(#1,#2)},rotate=#3] 
\macrorobinsonverticalreflectedbisgeneric{0}{-2.5}{180}{#4}{25}{#5}
\myrobinsoncrossreflectedbis{0}{0}{180}{black!10}
\macrorobinsonverticalreflectedbisgeneric{0}{2.5}{180}{#4}{25}{#5}
\end{scope}
}

\newcommand\VNEmacrotilengeneric[6]{
\begin{scope}[shift={(#1,#2)},rotate=#3] 
\macrorobinsonverticalreflectedbisgeneric{0}{-\deuxpuissancenmoinsun{#5}}{180}{#4}{#5}{#6}
\myrobinsoncrossreflectedbis{0}{0}{180}{black!10}
\macrorobinsonverticalreflectedbisgeneric{0}{\deuxpuissancenmoinsun{#5}}{180}{#4}{#5}{#6}
\end{scope}
}

\newcommand\topmacrotiletwogenericnew[5]{
\begin{scope}[shift={(#1,#2)},rotate=#3] 
\macrorobinsonemptygeneric{2.5}{0}{-90}{#4}{25}{#5}
\myrobinsonhorizontalbis{0}{0}{-90}{black!10}
\macrorobinsonemptygeneric{-2.5}{0}{-90}{#4}{25}{#5}
\end{scope}
}

\newcommand\macrorobinsoncrossgenericnew[6]{
\begin{scope}[shift={(#1,#2)},rotate=#3] 
\macrorobinsonverticalbisgeneric{0}{2.5}{0}{#4}{#5}{#6}
\myrobinsoncrossbis{0}{0}{0}{#4}
\macrorobinsonverticalbisgeneric{0}{-2.5}{0}{#4}{#5}{#6}
\end{scope}
}

\newcommand\HNEmacrotiletwogenericnew[5]{
\begin{scope}[shift={(#1,#2)},rotate=#3] 
\macrorobinsonverticalbisgeneric{-2.5}{0}{90}{#4}{25}{#5}
\myrobinsoncrossbis{0}{0}{90}{#4}
\macrorobinsonverticalbisgeneric{2.5}{0}{90}{#4}{25}{#5}
\end{scope}
}

\newcommand\letterRobidentrant{\trianglegauche}

\newcommand\lsortant{\nletterRobiH{0}}
\newcommand\lentrant{\nletterRobiV{180}}
\newcommand\lentrantdemigauche{\nletterRobiVter{0}}
\newcommand\lsortantdemigauche{\nletterRobiHbis{180}}
\newcommand\lentrantdemidroit{\nletterRobiVbis{0}}
\newcommand\lsortantdemidroit{\nletterRobiHter{180}}

\newcommand\leftrobinsonempty{\nletterRobiV{90}}
\newcommand\rightrobinsonempty{\letterRobidentrant}
\newcommand\leftrobinsonhorizontalbis{\demitriangledroithaut}
\newcommand\rightrobinsonhorizontalbis{\demitrianglegauchehaut}

\newcommand\leftrobinsoncrossbisqd{\demitriangledroithaut}

\newcommand\leftrobinsonverticalbisqd{\demitriangledroithaut}

\newcommand\rightrobinsonverticalbisqd{\demitriangledroithaut}
\newcommand\leftrobinsonemptyqd{\triangledroit}
\newcommand\leftrobinsonemptymqd{\trianglegauche}

\newcommand\rightrobinsonverticalreflectedbis{\trianglegauche}

\newcommand\rightrobinsoncornerdentedbis{\demitriangledroithaut}

\newcommand\leftrobinsonhorizontalbismq{\trianglegauche}

\newcommand\leftrobinsoncornerdentedbis{\trianglegauche}

\newcommand\rightrobinsoncrossreflectedbiscqv{\demitrianglegauchebas}

\newcommand\leftplusonemacrorobinsonempty[1]{$\patternplusone{\leftrobinsonempty}{#1}$} 
\newcommand\rightplusonemacrorobinsonempty[1]{$\patternplusone{\letterRobidentrant}{#1}$}

\def\facteurreduction{0.4}

\newcommand\Derivationd[6]{
\begin{tikzpicture}[baseline={(current bounding box.center)},scale=2,shorten >=1pt,>={Stealth[round]}]
\node[] at (-1,0) {\scalebox{1}{$\mathcal{T}^{#1}_{Robi} \Smodels$}}; 
\node[state,rectangle,rounded corners=3pt] (s) {#3};
\node[state,rectangle,rounded corners=3pt] at (2,0) (t) {#4};
\path[->] (s) edge[above] node[-]{\begin{tabular}{c} \scalebox{\facteurreduction}{#2}\\ $\begin{array}{l} #6\\#5 \end{array}$ \end{tabular}}
 (t);
\end{tikzpicture}
}

\newcommand\Derivationdd[6]{
\begin{tikzpicture}[baseline={(current bounding box.center)},scale=2,shorten >=1pt,>={Stealth[round]}]
\node[state,rectangle,rounded corners=3pt] (s) {#3};
\node[state,rectangle,rounded corners=3pt] at (2,0) (t) {#4};
\path[->] (s) edge[above] node[-]{\begin{tabular}{c} \scalebox{\facteurreduction}{#2}\\ $\begin{array}{l} #6\\#5 \end{array}$ \end{tabular}} (t);
\end{tikzpicture}
}

\newcommand\Boucled[5]{
\begin{tikzpicture}[baseline={(current bounding box.center)},scale=2,shorten >=1pt,>={Stealth[round]}]
\node[] at (-1,0) {\scalebox{1}{$\mathcal{T}^{#1} \Smodels$}}; 
\node[state,rectangle,rounded corners=3pt] (s) {#3};
\path  (s) edge[above,loop,-] node[-]{\begin{tabular}{c} \scalebox{\facteurreduction}{#2}\\ $\begin{array}{l} #5\\#4 \end{array}$ \end{tabular}} 
 ();
\end{tikzpicture}
}

\newcommand\Boucledd[5]{
\begin{tikzpicture}[baseline={(current bounding box.center)},scale=2,shorten >=1pt,>={Stealth[round]}]
\node[state,rectangle,rounded corners=3pt] (s) {#3};
\path  (s) edge[above,loop,-] node[-] {\begin{tabular}{c} \scalebox{\facteurreduction}{#2}\\ $\begin{array}{l} #5\\#4 \end{array}$ \end{tabular}} 
();
\end{tikzpicture}
}

\newcommand\myequation[2]{
\begin{equation}\label{#1}\text{
#2
}\end{equation}
}

\newcommand\ligne[4]{
#4{#1}{-2}
#3{#1}{-1}
#2{#1}{0}
}

\newcommand\tunun[2]{\myrobinsoncornerbumpybis{#1}{#2}{0}{white}}
\newcommand\tdeuxun[2]{\myrobinsoncornerdentedbis{#1}{#2}{0}{white}}
\newcommand\ttroisun[2]{\myrobinsoncrossbis{#1}{#2}{0}{white}}
\newcommand\tquatreun[2]{\myrobinsoncrossreflectedbis{#1}{#2}{0}{white}}
\newcommand\tcinqun[2]{\myrobinsonhorizontalbis{#1}{#2}{90}{white}}
\newcommand\tsixun[2]{\myrobinsonverticalbis{#1}{#2}{0}{white}}
\newcommand\tseptun[2]{\myrobinsonverticalreflectedbis{#1}{#2}{0}{white}}
\newcommand\thuitun[2]{\myrobinsonemptybis{#1}{#2}{0}{white}}

\newcommand\tundeux[2]{\myrobinsoncornerbumpybis{#1}{#2}{-90}{white}}
\newcommand\tdeuxdeux[2]{\myrobinsoncornerdentedbis{#1}{#2}{-90}{white}}
\newcommand\ttroisdeux[2]{\myrobinsoncrossbis{#1}{#2}{-90}{white}}
\newcommand\tquatredeux[2]{\myrobinsoncrossreflectedbis{#1}{#2}{-90}{white}}
\newcommand\tcinqdeux[2]{\myrobinsonhorizontalbis{#1}{#2}{0}{white}}
\newcommand\tsixdeux[2]{\myrobinsonverticalbis{#1}{#2}{-90}{white}}

\newcommand\thuitdeux[2]{\myrobinsonemptybis{#1}{#2}{-90}{white}}

\newcommand\tuntrois[2]{\myrobinsoncornerbumpybis{#1}{#2}{90}{white}}
\newcommand\tdeuxtrois[2]{\myrobinsoncornerdentedbis{#1}{#2}{90}{white}}
\newcommand\ttroistrois[2]{\myrobinsoncrossbis{#1}{#2}{90}{white}}
\newcommand\tquatretrois[2]{\myrobinsoncrossreflectedbis{#1}{#2}{90}{white}}
\newcommand\tcinqtrois[2]{\myrobinsonhorizontalbis{#1}{#2}{180}{white}}
\newcommand\tsixtrois[2]{\myrobinsonverticalbis{#1}{#2}{90}{white}}
\newcommand\tsepttrois[2]{\myrobinsonverticalreflectedbis{#1}{#2}{90}{white}}
\newcommand\thuittrois[2]{\myrobinsonemptybis{#1}{#2}{90}{white}}

\newcommand\tunquatre[2]{\myrobinsoncornerbumpybis{#1}{#2}{180}{white}}
\newcommand\tdeuxquatre[2]{\myrobinsoncornerdentedbis{#1}{#2}{180}{white}}
\newcommand\ttroisquatre[2]{\myrobinsoncrossbis{#1}{#2}{180}{white}}
\newcommand\tquatrequatre[2]{\myrobinsoncrossreflectedbis{#1}{#2}{180}{white}}
\newcommand\tcinqquatre[2]{\myrobinsonhorizontalbis{#1}{#2}{270}{white}}
\newcommand\tsixquatre[2]{\myrobinsonverticalbis{#1}{#2}{180}{white}}
\newcommand\tseptquatre[2]{\myrobinsonverticalreflectedbis{#1}{#2}{180}{white}}
\newcommand\thuitquatre[2]{\myrobinsonemptybis{#1}{#2}{180}{white}}

\newcommand\EXEMPLE{
\ligne{0}{\tuntrois}{\tsixdeux}{\tunquatre}
\ligne{1}{\thuitquatre}{\tdeuxquatre}{\tseptun}
\ligne{2}{\tunun}{\thuittrois}{\tundeux}
\ligne{3}{\tsixtrois}{\thuitquatre}{\tcinqtrois}
\ligne{4}{\tsixtrois}{\tundeux}{\ttroisdeux}
\ligne{5}{\tsixtrois}{\tcinqtrois}{\tdeuxquatre}
\ligne{6}{\tsixtrois}{\tunquatre}{\tcinqun}
\ligne{7}{\tsixtrois}{\thuitquatre}{\thuitquatre}
\ligne{8}{\tsixtrois}{\tundeux}{\tcinqquatre}
\ligne{9}{\tsixtrois}{\tcinqtrois}{\tdeuxdeux}
\ligne{10}{\tsixtrois}{\tunquatre}{\tquatretrois}
\ligne{11}{\tcinqdeux}{\thuittrois}{\tcinqtrois}
\ligne{12}{\tuntrois}{\thuittrois}{\tunquatre}
\ligne{13}{\thuitquatre}{\thuitquatre}{\thuitquatre}
\ligne{14}{\tcinqquatre}{\tunun}{\thuitdeux}
\ligne{15}{\tdeuxdeux}{\ttroisun}{\tcinqquatre}
\ligne{16}{\ttroisquatre}{\tuntrois}{\thuitdeux}
\ligne{17}{\tsixdeux}{\thuitquatre}{\tdeuxdeux}
\ligne{18}{\ttroisdeux}{\tunun}{\tsepttrois}
\ligne{19}{\tdeuxquatre}{\tquatreun}{\tquatretrois}
\ligne{20}{\tsixquatre}{\tuntrois}{\tsepttrois}
\ligne{21}{\tcinqquatre}{\tseptquatre}{\tquatrequatre}
\ligne{22}{\tcinqquatre}{\tunun}{\tsixdeux}
\ligne{23}{\tdeuxdeux}{\ttroisun}{\ttroisdeux}
\ligne{24}{\ttroisquatre}{\tuntrois}{\tsixdeux}
\ligne{25}{\tsixdeux}{\thuitquatre}{\tdeuxquatre}
\ligne{26}{\tsixdeux}{\tundeux}{\tsixun}
\ligne{27}{\tsixdeux}{\tcinqtrois}{\tdeuxdeux}
\ligne{28}{\tsixdeux}{\tunquatre}{\tquatretrois}
\ligne{29}{\tdeuxquatre}{\tseptun}{\tquatretrois}
\ligne{30}{\tseptquatre}{\tunun}{\tsepttrois}
\ligne{31}{\tdeuxun}{\tcinqdeux}{\tsepttrois}
\ligne{32}{\ttroistrois}{\tuntrois}{\tsepttrois}
\ligne{33}{\tcinqdeux}{\thuitun}{\tsepttrois}
\ligne{34}{\tquatredeux}{\tunun}{\tsepttrois}
\ligne{35}{\tdeuxtrois}{\tcinqdeux}{\tsepttrois}
\ligne{36}{\tcinqun}{\tuntrois}{\tsepttrois}
\ligne{37}{\thuitquatre}{\thuitquatre}{\tcinqtrois}
\ligne{38}{\thuitdeux}{\tundeux}{\ttroisdeux}
\ligne{39}{\thuitdeux}{\tcinqtrois}{\tdeuxquatre}
\ligne{40}{\thuitdeux}{\tunquatre}{\tcinqun}
\ligne{41}{\tcinqquatre}{\tseptquatre}{\tseptquatre}
\ligne{42}{\thuitdeux}{\tundeux}{\tcinqquatre}
\ligne{43}{\tcinqquatre}{\tquatrequatre}{\tdeuxun}
\ligne{44}{\thuitdeux}{\tunquatre}{\tquatreun}
\ligne{45}{\tdeuxdeux}{\tsixun}{\tquatredeux}
\ligne{46}{\tsepttrois}{\tundeux}{\tquatredeux}
}

\providecommand\EPURE[1]{}

\newcommand\NOCOLOR[1]{
#1
}
\newcommand\BW[1]{
}

\renewcommand\gg[1]{
  \ifnum#1=0	1
  \else \ifnum#1=1 2
  \else \ifnum#1=2 3
  \else \ifnum#1=3  5
  \else \ifnum#1=4 8
  \else \ifnum#1=5 13
  \else \ifnum#1=6 21
  \else \ifnum#1=7 34
   \else \ifnum#1=8 55
  \fi \fi \fi \fi \fi \fi \fi \fi
 }
 
  \newcommand\plusone[1]{%
  \ifnum#1=0	1%
  \else  \ifnum#1=1	2%
  \else \ifnum#1=2 3%
  \else \ifnum#1=3 4%
  \else \ifnum#1=4 5%
  \else \ifnum#1=5 6%
  \else \ifnum#1=6 7%
 \fi\fi\fi\fi\fi\fi\fi%
 }

\newcommand\plustwo[1]{%
  \ifnum#1=0	2%
  \else  \ifnum#1=1	3%
  \else \ifnum#1=2 4%
  \else \ifnum#1=3 5%
  \else \ifnum#1=4 6%
  \else \ifnum#1=5 7%
  \else \ifnum#1=6 8%
 \fi\fi\fi\fi\fi\fi\fi%
 }

\newcommand\plusthree[1]{%
  \ifnum#1=0	3%
  \else  \ifnum#1=1	4%
  \else \ifnum#1=2 5%
  \else \ifnum#1=3 6%
  \else \ifnum#1=4 7%
  \else \ifnum#1=5 8%
  \else \ifnum#1=6 9%
 \fi\fi\fi\fi\fi\fi\fi%
 }
 
\newcommand\plusfour[1]{%
  \ifnum#1=0	4%
  \else  \ifnum#1=1	5%
  \else \ifnum#1=2 6%
  \else \ifnum#1=3 7%
  \else \ifnum#1=4 8%
  \else \ifnum#1=5 9%
  \else \ifnum#1=6 10%
 \fi\fi\fi\fi\fi\fi\fi%
 }
 
 \def\ep{0.5}

\newcommand\lalpha[1]
{\gg{\plustwo{#1}}-3}
\newcommand\halpha[1]
{1}

\newcommand\gralpha[3]{
\EPURE{\node at (#1,#2) {$\alpha$};}
\draw (#1,#2) rectangle (#1+\lalpha{#3},#2+\halpha{#3});
}

\newcommand\lbeta[1]
{\gg{\plusone{#1}}+3}
\newcommand\hbeta[1]
{1}

\newcommand\grbeta[3]{
\EPURE{\node at (#1,#2) {$\beta$};}
\draw (#1,#2) rectangle (#1+\lbeta{#3},#2+\hbeta{#3});
\NOCOLOR{\draw[fill,color=red] (#1,#2+\hbeta{#3}-\ep) rectangle (#1+3,#2+\hbeta{#3});}
\BW            {\draw[fill,color=black] (#1,#2+\hbeta{#3}-\ep) rectangle (#1+3,#2+\hbeta{#3});}
}

\newcommand\lgamma[1]
{\gg{\plusthree{#1}}+3}
\newcommand\hgamma[1]
{1}

\newcommand\grgamma[3]{
\EPURE{\node at (#1,#2) {$\gamma$};}
\draw (#1,#2) rectangle (#1+\lgamma{#3},#2+\hgamma{#3});
\NOCOLOR{\draw[fill,color=yellow] (#1+\gg{\plustwo{#3}},#2+\hgamma{#3}-\ep) rectangle (#1+\gg{\plustwo{#3}}+3,#2+\hgamma{#3});}
\BW          {\draw[fill,color=black] (#1+\gg{\plustwo{#3}},#2+\hgamma{#3}-\ep) rectangle (#1+\gg{\plustwo{#3}}+3,#2+\hgamma{#3});}
}

\newcommand\ldelta[1]
{\gg{\plusthree{#1}}+3}
\newcommand\hdelta[1]
{1}

\newcommand\grdelta[3]{
\EPURE{\node at (#1,#2) {$\delta$};}
\draw (#1,#2) rectangle (#1+\ldelta{#3},#2+\hdelta{#3});
\NOCOLOR{\draw[fill,color=blue] (#1+\gg{\plusone{#3}},#2) rectangle (#1+\gg{\plusone{#3}}+3,#2+\ep);}
\BW{\draw[fill,color=black] (#1+\gg{\plusone{#3}},#2) rectangle (#1+\gg{\plusone{#3}}+3,#2+\ep);}

}

\newcommand\lepsilon[1]
{\gg{\plusone{#1}}+3}
\newcommand\hepsilon[1]
{1}

\newcommand\grepsilon[3]{
\EPURE{\node at (#1,#2) {$\epsilon$};}
\draw (#1,#2) rectangle (#1+\lepsilon{#3},#2+\hepsilon{#3});
\NOCOLOR{\draw[fill,color=black] (#1+\gg{\plusone{#3}},#2) rectangle (#1+\gg{\plusone{#3}}+3,#2+\ep);}
\BW{\draw[fill,color=black] (#1+\gg{\plusone{#3}},#2) rectangle (#1+\gg{\plusone{#3}}+3,#2+\ep);}
}

\newcommand\lA[1]
{\gg{\plustwo{#1}}-3}
\newcommand\hA[1]
{1}

\newcommand\lB[1]
{\gg{\plusone{#1}}+3}
\newcommand\hB[1]
{1}

\newcommand\lC[1]
{\gg{\plusthree{#1}}+3}
\newcommand\hC[1]
{1}

\newcommand\lD[1]
{\gg{\plusone{#1}}+3+\gg{\plustwo{#1}}}
\newcommand\hD[1]
{1}

\newcommand\lE[1]
{\gg{\plusone{#1}}+3}
\newcommand\hE[1]
{1}

\newcommand\grA[3]{
\EPURE{\node at (#1,#2) {A};}
\draw (#1,#2) rectangle (#1+\lA{#3},#2+\hA{#3});
}
\newcommand\grB[3]{
\EPURE{\node at (#1,#2) {B};}
\draw (#1,#2) rectangle (#1+\lB{#3},#2+\hB{#3});
\NOCOLOR{\draw[fill,color=black] (#1,#2+\hB{#3}-\ep) rectangle (#1+3,#2+\hB{#3});}
\BW{\draw[fill,color=black] (#1,#2+\hB{#3}-\ep) rectangle (#1+3,#2+\hB{#3});}
}

\newcommand\grC[3]{
\EPURE{\node at (#1,#2) {C};}
\draw (#1,#2) rectangle (#1+\lC{#3},#2+\hC{#3});
\NOCOLOR{\draw[fill,color=blue] (#1+\gg{\plustwo{#3}},#2+\hC{#3}-\ep) rectangle (#1+\gg{\plustwo{#3}}+3,#2+\hC{#3});}
\BW{\draw[fill,color=black] (#1+\gg{\plustwo{#3}},#2+\hC{#3}-\ep) rectangle (#1+\gg{\plustwo{#3}}+3,#2+\hC{#3});}
}

\newcommand\grD[3]{
\EPURE{\node at (#1,#2) {D};}
\draw (#1,#2) rectangle (#1+\lD{#3},#2+\hD{#3});
\NOCOLOR{\draw[fill,color=yellow] (#1+\gg{\plusone{#3}},#2) rectangle (#1+\gg{\plusone{#3}}+3,#2+\ep);}
\BW{\draw[fill,color=black] (#1+\gg{\plusone{#3}},#2) rectangle (#1+\gg{\plusone{#3}}+3,#2+\ep);}
}

\newcommand\grE[3]{
\EPURE{\node at (#1,#2) {E};}
\draw (#1,#2) rectangle (#1+\lE{#3},#2+\hE{#3});
\NOCOLOR{\draw[fill,color=red] (#1+\gg{\plusone{#3}},#2) rectangle (#1+\gg{\plusone{#3}}+3,#2+\ep);}
\BW          {\draw[fill,color=black] (#1+\gg{\plusone{#3}},#2) rectangle (#1+\gg{\plusone{#3}}+3,#2+\ep);}
}

\newcommand\gralphadeux[3]{
\EPURE{\node at (#1,#2) {$\alpha\alpha$};}
\edef\npu{\plusone{#3}}
\gralpha{#1}{#2}{\npu}
\grgamma{#1+\lalpha{\npu}}{#2}{\npu}
\edef\hc{\halpha{\npu}}
\grC{#1-3}{#2+\hc}{#3}
\grA{#1-3+\lC{#3}}{#2+\hc}{#3}
\grD{#1-3+\lC{#3}+\lA{#3}}{#2+\hc}{#3}
\edef\hhc{\halpha{\npu}+\hA{#3}}
\grdelta{#1-3}{#2+\hhc}{\npu}
\gralpha{#1-3+\lD{\npu}}{\hhc}{\npu}
}

\newcommand\grbetadeux[3]{
\EPURE{\node at (#1,#2) {$\beta\beta$};}
\edef\npu{\plusone{#3}}
\gralpha{#1}{#2}{\npu}
\grbeta{#1+\lalpha{\npu}}{#2}{\npu}
\edef\hc{\halpha{\npu}}
\grA{#1}{#2+\hc}{#3}
\grE{#1+\lA{#3}}{#2+\hc}{#3}
\grA{#1+\lA{#3}+\lE{#3}}{#2+\hc}{#3}
\edef\hhc{\halpha{\npu}+\hA{#3}}
\grbeta{#1-3}{#2+\hhc}{\npu}
\gralpha{#1-3+\lbeta{\npu}}{#2+\hhc}{\npu}
}

\newcommand\grepsilondeux[3]{
\EPURE{\node at (#1,#2) {$\epsilon\epsilon$};}
\edef\npu{\plusone{#3}}
\gralpha{#1}{#2}{\npu}
\grepsilon{#1+\lalpha{\npu}}{#2}{\npu}
\edef\hc{\halpha{\npu}}
\grA{#1}{#2+\hc}{#3}
\grB{#1+\lA{#3}}{#2+\hc}{#3}
\grA{#1+\lA{#3}+\lB{#3}}{#2+\hc}{#3}
\edef\hhc{\halpha{\npu}+\hA{#3}}
\grepsilon{#1-3}{#2+\hhc}{\npu}
\gralpha{#1-3+\lepsilon{\npu}}{#2+\hhc}{\npu}
}

\newcommand\grgammadeux[3]{
\EPURE{\node at (#1,#2) {$\gamma\gamma$};}
\edef\npu{\plusone{#3}}
\gralpha{#1}{#2}{\npu}
\grgamma{#1+\lalpha{\npu}}{#2}{\npu}
\gralpha{#1+\lalpha{\npu}+\lgamma{\npu}}{#2}{\npu}
\grbeta{#1+\lalpha{\npu}+\lgamma{\npu}+\lalpha{\npu}}{#2}{\npu}
\edef\hc{\hA{\npu}}
\grA{#1}{#2+\hc}{#3}
\grB{#1+\lA{#3}}{#2+\hc}{#3}
\grA{#1+\lA{#3}+\lB{#3}}{#2+\hc}{#3}
\grD{#1+\lA{#3}+\lB{#3}+\lA{#3}}{#2+\hc}{#3}
\grA{#1+\lA{#3}+\lB{#3}+\lA{#3}+\lD{#3}}{#2+\hc}{#3}
\grE{#1+\lA{#3}+\lB{#3}+\lA{#3}+\lD{#3}+\lA{#3}}{#2+\hc}{#3}
\grA{#1+\lA{#3}+\lB{#3}+\lA{#3}+\lD{#3}+\lA{#3}+\lE{#3}}{#2+\hc}{#3}
\edef\hhc{\halpha{\npu}+\hA{#3}}
\grepsilon{#1-3}{#2+\hhc}{\npu}
\gralpha{#1-3+\lepsilon{\npu}}{#2+\hhc}{\npu}
\grgamma{#1-3+\lepsilon{\npu}+\lalpha{\npu}}{#2+\hhc}{\npu}
\gralpha{#1-3+\lepsilon{\npu}+\lalpha{\npu}+\lgamma{\npu}}{#2+\hhc}{\npu}
}

\newcommand\grdeltadeux[3]{
\EPURE{\node at (#1,#2) {$\delta\delta$};}
\edef\npu{\plusone{#3}}
\gralpha{#1}{#2}{\npu}
\grdelta{#1+\lalpha{\npu}}{#2}{\npu}
\gralpha{#1+\lalpha{\npu}+\ldelta{\npu}}{#2}{\npu}
\grbeta{#1+\lalpha{\npu}+\ldelta{\npu}+\lalpha{\npu}}{#2}{\npu}
\edef\hc{\halpha{\npu}}
\grA{#1}{#2+\hc}{#3}
\grB{#1+\lA{#3}}{#2+\hc}{#3}
\grA{#1+\lA{#3}+\lB{#3}}{#2+\hc}{#3}
\grC{#1+\lA{#3}+\lB{#3}+\lA{#3}}{#2+\hc}{#3}
\grA{#1+\lA{#3}+\lB{#3}+\lA{#3}+\lC{#3}}{#2+\hc}{#3}
\grE{#1+\lA{#3}+\lB{#3}+\lA{#3}+\lC{#3}+\lA{#3}}{#2+\hc}{#3}
\grA{#1+\lA{#3}+\lB{#3}+\lA{#3}+\lC{#3}+\lA{#3}+\lE{#3}}{#2+\hc}{#3}
\edef\hhc{\halpha{\npu}+\hA{#3}}
\grepsilon{#1-3}{#2+\hhc}{\npu}
\gralpha{#1-3+\lepsilon{\npu}}{#2+\hhc}{\npu}
\grdelta{#1-3+\lepsilon{\npu}+\lalpha{\npu}}{#2+\hhc}{\npu}
\gralpha{#1-3+\lepsilon{\npu}+\lalpha{\npu}+\ldelta{\npu}}{#2+\hhc}{\npu}
}

\newcommand\hAdeux[1]{
\edef\npu{\plusone{#1}}
\hA{\npu}+\hgamma{#1}+\hD{\npu}
}

\newcommand\grAdeux[3]{
\EPURE{\node at (#1,#2) {AA};}
\edef\npu{\plusone{#3}}
\grA{#1}{#2}{\npu}
\grC{#1+\lA{\npu}}{#2}{\npu}
\edef\hc{\hA{\npu}}
\grgamma{#1-3}{#2+\hc}{#3}
\gralpha{#1-3+\lgamma{#3}}{#2+\hc}{#3}
\grdelta{#1-3+\lgamma{#3}+\lalpha{#3}}{#2+\hc}{#3}
\edef\hhc{\hA{\npu}+\halpha{#3}}
\grD{#1-3}{#2+\hhc}{\npu}
\grA{#1-3+\lD{\npu}}{#2+\hhc}{\npu}
}

\newcommand\grBdeux[3]{
\EPURE{\node at (#1,#2) {BB};}
\edef\npu{\plusone{#3}}
\grA{#1}{#2}{\npu}
\grB{#1+\lA{\npu}}{#2}{\npu}
\edef\hc{\hA{\npu}}
\gralpha{#1}{#2+\hc}{#3}
\grepsilon{#1+\lalpha{#3}}{#2+\hc}{#3}
\gralpha{#1+\lalpha{#3}+\lepsilon{#3}}{#2+\hc}{#3}
\edef\hhc{\hA{\npu}+\halpha{#3}}
\grB{#1-3}{#2+\hhc}{\npu}
\grA{#1-3+\lB{\npu}}{#2+\hhc}{\npu}
}

\newcommand\grEdeux[3]{
\EPURE{\node at (#1,#2) {EE};}
\edef\npu{\plusone{#3}}
\grA{#1}{#2}{\npu}
\grE{#1+\lA{\npu}}{#2}{\npu}
\edef\hc{\hA{\npu}}
\gralpha{#1}{#2+\hc}{#3}
\grbeta{#1+\lalpha{#3}}{#2+\hc}{#3}
\gralpha{#1+\lalpha{#3}+\lbeta{#3}}{#2+\hc}{#3}
\edef\hhc{\hA{\npu}+\halpha{#3}}
\grE{#1-3}{#2+\hhc}{\npu}
\grA{#1-3+\lE{\npu}}{#2+\hhc}{\npu}
}

\newcommand\grCdeux[3]{
\EPURE{\node at (#1,#2) {CC};}
\edef\npu{\plusone{#3}}
\grA{#1}{#2}{\npu}
\grC{#1+\lA{\npu}}{#2}{\npu}
\grA{#1+\lA{\npu}+\lC{\npu}}{#2}{\npu}
\grB{#1+\lA{\npu}+\lC{\npu}+\lA{\npu}}{#2}{\npu}
\edef\hc{\hA{\npu}}
\gralpha{#1}{#2+\hc}{#3}
\grbeta{#1+\lalpha{#3}}{#2+\hc}{#3}
\gralpha{#1+\lalpha{#3}+\lbeta{#3}}{#2+\hc}{#3}
\grdelta{#1+\lalpha{#3}+\lbeta{#3}+\lalpha{#3}}{#2+\hc}{#3}
\gralpha{#1+\lalpha{#3}+\lbeta{#3}+\lalpha{#3}+\ldelta{#3}}{#2+\hc}{#3}
\grepsilon{#1+\lalpha{#3}+\lbeta{#3}+\lalpha{#3}+\ldelta{#3}+\lalpha{#3}}{#2+\hc}{#3}
\gralpha{#1+\lalpha{#3}+\lbeta{#3}+\lalpha{#3}+\ldelta{#3}+\lalpha{#3}+\lepsilon{#3}}{#2+\hc}{#3}
\edef\hhc{\hA{\npu}+\halpha{#3}}
\grE{#1-3}{#2+\hhc}{\npu}
\grA{#1-3+\lE{\npu}}{#2+\hhc}{\npu}
\grC{#1-3+\lE{\npu}+\lA{\npu}}{#2+\hhc}{\npu}
\grA{#1-3+\lE{\npu}+\lA{\npu}+\lC{\npu}}{#2+\hhc}{\npu}
}

\newcommand\grDdeux[3]{
\EPURE{\node at (#1,#2) {DD};}
\edef\npu{\plusone{#3}}
\grA{#1}{#2}{\npu}
\grD{#1+\lA{\npu}}{#2}{\npu}
\grA{#1+\lA{\npu}+\lD{\npu}}{#2}{\npu}
\grB{#1+\lA{\npu}+\lD{\npu}+\lA{\npu}}{#2}{\npu}
\edef\hc{\hA{\npu}}
\gralpha{#1}{#2+\hc}{#3}
\grbeta{#1+\lalpha{#3}}{#2+\hc}{#3}
\gralpha{#1+\lalpha{#3}+\lbeta{#3}}{#2+\hc}{#3}
\grgamma{#1+\lalpha{#3}+\lbeta{#3}+\lalpha{#3}}{#2+\hc}{#3}
\gralpha{#1+\lalpha{#3}+\lbeta{#3}+\lalpha{#3}+\lgamma{#3}}{#2+\hc}{#3}
\grepsilon{#1+\lalpha{#3}+\lbeta{#3}+\lalpha{#3}+\lgamma{#3}+\lalpha{#3}}{#2+\hc}{#3}
\gralpha{#1+\lalpha{#3}+\lbeta{#3}+\lalpha{#3}+\lgamma{#3}+\lalpha{#3}+\lepsilon{#3}}{#2+\hc}{#3}
\edef\hhc{\hA{\npu}+\halpha{#3}}
\grE{#1-3}{#2+\hhc}{\npu}
\grA{#1-3+\lE{\npu}}{#2+\hhc}{\npu}
\grD{#1-3+\lE{\npu}+\lA{\npu}}{#2+\hhc}{\npu}
\grA{#1-3+\lE{\npu}+\lA{\npu}+\lD{\npu}}{#2+\hhc}{\npu}
}

\def\v{0}
\def\vv{0}

\newcommand\grAtrois[3]{
\EPURE{\node at (#1,#2) {AAA};}
\edef\npu{\plusone{#3}}
\grAdeux{#1}{#2}{#3}
\grCdeux{#1+13+\v}{#2}{#3}
\edef\hc{\hAdeux{\npu}}
\grgamma{#1-3}{#2+3+\vv}{2}
\gralpha{#1+13+\v}{#2+3+\vv}{2}
\grdelta{#1+18+\v+\v}{#2+3+\vv}{2}
\grDdeux{#1}{#2+4+\vv+\vv}{#3}
\grAdeux{#1+21+\v}{#2+4+\vv+\vv}{#3}
}

\usepackage{ifpdf}

\NOLICSSTYLE{
\ifCLASSINFOpdf
\else
\fi
}

\usepackage{amsmath}
\DeclareMathOperator{\lcm}{lcm}

\usepackage{amsthm}
\usepackage{amsfonts}
\usepackage{amssymb}
\usepackage{marvosym}

\usepackage{graphicx}

\makeatletter
\newcommand*{\shifttext}[2]{%
	\settowidth{\@tempdima}{#2}%
	\makebox[\@tempdima]{\hspace*{#1}#2}%
}
\makeatother

\newtheorem{theorem}{Theorem}
\newtheorem{example}[theorem]{Example}
\newtheorem{lemma}[theorem]{Lemma}
\newtheorem{proposition}[theorem]{Proposition}
\newtheorem{corollary}[theorem]{Corollary}
\newtheorem{definition}[theorem]{Definition}
\newtheorem{remark}[theorem]{Remark}
\newcommand\COULEUR[1]{#1}
\SANSCOMMENTAIRE{\renewcommand\COULEUR[1]{\textcolor{blue}{#1}}}

 \newcommand\R{\mathbb{R}}

\newcommand\N{\mathbb{N}}
 \newcommand\Z{\mathbb{Z}}

\newcommand{\nothalt}{\textit{{nonhalting}}}
\newcommand{\NotHalt}{\textit{\textbf{NotHalt}}}
\newcommand{\internalstate}{\textit{internalstate}}
\providecommand{\Smodels}{\shifttext{-1pt}{\rotatebox[origin=c]{-0}{ \RightScissors }}}

\usepackage{version}
\excludeversion{pasencore}

\begin{document}
%
\title{Bare Demo of IEEEtran.cls\\ for IEEE Conferences}

\title{The domino problem is decidable for robust tilesets}

\author{Nathalie Aubrun$^1$  \and Manon Blanc$^{1,2}$ \and Olivier Bournez$^{2}$}

\date{
$ ^1$ Université Paris-Saclay, 
		Laboratoire Interdisciplinaire des Sciences du Numérique (LISN), 
		91400 Orsay 
		France \\
$^2$ Institut Polytechnique de Paris (IP Paris), Laboratoire d'Informatique de l'X (LIX), Campus de l'Ecole Polytechnique, 
		91128 Palaiseau, France 
}


\maketitle

\begin{abstract}
One of the most fundamental problems in tiling theory is the domino problem: given a set of tiles and tiling rules, decide if there exists a way to tile the plane using copies of tiles and following their rules. The problem is known to be undecidable in general and even for sets of Wang tiles, which are unit square tiles wearing colours on their edges which can be assembled provided they share the same colour on their common edge, as proven by Berger in the 1960s.
In this paper, we focus on Wang tilesets.
We prove that the domino problem is decidable for robust tilesets, \emph{i.e.} tilesets that either cannot tile the plane or can but, if so, satisfy some particular invariant provably.
We establish that several famous tilesets considered in the literature are robust. We give arguments that this is true for all tilesets unless they are produced from non-robust Turing machines: a Turing machine is said to be non-robust if it does not halt and furthermore does so non-provably.
 As a side effect of our work, we provide a sound and relatively complete method for proving that a tileset can tile the plane.
 Our analysis also provides explanations for the observed similarities between proofs in the literature for various tilesets, as well as of phenomena that have been observed experimentally in the systematic study of tilesets using computer methods.
\end{abstract}


\renewcommand\subset{\subseteq}

\section{Introduction}


Wang tiles were introduced in the early '60s to study the undecidability of the $\forall\exists\forall$ fragment of first-order logic~\cite{Wang1961}. They are unit square tiles with colours on their edges. A tileset is a finite set of Wang tiles. A tiling of the infinite Euclidean plane is obtained using arbitrarily many copies of the tiles in the given tileset. Tiles are placed on the integer lattice points of the plane so that they cover the whole plane with no overlap. 
The tiling is \emph{valid} if everywhere the contiguous edges have the same colour. The \emph{domino problem} is the decision problem which takes as input $\tau$ a finite set of Wang tiles, and output \texttt{Yes} if there exists a valid tiling by $\tau$ and \texttt{No} otherwise.  In the Wang tiles model the tiles cannot be rotated otherwise this is a different model, called MacMahon squares, for which there trivially exists a valid tiling.  A famous result by Berger states that the Domino problem is undecidable~\cite{Berger:AMS}. Berger's original proof relies on the construction of an aperiodic tileset, that is used to perform Turing machine computations inside tilings. This proof has subsequently been simplified~\cite{Robinson:1971:UNT} and alternative proofs were exhibited~\cite{AanderaaLewis1974,Kari2007,DurandRomShen2012,fernique2017yet}. It is noteworthy that all known proofs rely on the existence of an aperiodic tileset, even if the techniques can be different. For most of the known aperiodic tilesets, once the tileset has been explicitly built, proving aperiodicity is not that complicated -- it was designed for! -- but proving a tiling exists may reveal more difficulty. Yet,  no general technique to prove an aperiodic tileset admits a valid tiling is known. One objective of this paper is to set a formalism unifying proofs of the existence of tilings for aperiodic tilesets. We also go one step further, studying several tilesets in this new light and proving they exhibit a particular form of recurrence invariant. 

\medskip

To achieve these objectives, we adopt the formalism of transducers to play with Wang tilesets.
We introduce two notions of \emph{robust} tilesets (Definition \ref{def:robust}), for which some semantic or provably invariant holds. This notion is motivated by our exploration of Robinson's tileset with transducers: our result on this particular tileset that inspires the rest of the paper is Theorem~\ref{thm:robinson:invariant:inductif}, which is important for two main reasons.  First, the transducer associated with the tileset $\tau_{Robi}$ is quite complex and things get worse as we compute compositions $\tau_{Robi}^n$. So a naive approach would give a complex and complicated proof.  Yet this theorem expresses in an extremely simple way the fact that a valid tiling exists: the recurrence relation~\eqref{eq:thm:robinson:invariant:inductif}, which can be read as an invariant, concerning coming discussions.
Second Theorem~\ref{thm:robinson:invariant:inductif} is not specific to Robinson's tileset: an analogous semantic statement can be written for every tileset for which one can prove it admits a valid tiling. This is expressed later in the paper in Propositions \ref{prop:direct} and \ref{prop:indirect}. We prove that the most famous tilesets considered in the literature are robust: they satisfy some invariant, which provably holds. We do so for Robinson's tileset (Theorem \ref{thm:robinson:invariant:inductif}), Jeandel-Rao's tileset (Theorem \ref{th:jeandelrao}) and Kari's tilesets (Proposition \ref{propImmortality}). We explain the phenomenon with a parallel with the theory of Turing machines, for which we can distinguish \emph{robust} Turing machines from \emph{non-robust} ones: a Turing machine is said to be non-robust if it does not halt and does so non-provably. The Halting problem for robust Turing machines is decidable, as it is computably enumerable and co-computably enumerable (Theorem \ref{th:mt}). A similar statement holds for tilesets: the domino problem for robust tilesets is decidable (Theorem \ref{th:domino:decidable}). We establish two results with the same flavour: tilesets that encode a Turing machine are robust iff the Turing machine is itself robust. Specifically, we prove this is the case if Turing machines are embedded into a tileset through piecewise affine maps (Section~\ref{section.KC} with the technique by Kari~\cite{Kari2007}) or through an adaptation of Robinson's tileset (Section~\ref{section.TM_Robinson}).  As a consequence of our work, we provide a sound and relatively complete method for proving that a tileset can tile the plane (Theorem \ref{th:domino:rc}).  Some experimental facts observed in the systematic computer-assisted exploration of Wang tilesets that have been conducted can be interpreted in a new light thanks to this setting (Section \ref{sec:experimental}). 

\medskip

The paper is organized as follows. The first section details how Wang tilesets can be seen as transducers and why this approach -- used for instance in~\cite{jeandel2021aperiodic} -- is particularly relevant. We propose in Section~\ref{section.transducers_logic} a way to use transducers as a logic. We also establish some basic facts about transducers and their relations to Wang tilesets. Then in Section~\ref{section.Robinson_transducer} we focus on Robinson's tileset $\tau_{Robi}$ to illustrate concepts from Section~\ref{section.tilesets_transducers}.  In Section~\ref{section.robustness_tileset}, we present two notions of robustness for tilesets, distinguishing the cases where an invariant semantically holds and provably holds. We then prove in Section~\ref{subsection.examples_robust_tilesets} that Robinson's tileset satisfies the strongest of these notions and so does Jeandel-Rao's tileset (Section~\ref{section.JR}). We also make a parallel with Turing machines for which we can distinguish \emph{robust} Turing machines from \emph{non-robust} ones in Section \ref{sec:mt}. In Sections~\ref{section.KC} and~\ref{section.TM_Robinson} we establish a  correspondence  between robustness for tilesets and robustness for Turing machines. We conclude with a discussion of our results in Section~\ref{sec:conclusion}, where we explain how they relate to some experimental facts and present some perspectives. 

\section{Wang tilesets as transducers}
\label{section.tilesets_transducers}

\subsection{Transducers and meta-transducers}
\label{subsection.transducers}

\begin{definition}[Transducer]
Let $\Sigma$ be a finite alphabet.  A \emph{transducer} is a triple $(Q, \Sigma,\delta)$ with
\begin{itemize}
\item $Q$ a finite set of states, $\Sigma$ a finite alphabet;
\item $\delta \subseteq Q \times Q \times \Sigma \times \Sigma$ a set of transitions.
\end{itemize}
\end{definition}

\newcommand\V{V}
\newcommand\B{B}
\newcommand\RR{R}

For example the transducer $\mathcal{T}_{1}$  with $Q=\{\V,\B,\RR\}$, $\Sigma=Q$ is pictured below. The fact that  $(q, r, w, w') \in \delta$ is represented by an edge between $q$ and $r$ labelled by $w|w'$.

\begin{center}
\begin{tikzpicture}[scale=1,shorten >=1pt,>={Stealth[round]}]
	\node[state] (vert) {\V};
    \node[state] at (1.5,2) (bleu) {\B};
    \node[state] at (3,0) (rouge) {\RR};
    \path[->]   (rouge) edge[above right] node {\RR$\mid$\RR} (bleu)
	            		(bleu) edge[above left] node{\RR$\mid$\V} (vert)
		             (vert) edge[bend left, below] node{\RR$\mid$\RR} (rouge)
		             (rouge) edge[bend left, below] node{\V$\mid$\RR} (vert);
\end{tikzpicture}
\end{center}

To simplify notations and the writing of proofs, it will often be more convenient to use \emph{meta-transducers} rather than transducers. Meta-transducers allow us to deal not only with single letters but also with words of arbitrary finite length as labels of transitions.

\begin{definition}[Meta-transducer]\label{def:meta:transducer}
Let $\Sigma$ be a finite alphabet.  A \emph{meta-transducer} is a triple $(Q, \Sigma,\delta)$ with
\begin{itemize}
\item   $Q$ a finite set of states, $\Sigma$ a finite alphabet;
\item $ \delta \subseteq Q \times Q \times \Sigma^{*} \times \Sigma^{*}$ a set of transitions -- or edges --
with 
$|w|,|w'| \ge 1$ where $|w|$ denotes the length of $w$.
\end{itemize}
\end{definition}

We consider the extended transition relation $\delta^*$ as the smallest relation, containing $\delta$,  such that whenever $(q, r, x, y) \in \delta^*$ and $(r, s, w, w', s) \in \delta$ then $(q, s, x w, y w') \in \delta^*$.

If $\mathcal{T}$ is a (meta-)transducer, we write $w \mathcal{T} w'$ if there exist $q,s\in Q$  and $w,w'\in \Sigma^*$ s.t. $(q, s,  w,  w') \in \delta^*$. This is equivalent to say that there exists a sequence $q_{0} , \dots, q_{n}$ s.t. $(q_{i}, q_{i+1},w_{i}, w'_{i})$ for every $i=0\dots n-1$, where $w=w_{1}\dots w_{n}$, $w'=w'_{1}\dots w'_{n}$ and the $w_i,w'_i$ are words on $\Sigma$.
By definition,  a transducer is a meta-transducer: consider every letter as a word of length $1$.
%

\subsection{Loops}

Consider $\mathcal{T}=(Q, \Sigma,\delta)$ a meta-transducer.  A transition $(q, q, w, w') \in \delta$ is called a \emph{loop}. Moreover, if the word $w'$ is a cyclic permutation of $w$, i.e. if there exist two non-trivial words $u,v$ such that $w=uv$ and $w'=vu$, we say that the loop $(q, q, w, w')$ is a \emph{cyclic loop}. Finally, a loop $(q, q, w, w) \in \delta$ is called a \emph{periodic loop}. Note that every periodic loop is a cyclic loop, but the converse is not true.

%
%

We say that a transducer \emph{contains a loop} if there exists some $h: [[0\dots m]] \to Q$ such that  $({h(i)},{h(i+1)},w_{i},w'_{i}) \in \delta$ with $h(m+1)$ abusively defined as $h(0)$.

%
%
%
%
%
%

\subsection{Tiles and transducers}

A \emph{Wang tileset} $T$ is a triple $(H,V,\tau)$ where:
\begin{itemize}
\item $H$ is a finite set that corresponds to the horizontal colours;
\item $V$ is a finite set that corresponds to vertical colours;
\item $\tau\subset H^2\times V^2$ is the set of tiles.
\end{itemize}
We denote $t = (w, e, s, n)\in \tau$ \emph{a tile} in $T$ where $w$ (resp. $e$, $s$, $n$) is the colour on the left (resp. right, bottom, top) edge. For a tile $t$, we also write $t_w$, $t_e$, $t_s$ and $t_n$ the colours of its edges.

A Wang tileset expressed as triple $(H,V,\tau)$ exactly matches the definition of a transducer of Section~\ref{subsection.transducers} tiles are transitions, horizontal colours $H$ are states $Q$ and vertical colours $V$ are letters of the alphabet $\Sigma$.

\begin{figure}[!ht]
\begin{center}
\begin{tikzpicture}[scale=1,shorten >=1pt,>={Stealth[round]},transform shape]
\wang{-2.5}{-0.5}{vert}{vert}{rouge}{bleu}
	\node[state] (bleu) {\lettredT{bleu}};
    \node[state] at (3,0) (vert) {\lettredT{vert}};
    \path[->] (bleu) edge[above] node{\lettrebT{rouge}$\mid$\lettreuT{vert}} (vert);

	\node[state] at (5,0)(bleubis) {\lettredT{bleu}};
    \node[state] at (8,0) (vertbis) {\lettredT{vert}};
    \path[->] (bleubis) edge[above] node{\wangEdge{vert}{vert}{rouge}{bleu}} (vertbis);

\end{tikzpicture}
\end{center}
\caption{Two graphical representations of a Wang tile as a transition in the associated transducer.}
\label{figure:tile_as_transition}
\end{figure}
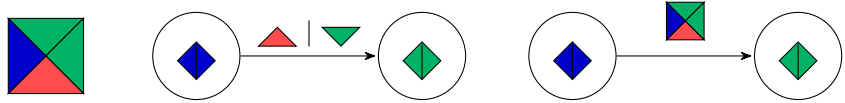

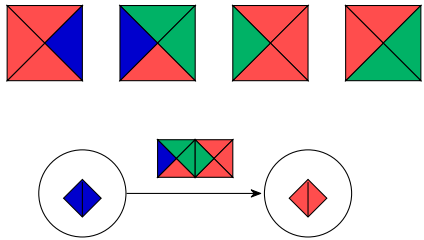
\begin{figure}[!ht]
\begin{center}
\begin{tikzpicture}[scale=1,,shorten >=1pt,>={Stealth[round]},transform shape]
\begin{scope}[scale=1]
	\wang{0}{0}{rouge}{bleu}{rouge}{rouge}
	\wang{1.5}{0}{vert}{vert}{rouge}{bleu}
	\wang{3}{0}{rouge}{rouge}{rouge}{vert}
	\wang{4.5}{0}{rouge}{vert}{vert}{rouge}
\end{scope}

\begin{scope}[shift={(0,0.5)}]
	\node[state] at (1,-2)(bleubis) {\lettredT{bleu}};
    \node[state] at (4,-2) (vertbis) {\lettredT{rouge}};
    \path[->] (bleubis) edge[above] node{\wangEdgeEdge{vert}{vert}{rouge}{bleu}{rouge}{rouge}{rouge}{vert}} (vertbis);
\end{scope}

\end{tikzpicture}
\end{center}
\caption{An example a of meta-transducer derived from the tileset $\tau$ composed by the four tiles pictured on the top.}
\label{figure:example_meta-transducer_tau}
\end{figure}

 \begin{remark}
The choice of $H$ as states and $V$ as colours is arbitrary.  We could also have considered the opposite. The transducer that we would obtain will be called the \emph{vertical transducer}, as opposed to the (horizontal) transducer considered here.
 \end{remark}

  \begin{remark} The existence of a path of length two between two vertices of the transducer tests the horizontal compatibility between two tiles.
 \end{remark}

\begin{remark}
There is a bijection between the tiles of a Wang tileset and the edges of its associated transducer. From now on we will use tileset or transducer interchangeably without any possible confusion.
\end{remark}

A natural operation on transducers, denoted by $\cup$, is the union: given $\mathcal{T}_{1}=(Q_{1}, \Sigma_1,\delta_{1})$ and $\mathcal{T}_{2}=(Q_{2}, \Sigma_2,\delta_{2})$, their union $\mathcal{T}_{1} \cup \mathcal{T}_{2}$ is given by $(Q,\Sigma,\delta)$ where $Q= Q_{1} \cup Q_{2}$, $\Sigma= \Sigma_{1} \cup \Sigma_{2}$, and $(q,q',w,w') \in \delta$ iff $(q,q',w,w') \in \delta_{1}$ or $(q,q',w,w') \in \delta_{2}$.

Any transducer is thus the union of its edges.  Combined with the remark immediately above, the transducer associated with a Wang tileset can be seen as the union of its tiles.

%
%
%
%
%
%
%

A \emph{tiling} $x$ by $\tau$ (or by $T$) is a mapping $x:\Z^2\to \tau$ such that adjacent edges share the same colour, i.e. for every $(i,j)\in\Z^2$
\begin{itemize}
\item $x(i,j)_n=x(i,j+1)_s$;
\item $x(i,j)_e=x(i+1,j)_w$.
\end{itemize}

The tiling $x$ is \emph{periodic} with period $(a, b) \in \mathbb{Z}^2 \backslash\{(0,0)\}$ iff $x(u, v)=x(u+a, v+b)$ for every $(u, v) \in \mathbb{Z}^2$. If there exists a periodic valid tiling with tiles from $\tau$, then there exists a \emph{doubly periodic} valid tiling, i.e. a tiling $x$ such that, for some $a, b>0, x(u, v)=x(u+a, v)=x(u, v+b)$ for all $(u, v) \in \mathbb{Z}^2$  (see e.g. \cite{kari1996small} or \cite[Lemma 1]{jeandel2021aperiodic}). A tileset $\tau$ is called \emph{aperiodic} iff the two following conditions are satisfied (i) there exists a valid tiling and (ii) there does not exist any periodic valid tiling.

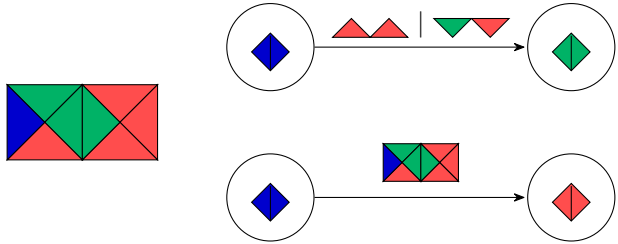
\begin{figure}[!ht]
\begin{center}
\begin{tikzpicture}[scale=1,shorten >=1pt,>={Stealth[round]},transform shape]
\wang{-2.5}{-0.5}{vert}{vert}{rouge}{bleu}
\wang{-1.5}{-0.5}{rouge}{rouge}{rouge}{vert}

	\node[state] at (1 ,1) (bleu) {\lettredT{bleu}};
    \node[state] at (5,1) (vert) {\lettredT{vert}};
    \path[->] (bleu) edge[above] node{\lettrebbT{rouge}{rouge}$\mid$\lettreuuT{vert}{rouge}} (vert);

	\node[state] at (1,-1)(bleubis) {\lettredT{bleu}};
    \node[state] at (5,-1) (vertbis) {\lettredT{rouge}};
    \path[->] (bleubis) edge[above] node{\wangEdgeEdge{vert}{vert}{rouge}{bleu}{rouge}{rouge}{rouge}{vert}} (vertbis);

\end{tikzpicture}
\end{center}
\caption{Two graphical representations of a finite horizontal valid tiling as a transition in a meta-transducer derived from the tileset pictured on the left.}
\label{figure:finite_H_pattern_as_transition}
\end{figure}

\subsection{Horizontal patterns: compatibility}

Since we are interested in tilings of the plane, we need a way to go from finite horizontal patterns, i.e. transitions in a meta-transducer $\mathcal{T}$ derived from $\tau$, to bi-infinite tilings with the formalism of meta-transducers. The notion of compatibility is an intermediate step that allows us to speak of infinite horizontal patterns.

A biinfinite word (or biinfinite sequence) on the alphabet $V$ is a sequence $\left(w_i\right)_{i \in \mathbb{Z}}$ such that, for every $i \in \mathbb{Z}, w_i \in V$.
If $w=(w_i)_{i\in\Z}\in V^\Z$ is a bi-infinite word, we say $w'=(w'_i)_{i\in\Z}\in V^\Z$ is the image of $w$ by the transducer $\mathcal{T}$ if there exists a bi-infinite sequence of states $q=(q_i)_{i\in\Z}\in H^\Z$ such $(q_i,q_{i+1},w_i,w'_i)\in \mathcal{T}$ that for every $i\in\Z$.
We denote this by $w\mathcal{T}w'$. This corresponds to the existence of an infinite horizontal stripe of Wang tiles from $\tau$ with colours $w$ on the bottom and $w'$ on the top.

\begin{remark}
The relation induced by $\mathcal{T}$ on bi-infinite words is usually non-deterministic, so this is a partial relation, not a function.
\end{remark}

We say that $\mathcal{T}$ is \emph{compatible} if there exist $\omega,\omega'\in V^\Z$ such that $\omega\mathcal{T}\omega'$.


\begin{proposition}[Folklore] \label{prop:compatible_iff_loop}
A tileset $\tau$ is compatible if and only if the associated transducer $\mathcal{T}$ contains a loop.
\end{proposition}

%
%
%
%
%
%
%
%
%
%
%
%
%
%
%
%
%
%
%

\subsection{Example}

In this subsection, we focus on a very simple Wang tileset to illustrate how transducers and meta-transducers are helpful in the study of Wang tilings (more details and pictures are provided in Appendix \ref{more:example:periodic_tileset}).

\begin{example}\label{example:periodic_tileset}
Consider $\tau$ the following set of four Wang tiles
\begin{center}
\begin{tikzpicture}[scale=1]
\wang{0}{0}{rouge}{bleu}{rouge}{rouge}
\wang{1.5}{0}{vert}{vert}{rouge}{bleu}
\wang{3}{0}{rouge}{rouge}{rouge}{vert}
\wang{4.5}{0}{rouge}{vert}{vert}{rouge}
\end{tikzpicture}
\end{center}
The tileset $\tau$ can equivalently be represented by the following transducer $\mathcal{T}$
\begin{center}
\begin{tikzpicture}[scale=1,shorten >=1pt,>={Stealth[round]}]
	\node[state] (vert) {\lettredT{vert}};
    \node[state] at (1.5,2) (bleu) {\lettredT{bleu}};
    \node[state] at (3,0) (rouge) {\lettredT{rouge}};
    \path[->]   (rouge) edge[above right] node {\wangEdge{rouge}{bleu}{rouge}{rouge}} (bleu)
	            		(bleu) edge[above left] node{\wangEdge{vert}{vert}{rouge}{bleu}} (vert)
		             (vert) edge[bend left, below] node{\wangEdge{rouge}{rouge}{rouge}{vert}} (rouge)
		             (rouge) edge[bend left, below] node{\wangEdge{rouge}{vert}{vert}{rouge}} (vert);
\end{tikzpicture}
\end{center}



Compatibility of the transducer $\mathcal{T}$ is established by the fact that we can tile a line like

\begin{center}
\begin{tikzpicture}[scale=1]
\node at (-1,0.5) {\scalebox{1}{$\dots$}};
\wang{0}{0}{rouge}{rouge}{rouge}{vert}
\wang{1}{0}{rouge}{vert}{vert}{rouge}
\wang{2}{0}{rouge}{rouge}{rouge}{vert}
\wang{3}{0}{rouge}{vert}{vert}{rouge}
\wang{4}{0}{rouge}{rouge}{rouge}{vert}
\wang{5}{0}{rouge}{vert}{vert}{rouge}
\node at (7,0.5) {\scalebox{1}{$\dots$}};
\end{tikzpicture}
\end{center}

which follows from the fact that it contains the following loop

\begin{center}
\begin{tikzpicture}[scale=0.9,shorten >=1pt,>={Stealth[round]}]
	\node[state] (V) {\lettredT{vert}};
	\path[->] (V) edge[out=30,in=330,looseness=6, right] node{\wangEdgeH{rouge}{rouge}{rouge}{vert}{rouge}{vert}{vert}{rouge}} (V);
\end{tikzpicture}
\end{center}

and the fact that we can tile a line like
\begin{center}
\begin{tikzpicture}[scale=1]
\node at (-1,-1) {\scalebox{1}{$\dots$}};
\wang{0}{-1.5}{rouge}{bleu}{rouge}{rouge}
\wang{1}{-1.5}{vert}{vert}{rouge}{bleu}
\wang{2}{-1.5}{rouge}{rouge}{rouge}{vert}
\wang{3}{-1.5}{rouge}{bleu}{rouge}{rouge}
\wang{4}{-1.5}{vert}{vert}{rouge}{bleu}
\wang{5}{-1.5}{rouge}{rouge}{rouge}{vert}
\node at (7,-1) {\scalebox{1}{$\dots$}};
\end{tikzpicture}
\end{center}

which follows from the fact that it contains the following loop

\begin{center}
\begin{tikzpicture}[scale=1,shorten >=1pt,>={Stealth[round]}]
	\node[state] (V) {\lettredT{vert}};
	\path[->] (V) edge[out=30,in=330,looseness=6, right] node{\wangEdgeHH} (V);
\end{tikzpicture}
\end{center}
\end{example}


\subsection{Composition}


We present here how the notion of composition allows us to make these horizontal patterns grow in the vertical direction. Combining both concepts of meta-transducer and composition, we obtain a formalism in terms of meta-transducers to deal with rectangular patterns valid for $\tau$ of arbitrary sizes.

%

\begin{definition}[Notation $\mathcal{T} \circ \mathcal{T'}$]
If $(Q,\Sigma,\delta)$ and $(Q',\Sigma',\delta')$ are two meta-transducers, then the \emph{composition of $\mathcal{T}$ and $\mathcal{T}'$},  is the meta-transducer $(Q \times Q',\Sigma\times\Sigma',\delta''$) where
$\left((w,w'),(e,e'),s,n',\right) \in \delta''$ iff  $(w,e,s,n)\in \delta$ and
$ (w',e',s',n')\in \delta'$ and $n=s'$.
\end{definition}


\begin{proposition}
The composition of two meta-transducers is a meta-transducer.
\end{proposition}

As a corollary, if we start with two Wang tilesets, their composition is still a Wang tileset.

\begin{remark}
The states of the compositions are pairs of horizontal colours and the edges correspond to vertical compatibility between the top and bottom vertical colours of tiles associated with the corresponding horizontal colours.
\end{remark}
\begin{remark}
If we were working with vertical transducers, then the composition of two vertical transducers would be about pairs of vertical colours and edges would correspond to some horizontal compatibility  (between the left and right horizontal colour of tiles).
\end{remark}
\begin{remark}
Recall that horizontal compatibility is tested by the existence of paths in the (horizontal) transducer. The existence of paths would correspond to the vertical compatibility in the vertical transducer.
\end{remark}


Fix some meta-transducer $\mathcal{T}$.
In particular we can compose $\mathcal{T}$ with itself and iterate to get a sequence of transducers $(\mathcal{T}^n)_{n\in\N^*}$ given by:
\begin{itemize}
\item $\mathcal{T}^1=\mathcal{T}$
\item $\mathcal{T}^{n+1}=\mathcal{T}^n\circ \mathcal{T}$ for $n\geq0$.
\end{itemize}

As a direct consequence of the definition, we relate the compatibility of the composition $\mathcal{T}^n$ with the existence of a valid tiling of a horizontal strip.

\begin{proposition} \label{proposition:infinite_path_horizontal_strip}
Consider a transducer $\mathcal{T}$.
There exist $w,w'\in V^\Z$ such that $w\mathcal{T}^n w'$ iff there exists a valid tiling by $\tau$ of  $\Z\times[0;n-1]$ the horizontal strip of height $n$.
\end{proposition}

If we start from a Wang tileset, then the set of states of $\mathcal{T}^{n}$ can be identified with $H^{n}$.
We say then that $\mathcal{T}^{n}$ is of \emph{height $n$}.

\begin{example}
Consider $\mathcal{T}^2$ the composition of $\mathcal{T}$ from Example~\ref{example:periodic_tileset} with itself.  Then $\mathcal{T}^2$ is as pictured below.
\begin{center}
\begin{tikzpicture}[scale=1,shorten >=1pt,>={Stealth[round]}]
	\node[state, rectangle, rounded corners=3mm] (RV) {\lettreddT{rouge}{vert}};
    \node[state, rectangle, rounded corners=3mm] at (4,0) (BR) {\lettreddT{bleu}{rouge}};
    \node[state, rectangle, rounded corners=3mm] at (8,0) (VV) {\lettreddT{vert}{vert}};
    \node[state, rectangle, rounded corners=3mm] at (2,4) (VR) {\lettreddT{vert}{rouge}};
    \node[state, rectangle, rounded corners=3mm] at (6,4) (RB) {\lettreddT{rouge}{bleu}};
    \node[state, rectangle, rounded corners=3mm] at (2,-4) (VB) {\lettreddT{vert}{bleu}};
    \node[state, rectangle, rounded corners=3mm] at (6,-4) (RR) {\lettreddT{rouge}{rouge}};
    \path[->]   (VV) edge[right] node {\wangEdgeV{white}{rouge}{rouge}{vert}{rouge}{rouge}{white}{vert}} (RR)
	            		(RR) edge[above] node{\wangEdgeV{white}{vert}{vert}{rouge}{rouge}{bleu}{white}{rouge}} (VB)
		             (VB) edge[left] node{\wangEdgeV{white}{rouge}{rouge}{vert}{vert}{vert}{white}{bleu}} (RV)
		             (RV) edge[] node{\wangEdgeV{white}{bleu}{rouge}{rouge}{rouge}{rouge}{white}{vert}} (BR)
		             (RV) edge[above left] node{\wangEdgeV{white}{vert}{vert}{rouge}{rouge}{rouge}{white}{vert}} (VR)
		             (VR) edge[above] node{\wangEdgeV{white}{rouge}{rouge}{vert}{rouge}{bleu}{white}{rouge}} (RB)
		             (RB) edge[above right] node{\wangEdgeV{white}{vert}{rouge}{bleu}{rouge}{vert}{white}{rouge}} (VV)
		             (BR) edge[] node{\wangEdgeV{white}{vert}{rouge}{bleu}{rouge}{vert}{white}{rouge}} (VV);
\end{tikzpicture}
\end{center}

\end{example}

\subsection{Inclusion between transducers}
\label{section.transducers_logic}

A meta-transducer can also be seen as a labelled multi-graph whose edges are labelled by pairs of words on $\Sigma$.  Its vertices are elements of $Q$ (states), and there is an edge between $q$ and $r$ iff $(q, r, w, w') \in \delta$ for some $w,w'$.

\begin{definition}[Notation $\mathcal{T} \subseteq \mathcal{T'}$]
We write $\mathcal{T} \subseteq \mathcal{T'}$ when the first is a sub-multi-graph of the second. Formally, we have $\mathcal{T}=(Q,\Sigma,\delta)$, $\mathcal{T'}=(Q',\Sigma',\delta')$, with $Q \subset Q'$, $\Sigma \subset \Sigma'$ and whenever $(q, r, w, w') \in \delta$ then $(q, r, w, w') \in \delta'$. 
\end{definition}

\newcommand\Gun{\begin{tikzpicture}[baseline={(current bounding box.center)},scale=1,shorten >=1pt,>={Stealth[round]}]
	\node[state] (vert) {\V};
    \node[state] at (3,0) (rouge) {\RR};
    \path[->]                (vert) edge[bend left, below] node{\RR$\mid$\RR} (rouge)
		             (rouge) edge[bend left, below] node{\V$\mid$\RR} (vert);
\end{tikzpicture}}
\newcommand\Gdeux{
\begin{tikzpicture}[baseline={(current bounding box.center)},scale=1,shorten >=1pt,>={Stealth[round]}]
	\node[state] (vert) {\V};
    \node[state] at (1.5,2) (bleu) {\B};
    \node[state] at (3,0) (rouge) {\RR};
    \path[->]   (rouge) edge[above right] node {\RR \RR $\mid$\RR \V} (bleu)
	            		(bleu) edge[above left] node{\RR$\mid$\V} (vert)
		             (vert) edge[bend left, below] node{\RR$\mid$\RR} (rouge)
		             (rouge) edge[bend left, below] node{\V$\mid$\RR} (vert);
\end{tikzpicture}
}

\begin{example} \label{ex:ttp}
\ \\
\begin{center}
\Gun
$\;\subseteq$
\Gdeux
\end{center}
\end{example}


Transducers and meta-transducers are a natural logics for tilings:
\begin{definition}[Notation $\mathcal{T} \Smodels \mathcal{T}'$]
Given two transducers $\mathcal{T}$ and $\mathcal{T}'$, we write
$\mathcal{T} \Smodels \mathcal{T'}$ iff every edge of the transducer $\mathcal{T'}$ is an edge of~$\mathcal{T}$.
\end{definition}


This just means that the local constraints associated with $\mathcal{T}'$ fulfil all the constraints associated with $\mathcal{T}$. We use a scissor symbol, as this corresponds to deleting (cutting) some edges.

\begin{example}
Consider the two transducers $\mathcal{T}$ and $\mathcal{T}'$ pictured below in Example \ref{ex:ttp} on respectively right and left.

Then we have
\[
\mathcal{T}\Smodels\mathcal{T}',
\]
or said otherwise
\begin{center}
\Gdeux
$\;\Smodels$
\Gun
\end{center}
\end{example}



As such, when we write $\mathcal{T} \Smodels \mathcal{T}'$ and the states of $\mathcal{T}$ are of height $n$, then $\mathcal{T}'$ is necessarily made of states of height $n$.



\subsection{Loops}
\label{sub:loop}






\begin{proposition} \label{proposition:circular_loop_periodic_loop}
Let $\tau$ be a tileset and $\mathcal{T}$ the associated transducer.  If $\mathcal{T}$ contains a cyclic loop of order $N$ then $\mathcal{T}^{m}$ contains a periodic loop for some $1 \le m \le N$.
\end{proposition}

\begin{proof}
Assume that $\mathcal{T}$ contains a cyclic loop $(w,w,uv,vu)$ with $|uv|=N$.  Said otherwise, there exists a sequence of states $w_1=w,\dots w_N$ such that
\begin{center}
\begin{tikzpicture}[scale=1,shorten >=1pt,>={Stealth[round]}]
	\node[] at (-1.25,0) {\scalebox{1}{$\mathcal{T} \Smodels$}};
	\node[state] (v1) {$w_1$};
	\node[state] at (3,0) (v2) {$w_2$};
	\node[state] at (6,0) (v3) {$w_3$};
	\node at (7,0) {$\dots$};
	\node[state] at (8.2,0) (vN1) {$w_{N-1}$};
	\node[state] at (6.5,-2.5) (vN) {$w_N$};

	\path[->] (v1) edge[above] node{\scalebox{0.8}{$u^{(1)}|v^{(1))}$}} (v2);
	\path[->] (v2) edge[above] node{\scalebox{0.8}{$u^{(2)}|v^{(2)}$}} (v3);
	\path[->] (vN1) edge[right] node{\scalebox{0.8}{$v^{(|v|-1)}|u^{(|u|-1)}$}} (vN);
	\path[->] (vN) edge[bend left, below] node{\scalebox{0.8}{$v^{(|v|)}|u^{(|u|)}$}} (v1);
\end{tikzpicture}
\end{center}Thus, we have a sequence of loops
\[
\mathcal{T}\Smodels (w_i,w_i,u^{(i)}v^{(i)},v^{(i)}u^{(i)})
\]
for every $i=1\dots N$. Since there are $N$ cyclic permutations of $uv$ and we have exactly $N$ words $b_i=u^{(i)}v^{(i)}$ and $N$ words $t_i=v^{(i)}u^{(i)}$, each $t_i$ must also be some $b_{\sigma(i)}$ where $\sigma: [1;N]\to[1;N]$ is a permutation. Denote $m$ the size of the orbit of $1$ by $\sigma$, it is the size of the cycle $\left(1,\sigma(1),\dots,\sigma^{m-1}(1)\right)$ with $\sigma^m(1)=1$. Using iteratively the rule for composition on the transducer $\mathcal{T}$, we have that :
\[
\mathcal{T}^m\Smodels (w,w,b_1,t_m)
\]
and since $t_m=b_{\sigma^m(1)}=b_1$ this is a periodic loop.
\end{proof}

\begin{proposition} \label{proposition:loop_horizontal_strip}
Consider a transducer $\mathcal{T}$.
Then $\mathcal{T}^{n}$ contains a  loop iff $\tau$ there exists a valid tiling by $\tau$ of  $\Z\times[0;n-1]$ the horizontal strip of height $n$.
\end{proposition}

\begin{proof}
If $\mathcal{T}^{n}$ contains a  loop, then there is an infinite path in his loop and hence this follows from Proposition \ref{proposition:infinite_path_horizontal_strip}.

Conversely, by Proposition  \ref{proposition:infinite_path_horizontal_strip}, there exists some $w,w'\in V^\Z$ such that $w\mathcal{T}^n w'$.
Hence, there exists a bi-infinite sequence of states $q=(q_i)_{i\in\Z}\in H^\Z$ such $(q_i,q_{i+1},w_i,w'_i)\in \mathcal{T}$ that for every $i\in\Z$. As there are finitely many states, there exists some $q$ not appearing finitely many times: we must have $q_{i}=q_{j}$ with $i<j$ for some $i,j$. Then $(q,q,w_{i} w_{i+1} \dots w_{j-1}, w'_{i} w'_{i+1} \dots w'_{j-1}) \in \mathcal{T}$. In other words, $\mathcal{T}$ has a loop in state $q$.
\end{proof}

This can be reinforced in:

\begin{proposition} \label{proposition:loop_horizontal_strip_reinforced}
Consider a transducer $\mathcal{T}$.
Then $\mathcal{T}^{n}$ contains a  loop of order $N$ iff $\tau$ there exists a valid tiling by $\tau$ of  $\Z\times[0;n-1]$ the horizontal strip of height $n$ using a pattern repeated infinitely with period $N$.
\end{proposition}

\begin{proof}
This follows from the previous proof, using the fact that a loop is associated with a rectangular pattern, whose width corresponds to its order.
\end{proof}

%
%
%
%

\section{Robinson's tileset as a transducer to prove tileability}
\label{section.Robinson_transducer}

\subsection{Robinson's tileset}

\begin{figure}[!ht] 
\begin{center}
\includegraphics[width=0.99\linewidth]{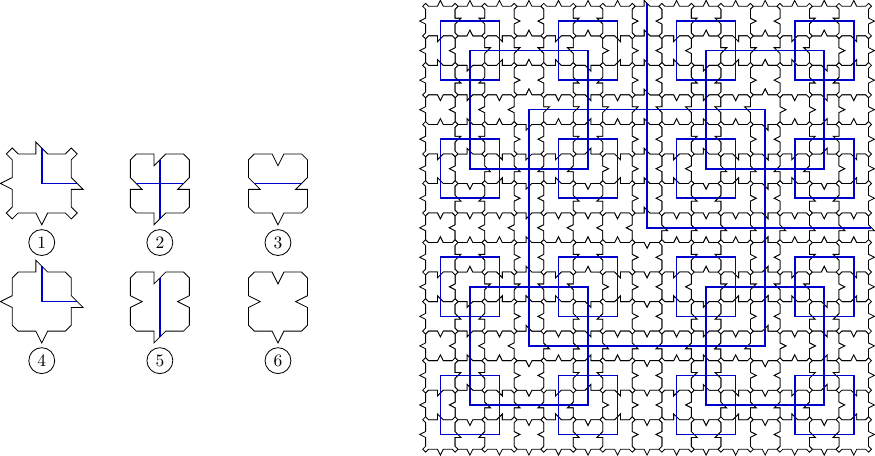}
%
%
\end{center}
\caption{\label{figure:Robinson_tileset} To the left Robinson's tileset, where tiles can be rotated and reflected. To the right a 
 pattern that appears in every tiling by Robinson's tileset.}
\end{figure}

A famous tileset is Robinson's tileset~\cite{Robinson:1971:UNT} which simplifies the construction of Berger~\cite{Berger:AMS}.  This aperiodic Wang tileset can be represented more compactly with the geometric tiles pictured in Figure~\ref{figure:Robinson_tileset}, which may be rotated and reflected. Formally, they are not Wang tiles but they can be transformed in such, using Robinson's arguments \cite{Robinson:1971:UNT}. An example of tiling of the plane using Robinson's tileset is given in Figure~\ref{figure:Robinson_tileset}.

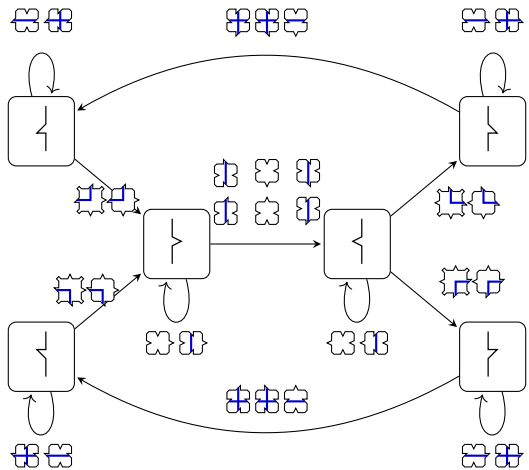
\begin{figure}[!ht]
\newcommand\edgename[1]{}
\begin{center}
\providecommand\aide[1]{}
\providecommand\edgename[1]{#1}
\def\sf{1}

\newcommand\COMMUN{
	\aide{\draw[help lines,step=1] (-10,-10) grid (10,10);}
	\node[state,rectangle,rounded corners=3pt] (rmi) at (-1.6,3) {\scalebox{1.2}{\aide{rmi}\nletterRobiV{90}}};
	\node[state,rectangle,rounded corners=3pt] (lmi) at (1.6,3) {\scalebox{1.2}{\aide{lmi}\nletterRobiV{-90}}};
	
	\node[state,rectangle,rounded corners=3pt] (rbi) at (4,5) {\scalebox{1.2}{\aide{rbi}\nletterRobiVter{-90}}};
	\node[state,rectangle,rounded corners=3pt] (rti) at (4,1) {\scalebox{1.2}{\aide{rti}\nletterRobiVbis{-90}}};

	\node[state,rectangle,rounded corners=3pt] (lbi) at (-4,5) {\scalebox{1.2}{\aide{lbi}\nletterRobiVbis{90}}};
	\node[state,rectangle,rounded corners=3pt] (lti) at (-4,1) {\scalebox{1.2}{\aide{lti}\nletterRobiVter{90}}};

}

\begin{tikzpicture}[baseline={(current bounding box.center)},scale=0.75]
	\COMMUN
	\path[->,-stealth,shorten >=1pt] (rmi) edge[loop below] node{\scalebox{\sf}{\tRobiEmpty{90}{white} \hspace{-0.4cm} \tRobiH{90}{white}}}  node[above right] {\edgename{$e_{6}$}} 	(rmi); 
	\path[->,-stealth,shorten >=1pt] (lmi) edge[loop below] node{\scalebox{\sf}{\tRobiEmpty{-90}{white} \hspace{-0.4cm} \tRobiH{-90}{white}}} node[above left] {\edgename{$e_{7}$}}  (lmi);
	
	\path[->,-stealth,shorten >=1pt] (lti) edge[above left, pos = -0.1] node{\scalebox{\sf}{
	\tikz[scale=0.15]{\tRobiCorner{180}{white}  \tRobiCornerDented{180}{white}}}} node[right,pos=0.3] {\edgename{$e_{-3}$}}  (rmi);	
	
	\path[->,-stealth,shorten >=1pt] (rmi) edge[above] node{\scalebox{\sf}{\begin{tabular}{c} \tRobiVbis{180}{white}\tRobiEmpty{0}{white}\tRobiVbis{0}{white}\\ \tRobiV{180}{white}\tRobiEmpty{180}{white}\tRobiV{0}{white}\end{tabular}}} node[below] {\edgename{$e_{0}$}} (lmi);
	
	\path[->,-stealth,shorten >=1pt] (lbi) edge[below left, pos = 0.2] node{\scalebox{\sf}{
	\tikz[scale=0.15]{\tRobiCorner{90}{white} \tRobiCornerDented{90}{white}}}} node[right, pos=0.3] {\edgename{$e_{3}$}}   (rmi);
	
	\path[->,-stealth,shorten >=1pt] (lmi) edge[right, pos=0.3] node{\scalebox{\sf}{
	\tikz[scale=0.15]{\tRobiCorner{0}{white} \tRobiCornerDented{0}{white}}
	}}  node[left, pos=0.5] {\edgename{$e_{1}$}} (rbi);

	\path[->,-stealth,shorten >=1pt] (lmi) edge[right,pos=0.5] node{\scalebox{\sf}{\tikz[scale=0.15]{\tRobiCorner{-90}{white} \tRobiCornerDented{-90}{white}}}}  node[left, pos=0.6] {\edgename{$e_{-1}$}} (rti);

	\path[->,-stealth,shorten >=1pt] (lti) edge[loop below] node{\scalebox{\sf}{\tRobiCross{-90}{white} \hspace{-0.4cm} \tRobiV{-90}{white}}} node[above] {\edgename{$e_{-5}$}}  (lti);
	
	\path[->,-stealth,shorten >=1pt] (rbi) edge[bend right,above] node{\scalebox{\sf}{
	\tRobiCross{0}{white} \hspace{-0.4cm} \tRobiCrossBis{0}{white} \hspace{-0.4cm} \tRobiH{0}{white}
	}} node[below] {\edgename{$e_{2}$}} (lbi);
	
	\path[->,-stealth,shorten >=1pt] (rti) edge[bend left,above] node{\scalebox{\sf}{
	\tRobiCross{180}{white} \hspace{-0.4cm} \tRobiCrossBis{180}{white} \hspace{-0.4cm} \tRobiH{180}{white}
	}}  node[below] {\edgename{$e_{-2}$}}  (lti);

	\path[->,-stealth,shorten >=1pt] (rti) edge[loop below] node{\scalebox{\sf}{\tRobiVbis{90}{white} \hspace{-0.4cm} \tRobiCrossBis{90}{white}}} node[above] {\edgename{$e_{-4}$}}  (rti);

	\path[->,-stealth,shorten >=1pt] (lbi) edge[loop above] node{\scalebox{\sf}{\tRobiVbis{-90}{white} \hspace{-0.4cm} \tRobiCrossBis{-90}{white}}} node[above] {\edgename{$e_{5}$}}  (lbi);
	
	\path[->,-stealth,shorten >=1pt] (rbi) edge[loop above] node{\scalebox{\sf}{\tRobiV{90}{white} \hspace{-0.4cm} \tRobiCross{90}{white}}}  node[above] {\edgename{$e_{4}$}}  (rbo);
		
\end{tikzpicture}
\end{center}
\caption{The transducer $\mathcal{T}_{Robi}$ corresponding to Robinson's tileset.}
\label{figure.Robinson_transducer}
\end{figure}

We convert this tileset into a transducer by defining two alphabets for horizontal and vertical colours
\[
H:=\left\{\Hun,\Hdeux,\Htrois,\Hquatre, \Hcinq,\Hsix\right\};
\]
\[
V:=\left\{\nletterRobiV{90},\nletterRobiV{-90},\nletterRobiVter{-90},\nletterRobiVbis{-90},\nletterRobiVbis{90},\nletterRobiVter{90}\right\}.
\]
Note that with this choice for $H$ and $V$, we forget about the corner types (bumpy or dented).  However, this will not be an issue later on,  since no two tiles are wearing the same colours from $H$ and $V$ with different corner types. Robinson's tileset can be associated with the transducer $\mathcal{T}_{Robi}$ depicted in Figure~\ref{figure.Robinson_transducer}.  For more readability, in this figure, we label a transition from a state $q$ to a state $q'$ by the corresponding tile from Figure~\ref{figure:Robinson_tileset} with the proper orientation.

\subsection{Patterns inside Robinson's tilings}

We set $g(n):=2^n-1$ for every integer $n\geq0$, so that in the sequel we are only interested in the meta-transducers $\mathcal{T}_{n}:=\mathcal{T}_{Robi}^{g(n)}$. Note that the sequence $(g(n))_{n\in\mathbb{N}}$ satisfies the recurrence relation $g(n)=2g(n-1)+1$. We also define notations for some horizontal and vertical patterns

\begin{itemize}
\item $\pattern{u}{n}:=\scalebox{0.75}{\trianglehaut}^{g(n-1)}u\scalebox{0.75}{\trianglehaut}^{g(n-1)}$ for $u\in\left\{\Hun,\Hdeux,\Htrois\right\}$;
\item $\pattern{b}{n}:=\scalebox{0.75}{\trianglebas}^{g(n-1)}b\scalebox{0.75}{\trianglebas}^{g(n-1)}$ for $b\in\left\{\Hquatre, \Hcinq,\Hsix\right\}$.
%
%
%
%
%
%
%

\end{itemize}

We also write 
\begin{tikzpicture}[baseline={(current bounding box.center)},scale=0.3]
\node[anchor=center,scale=1]  at (0,1.5) {a};
\node[anchor=center,scale=0.5]  at (0,0) {n};
\node[anchor=center,scale=1]  at (0,-1.5) {a};
\myrobinsonemptynotdotted{0}{0}{0}{n}
\draw[shift={(-0.5,+0.5)},color=black,fill=white,dotted,-] 
(1,0.5) 
-- (1,0);
\draw[shift={(-0.5,+0.5)},color=black,fill=white,dotted,-] 
(0,0.5) 
-- (0,0);
\draw[shift={(-0.5,-1)},color=black,fill=white,dotted,-] 
(1,0.5) 
-- (1,0);
\draw[shift={(-0.5,-1)},color=black,fill=white,dotted,-] 
(0,0.5) 
-- (0,0);
\end{tikzpicture}
(for example, \tikz[baseline={(current bounding box.center)},scale=0.3] {\macrorobinsonverticalbisgeneric{2}{-1}{0}{white}{25}{n}})
for a pattern made of  the tile $a$ repeated $g(n)$ times vertically, where $a$ can be either \tikz[scale=0.3]{\mytRobiV{0}{white}}, \mytRobiVbis{0}{white},\mytRobiV{180}{white}, \mytRobiVbis{180}{white},  \mytRobiEmpty{180}{white} or \mytRobiEmpty{0}{white}. 

\begin{figure}[!ht]
\centerline{
\begin{tikzpicture}[scale=0.27]
\EXEMPLE
\end{tikzpicture}
}
\caption{One of the many possibilities to tile a $46\times3$ rectangle with Robinson's tileset. All vertical patterns that appear in this pattern are different. }
\label{figure:motif_46x3_Robi}
\end{figure}
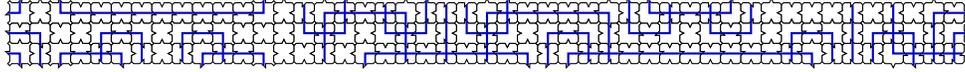

\begin{figure}
\newcommand\edgename[1]{#1}
\begin{center}
\providecommand\edgename[1]{}
\providecommand\aide[1]{}

\def\sf{1}
\begin{tikzpicture}[baseline={(current bounding box.center)},scale=0.75]
	\hspace{-0.5cm}
	\aide{\draw[help lines,step=1] (-10,-10) grid (10,10);}
	\node[state,rectangle,rounded corners=3pt] (rmi) at (-2,5) {\scalebox{1.2}{\aide{rmi}\scalebox{0.5}{\rightmymacrotileoneqd{n}}
	}};
	\node[state,rectangle,rounded corners=3pt] (lmi) at (2,5) {\scalebox{1.2}{\aide{lmi}\scalebox{0.6}{\leftmymacrotileone{n}}
	}};
	
	\node[state,rectangle,rounded corners=3pt] (rbi) at (5,9) {\scalebox{1.2}{\aide{rbi}	 \scalebox{0.5}{\rightmymacrotileone{n}}
	}};
	\node[state,rectangle,rounded corners=3pt] (rti) at (5,1) {\scalebox{1.2}{\aide{rti}\scalebox{0.5}{\rightmymacrotileonemqd{n}}
	}};

	\node[state,rectangle,rounded corners=3pt] (lbi) at (-5,9) {\scalebox{1.2}{\aide{lbi}
	\scalebox{0.5}{\leftmymacrotileoneqd{n}}
	}};
	\node[state,rectangle,rounded corners=3pt] (lti) at (-5,1) {\scalebox{1.2}{\aide{lti}	\scalebox{0.5}{\leftmymacrotileonecqv{n}}
	}};

	
	\path[->,-stealth,shorten >=1pt] (lti) edge[pos=0.5] node{\scalebox{\sf}{
	\tikz[scale=0.2]{\mymacrotiletwogeneric{-1}{3}{180}{black!10}{n}}}} node[above,pos=0.02] {\edgename{$e^{n}_{-3}$}}  (rmi);	
	
	\path[->,-stealth,shorten >=1pt] (rmi) edge[above] node{\scalebox{\sf}{\begin{tabular}{c} 
	\tikz[scale=0.26]{
	\macrorobinsonverticalreflectedbisgeneric{0}{0}{180}{white}{25}{n}
	\macrorobinsonemptygeneric{2}{0}{0}{white}{25}{n}
	\macrorobinsonverticalreflectedbisgeneric{4}{0}{0}{white}{25}{n}
	}\\
	\\
	\tikz[scale=0.26]{
	\macrorobinsonverticalbisgeneric{0}{0}{180}{white}{25}{n}
	\macrorobinsonemptygeneric{2}{0}{180}{white}{25}{n}
	\macrorobinsonverticalbisgeneric{4}{0}{0}{white}{25}{n}
	}
	\end{tabular}}} node[below] {\edgename{$e^{n}_{0}$}}  (lmi);
	\path[->,-stealth,shorten >=1pt] (lbi) edge[pos=0.5] node{\scalebox{\sf}{
	\tikz[scale=0.2]{\mymacrotiletwogeneric{-1}{3}{90}{black!10}{n}}}}  node[below,pos=0.02] {\edgename{$e^{n}_{3}$}}  (rmi);	
	\path[->,-stealth,shorten >=1pt] (lmi) edge[pos=0.5] node{\scalebox{\sf}{
	\tikz[scale=0.2]{\mymacrotiletwogeneric{-1}{3}{0}{black!10}{n}}}}  node[left,pos=0.15] {\edgename{$e^{n}_{1}$}}   (rbi);

	\path[->,-stealth,shorten >=1pt] (lmi) edge[pos=0.5] node{\scalebox{\sf}{	\tikz[scale=0.2]{\mymacrotiletwogeneric{-1}{3}{-90}{black!10}{n}}}}  node[left,pos=0.15] {\edgename{$e^{n}_{-1}$}}   (rti);

%
	\path[->,-stealth,shorten >=1pt] (rbi) edge[bend right,above] node{\scalebox{\sf}{
	\tikz[scale=0.3]{\macrorobinsoncrossgenericnew{0}{0}{0}{white}{25}{n-1}}
	\tikz[scale=0.3]{\macrorobinsoncrossbisgenericnew{0}{0}{0}{white}{25}{n-1}}
	\tikz[scale=0.3]{\macrorobinsonhorizontalgenericnew{0}{0}{0}{white}{25}{n-1}}
	}} node[below,pos=0.2] {\edgename{$e^{n}_{2}$}}  (lbi);
	
	\path[->,-stealth,shorten >=1pt] (rti) edge[bend left,above] node{\scalebox{\sf}{
	\tikz[scale=0.3]{\macrorobinsoncrossgenericnew{0}{0}{180}{white}{25}{n-1}	}
	\tikz[scale=0.3]{\macrorobinsoncrossbisgenericnew{0}{0}{180}{white}{25}{n-1}}
	\tikz[scale=0.3]{\macrorobinsonhorizontalgenericnew{0}{0}{180}{white}{25}{n-1}}
	}} node[below,pos=0.2] {\edgename{$e^{n}_{-2}$}}  (lti);

\end{tikzpicture}
\end{center}
\caption{\label{fig:Hn} The meta-transducer $H_{n}$ for $n\geq1$.}
\end{figure}
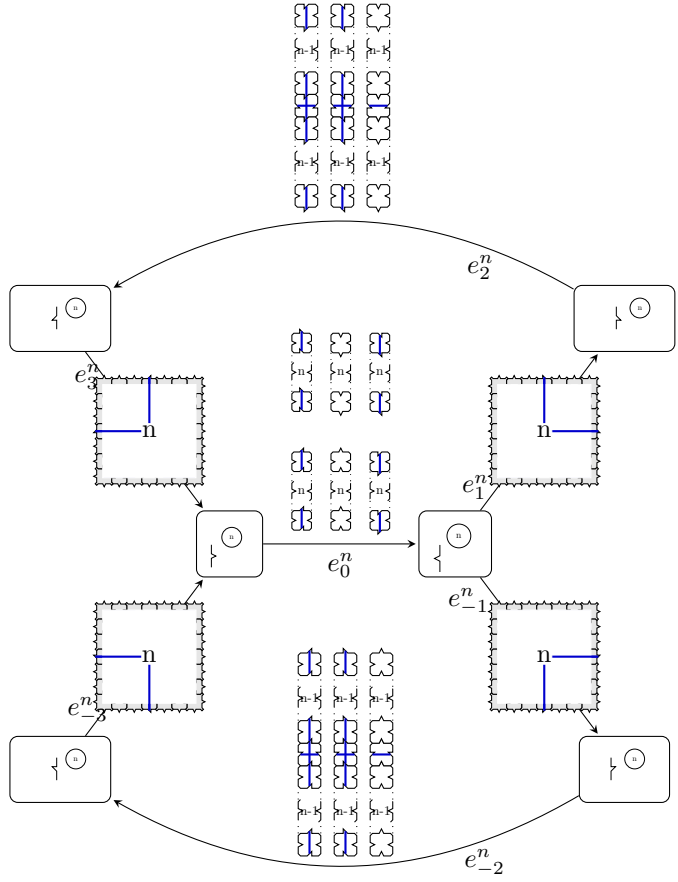

Studying the entire transducer $\mathcal{T}_n$ would be too tedious (see Figure~\ref{figure:motif_46x3_Robi} for an insight into the combinatorics of $\mathcal{T}_3$). We rather consider the meta-transducer $H_{n}$ defined in Figure \ref{fig:Hn}. As we will see (Theorem \ref{thm:robinson:invariant:inductif}), this transducer $H_{n}$ satisfies that  $\mathcal{T}_n\Smodels H_{n}$ for every $n\in\N^*$. For the sake of clarity we choose to label the various edges $e_{j}^{n}$ with patterns.

%
%
For instance the label  \tikz[baseline={(current bounding box.center)},scale=0.3]{\macrorobinsonverticalbisgeneric{2}{-1}{0}{white}{25}{n}} for edge $e_{0}^{n}$ of Figure \ref{fig:Hn} corresponds to \demitrianglebasdroit $|$ \demitrianglebasdroit.  In the same spirit, the pattern
\tikz[baseline={(current bounding box.center)},scale=0.4]{\mymacrotiletwogeneric{-1}{3}{0}{black!10}{n}} corresponds to the pair
$\pattern{\trianglebas}{n}|\macrodemitrianglehautdroit{n}$ and similarly for other patterns.

\begin{example}
For $n=2$ the label pictured below to the left is realised by the pattern pictured below to the right.  For $n=3$ this label is realised by the pattern pictured to the right in Figure~\ref{figure:Robinson_tileset}.
\begin{center}
\begin{tikzpicture}[scale=0.4]
\mymacrotiletwogeneric{-1}{3}{0}{black!10}{n}
\mymacrotiletwo{10}{3}{0}{black!10}
\end{tikzpicture}
\end{center}
\end{example}

%
%
%
%
%
%
%
%
%
%
%
%
%

%
%
%

\subsection{Robinson's tilings described by a recurrence relation}

\begin{theorem}[Key property of Robinson tiling] \label{ex:nr} \label{thm:robinson:invariant:inductif}
For all $n\in\N$,
\begin{equation} \label{eq:thm:robinson:invariant:inductif}
H_{n} \circ \mathcal{T}_{Robi} \circ H_{n} \Smodels H_{n+1}
\end{equation}
\end{theorem}


We will see in Section~\ref{sec:mt} on Turing machines why this is natural to call such a recurrence relation \emph{an invariant}.

 \begin{remark}
We just need to prove that any edge of $H_{n+1}$ is among the edges of $H_{n} \circ \mathcal{T}_{Robi} \circ H_{n}$. We start by the edges of $e_{0}^{n+1}$.
\end{remark}

\begin{lemma} \label{lemmaProofMain1}
	For all $n$, we have
	\scalebox{0.8}{
	$e_{0}^{n} \circ  \tikz[scale=1,baseline={(current bounding box.center)}]{\robinsonemptybis{0}{0}{0}{white}} \circ e_{0}^{n}
	\Smodels \tikz[scale=1,baseline={(current bounding box.center)}]{\macrorobinsonemptygeneric{0}{0}{0}{white}{25}{n+1}}$.}
\end{lemma}

As already discussed, when drawing a meta-transducer, we may either label an edge by a pair $w|w'$, or by a graphical pattern encoding it. We actually label every edge by both: a graphical pattern, and then below $w'$, and below $w$.

\def\facteurreduction{1}

\begin{proof}
We know
\myequation{eq:rec:un}{
$e_{0}^{n} \Smodels$
\Derivationdd{g(n)}
	{\tikz{\macrorobinsonemptygenericnew{0}{0}{0}{black!10}{25}{n}}}
	{\leftmacrorobinsonempty{n}}
	{\rightmacrorobinsonempty{n}}
	{\lsortant}
	{\lsortant}
}

We have:

\begin{center}
\tikz[baseline={(current bounding box.center)}]{ \robinsonemptybis{0}{0}{0}{black!10}}  $\Smodels$
\Derivationdd{1}
	{\begin{tikzpicture}\myrobinsonempty{0}{0}{0}{black!10}\end{tikzpicture}}
	{\leftrobinsonempty}
	{\rightrobinsonempty}
	{\lsortant}
	{\lsortant}
	\end{center}

By composition, we deduce:

\begin{equation}\label{eq:rec:unp}
e_{0}^{n} \circ  \tikz[baseline={(current bounding box.center)}]{\robinsonemptybis{0}{0}{0}{black!10}}
\Smodels
\Derivationdd{g(n)+1}
	{\begin{tikzpicture}\macrorobinsonemptygeneric{0}{0}{0}{black!10}{25}{n}\myrobinsonempty{0}{2.5}{0}{black!10}\end{tikzpicture}}
	{\leftplusonemacrorobinsonempty{n}}
	{\rightplusonemacrorobinsonempty{n}}
	{\lsortant}
	{\lsortant}
\end{equation}

By composition with \eqref{eq:rec:un} we obtain the lemma.
\end{proof}

\begin{lemma}\label{lemmaProofMain2}
	For all $n$, we have
	\begin{align*}	
	e_{0}^{n }\circ \mathcal{T}_{Robi} \circ e^{n}_{0} &\Smodels  e^{n+1}_{0};\\
	e_{5}^{n }\circ \mathcal{T}_{Robi} \circ e^{n}_{5} &\Smodels  e^{n+1}_{5}.
	\end{align*}
\end{lemma}

\begin{proof}
Use exactly the same reasoning, but replacing the role of \tikz{\myrobinsonempty{0}{0}{0}{black!10}} by \tikz{\myrobinsonempty{0}{0}{180}{black!10}}, \tikz{\myrobinsonverticalbis{0}{0}{0}{black!10}}, \tikz{\myrobinsonverticalbis{0}{0}{180}{black!10}}, \tikz{\myrobinsonverticalreflectedbis{0}{0}{0}{black!10}} and  \tikz{\myrobinsonverticalreflectedbis{0}{0}{180}{black!10}} in
 previous lemma, for the first assertion.

 Replace the role of $e_{0}^{n }$ by $e_{5}^{n }$ in all previous arguments, to get the second assertion.
\end{proof}

\begin{lemma}\label{lemmaProofMain3}
	For all $n$, we have
	\begin{align*}	
	e_{0}^{n }\circ \mathcal{T}_{Robi} \circ e^{n}_{0} &\Smodels  e^{n+1}_{2};\\
	e_{0}^{n }\circ \mathcal{T}_{Robi} \circ e^{n}_{0} &\Smodels  e^{n+1}_{-2}.
	\end{align*}
\end{lemma}

\begin{proof}
From \eqref{eq:rec:un},

\centerline{
\Derivationd{1}
	{\begin{tikzpicture}\myrobinsonhorizontalbis{0}{0}{0}{black!10}\end{tikzpicture}}
	{\leftrobinsonhorizontalbis}
	{\rightrobinsonhorizontalbis}
	{\lsortant}
	{\lsortant}
}

and \eqref{eq:rec:un} again at rank $n-1$ , we obtain

$$e_{0}^{n} \circ  \tikz[baseline={(current bounding box.center)}]{\myrobinsonhorizontalbis{0}{0}{0}{black!10}} \circ e_{0}^{n}
\Smodels \tikz[baseline={(current bounding box.center)}]
{\atrightmacrotiletwogenericnew{0}{0}{0}{white}{n}}$$

Similarly, we can obtain:

$$e_{0}^{n} \circ  \tikz[baseline={(current bounding box.center)}]{\myrobinsoncrossreflectedbis{0}{0}{180}{black!10}} \circ e_{0}^{n}
\Smodels \tikz[baseline={(current bounding box.center)}]
{\VNEmacrotiletwogenericnew{0}{0}{0}{white}{n}}$$

and all the other patterns of $e^{n+1}_{2}$ and $e^{n+1}_{-2}$.
\end{proof}

Consequently we only need to prove the following lemma to complete the proof of Theorem~\ref{thm:robinson:invariant:inductif}.

\begin{lemma} \label{lemmaProofMain4}
	For all $n$, we have
	\begin{align*}	
	H_{n} \circ \mathcal{T}_{Robi} \circ H_{n} &\Smodels e_{1}^{n+1};\\
	H_{n} \circ \mathcal{T}_{Robi} \circ H_{n} &\Smodels e_{3}^{n+1};\\
	H_{n} \circ \mathcal{T}_{Robi} \circ H_{n} &\Smodels e_{-1}^{n+1};\\
	H_{n} \circ \mathcal{T}_{Robi} \circ H_{n} &\Smodels e_{-3}^{n+1}.
	\end{align*}
\end{lemma}

\begin{proof}
We only prove the first assertion as the second is obtained by replacing the role of right and left
in the following reasoning.

The third and fourth assertions can be obtained by inverting the role of height $1$ and height $g(n)$ in the coming reasoning and by turning a rectangular pattern of some right angle.

So, we only need to prove:

\def\facteurreduction{0.5}

\begin{equation}\label{eq:lem:proof}
\begin{split}
H_{n} \circ \mathcal{T}_{Robi} \circ H_{n} \Smodels  \\
\Derivationdd{g(n+1)}
	{\tikz{\mymacrotilengeneric{0}{0}{0}{black!10}{3}{n+1}}}
	{\leftmymacrotileone{n+1}}
	{\rightmymacrotileone{n+1}}
	{\pattern{\lsortant}{n+1}}
	{\patterntriplet{\lentrant}{\lentrantdemidroit}{n}}
	\end{split}
\end{equation}
\end{proof}

This follows from the following arguments:

\begin{itemize}
\item
We have

$$\{e^{n}_1,e^{n}_2\}    \Smodels $$
$$
\Derivationdd{g(n)}
	{\tikz[scale=0.5]{\mymacrotilengeneric{0}{0}{0}{black!10}{3}{n}\atrightmacrotilengeneric{8}{0}{0}{white}{3}{n-1}}}
	{\leftmymacrotileone{n}}
	{\rightatrightmacrotiletwo{n}}
	{\pattern{\lsortant}{n} \lentrant}
	{\patterntriplet{\lentrant}{\lentrantdemidroit}{n} \lentrant}
$$

Then

\def\facteurreduction{0.4}

\begin{equation} \label{item:un}
\begin{split}
\{e^{n}_1,e^{n}_2,e^{n}_{3}\}    \Smodels \\
\Derivationdd{g(n)}
	{\tikz[scale=0.5]{\mymacrotilengeneric{0}{0}{0}{black!10}{3}{n}\atrightmacrotilengeneric{8}{0}{0}{white}{3}{n-1}\mymacrotilengeneric{16}{0}{90}{black!10}{3}{n}}}
	{\leftmymacrotileone{n}}
	{\rightmymacrotileoneqd{n}}
	{\pattern{\lsortant}{n+1}}
	{\patterntriplet{\lentrant}{\lentrantdemidroit}{n}\lsortant\patterntriplet{\lentrant}{\lentrantdemigauche}{n}}
	\end{split}
\end{equation}

\def\facteurreduction{1}

\item
As we have $\mathcal{T}_{Robi} \Smodels \Boucledd{1}
	{\tikz{\myrobinsonempty{0}{0}{-90}{black!10}}}
	{\leftmacrorobinsonemptymqd}
	{\lentrant}
	{\lsortant}$
we can repeat the loop any finite number of times and in particular $g(n)$ times to obtain
\myequation{eq:un:robinsonempty}{
\Boucled{1}
	{\tikz{\macrorobinsonemptygeneric{0}{0}{-90}{black!10}{25}{n}}}
	{\leftmacrorobinsonemptymqd}
	{\pattern{\lentrant}{n}}
	{\pattern{\lsortant}{n}}
	}

Similarly, as we have

\centerline{
$\mathcal{T}_{Robi} \Smodels \Boucledd{1}
	{\tikz{\myrobinsonverticalbis{0}{0}{90}{black!10}}}
	{\leftrobinsonemptyqd}
	{\lentrant}
	{\lsortant}$}

we have:
\begin{equation}\label{eq:gen:deux}\text{
\Boucled{1}
	{\tikz{\macrorobinsonverticalbisgeneric{0}{0}{90}{black!10}{25}{n}}}
	{\leftrobinsonverticalbisqd}
	{\pattern{\lentrant}{n}}
	{\pattern{\lsortant}{n}}
}\end{equation}

From \eqref{eq:un:robinsonempty} and   the fact that

\centerline{
\Boucled{1}
	{\begin{tikzpicture}\myrobinsonhorizontalbis{0}{0}{-90}{black!10}\end{tikzpicture}}
	{\leftrobinsonhorizontalbismq}
	{\lentrantdemidroit}
	{\lsortantdemidroit}
} 

and reusing  \eqref{eq:un:robinsonempty} we obtain

\centerline{
\Derivationd{1}
	{\tikz{\topmacrotiletwogenericnew{0}{0}{0}{white}{n-1}}}
	{\leftrobinsonhorizontalbismq}
	{\leftrobinsonhorizontalbismq}
	{\patterntriplet{\lentrant}{\lentrantdemidroit}{n}}
	{\patterntriplet{\lsortant}{\lsortantdemidroit}{n}}
}

We have

\centerline{
\Derivationd{1}
	{\begin{tikzpicture}\myrobinsoncornerdentedbis{0}{0}{0}{black!10}\end{tikzpicture}}
	{\leftrobinsoncornerdentedbis}
	{\rightrobinsoncornerdentedbis}
	{\lsortant}
	{\lentrantdemidroit}
}

From \eqref{eq:gen:deux},
\centerline{
\Boucled{1}
	{\begin{tikzpicture}\myrobinsoncrossbis{0}{0}{90}{black!10}\end{tikzpicture}}
	{\leftrobinsoncrossbisqd}
	{\lentrantdemigauche}
	{\lsortantdemigauche}

} and \eqref{eq:gen:deux}  again , we have

\def\facteurreduction{0.5}

\centerline{
\Boucled{1}
	{\tikz{\HNEmacrotiletwogenericnew{0}{0}{0}{white}{n-1}}}
	{\leftrobinsonverticalbisqd}
	{\patterntriplet{\lentrant}{\lentrantdemigauche}{n}}
	{\patterntriplet{\lsortant}{\lsortantdemigauche}{n}}
}

We deduce:

\myequation{item:deux}{
\Derivationd{1}
	{\tikz[scale=1]{\topmacrotiletwogenericnew{-5}{0}{0}{white}{n-1} \myrobinsoncornerdentedbis{0}{0}{0}{black!10}  \HNEmacrotiletwogenericnew{5}{0}{0}{white}{n-1}}}
	{\leftmacrorobinsonemptymqd}
	{\rightrobinsonverticalbisqd}
	{\patterntriplet{\lentrant}{\lentrantdemidroit}{n}\lsortant \patterntriplet{\lentrant}{\lentrantdemigauche}{n}}
	{\patterntriplet{\lsortant}{\lsortantdemidroit}{n} \lentrantdemidroit \patterntriplet{\lsortant}{\lsortantdemigauche}{n} }
}

\def\facteurreduction{0.7}

\item
We have

$$
\{e_{-1}^{n},e_{-2}^{n}\} \models
\Derivationdd{g(n)}
	{\tikz[scale=0.8]{\mymacrotilengeneric{0}{0}{-90}{black!10}{3}{n}\VNEmacrotilengeneric{8}{0}{0}{white}{3}{n-1}}}
	{\leftmymacrotileonemqd{n}}
	{\rightVNEmacrotiletwo{n}}
	{\patterntriplet{\lsortant}{\lsortantdemidroit}{n}    \lentrantdemidroit}
	{\pattern{\lentrant}{n} \lentrantdemidroit}
$$

Consequently

\begin{equation} \label{item:trois}
\begin{split}
\{e_{-1}^{n},e_{-2}^{n},e_{-3}^{n}\} \models \\
\Derivationdd{g(n)}
	{\tikz[scale=0.5]{\mymacrotilengeneric{0}{0}{-90}{black!10}{3}{n}\VNEmacrotilengeneric{8}{0}{0}{white}{3}{n-1}\mymacrotilengeneric{16}{0}{180}{black!10}{3}{n}}}
	{\leftmymacrotileonemqd{n}}
	{\rightmymacrotileonecqv{n}}
	{\patterntriplet{\lsortant}{\lsortantdemidroit}{n}   \lentrantdemidroit \patterntriplet{\lsortant}{\lsortantdemigauche}{n}  }
	{\patterntriplet{\lentrant}{\lentrantdemidroit}{n+1} }
	\end{split}
\end{equation}

\item Combining \eqref{item:un}, \eqref{item:deux} and \eqref{item:trois}, we obtain \eqref{eq:lem:proof}. This ends the proof.

\end{itemize}


\section{Two notions of robustness for tilesets}
\label{section.robustness_tileset}

\begin{definition}[Composition pattern]
Consider a finite set $S$ of transducers.
Consider the signature made of the symbols of $S$ (with arity $0$) and the symbol $\circ$ (of arity $2$).
A \emph{composition pattern} $F$ over~$S$ is a term over this signature. \end{definition}

For example $H \circ \tau \circ H$ is a composition pattern over $\{H,\tau\}$.  Such a composition pattern is interpreted as expected: $\circ$ corresponds to composition.

\subsection{Semantically robust tilesets}

\begin{definition}\label{defSemRobust}
A tileset $\tau$ is \emph{semantically robust} (or \emph{inductive})  if there exists a
family of transducers $(\mathcal{T}_{n})_{n}$,  of respective heights $(g(n))_{n}$, with $g:\N^*\to\N^*$ increasing and injective with $g(1)=1$ and some (fixed) composition pattern $F=F(\mathcal{T}_{1}, \mathcal{T}_{n-k},\dots, \mathcal{T}_{n})$ over $\mathcal{T}_{1}, \mathcal{T}_{n-k}, \dots, \mathcal{T}_{n} $ for some integer $k$, such that:
\begin{enumerate}
\item Initialisation:
$\tau \Smodels \mathcal{T}_{1}$, \dots, $\tau^{g(k+1)} \Smodels \mathcal{T}_{k+1}$ 
\item  $F$ provides an invariant:
\begin{equation} \label{eq:recurrence}
\forall n, \quad
F(\mathcal{T}_{1}, \mathcal{T}_{n-k},\dots, \mathcal{T}_{n}) \Smodels \mathcal{T}_{n+1}\end{equation}
\item For all $n$, $\mathcal{T}_{n}$ contains a loop.
\end{enumerate}
\end{definition}

\subsection{Properties}

\begin{proposition}\label{prop:direct}
If a tileset $\tau$ admits a tiling, then it is semantically robust.
\end{proposition}

\begin{proof}
Consider  $g(n)=n$ and $\mathcal{T}_{n} = \tau^{g(n)}$ for all $n$. We have 1) from definition, 
Condition 2) is satisfied, as we have $\mathcal{T}^{n+1}=\mathcal{T}^n\circ \mathcal{T}$ for all $n\geq0$. Condition 3) comes from Proposition \ref{proposition:loop_horizontal_strip}.
\end{proof}

\begin{proposition}\label{prop:indirect}
If a tileset $\tau$ is semantically robust (inductif), then it admits a tiling.
\end{proposition}

\begin{proof}
If $\tau$ is semantically robust, then by induction and combining conditions 1) and 2, we have that for all $n$, $\tau^{g(n)} \Smodels \mathcal{T}_{n}$.  By Proposition \ref{proposition:loop_horizontal_strip}, we can tile horizontal stripe of height $g(n)$ for arbitrary large $n$. By a compactness argument, we conclude that $\tau$  tile the plane.
\end{proof}

\subsection{Provably robust tilesets}

Their might be a strong difference between the fact that
$$
\forall n, \quad
F(\mathcal{T}_{1}, \mathcal{T}_{n-k},\dots, \mathcal{T}_{n}) \Smodels \mathcal{T}_{n+1}
$$ holds
for some composition pattern $F$,  written as
\[
\models  \exists F, \quad
\forall n, \quad
F(\mathcal{T}_{1}, \mathcal{T}_{n-k},\dots, \mathcal{T}_{n}) \Smodels \mathcal{T}_{n+1}
\]
and the fact that we can prove it holds, written
\[
\vdash \exists F, \quad
\forall n, \quad
F(\mathcal{T}_{1}, \mathcal{T}_{n-k},\dots, \mathcal{T}_{n}) \Smodels \mathcal{T}_{n+1}.
\]
This is for instance the case if a tileset admits a tiling, but we cannot prove the existence of any valid tiling. The same phenomenon happens for Turing machines, for which non-termination may hold while there is no proof of it. We will discuss these issues in Section \ref{sec:mt}. In the meantime we introduce the concept of \emph{provably robust} tileset to add this condition.


\begin{definition}\label{def:robust}
A tileset $\tau$ is \emph{provably robust} if there exists a family of transducers $(\mathcal{T}_{n})_{n}$, of respective heights $(g(n))_{n}$, with $g:\N^*\to\N^*$ increasing and injective, $g(1)=1$ and some (fixed) composition pattern $F=F(\mathcal{T}_{1}, \mathcal{T}_{n-k},\dots, \mathcal{T}_{n})$ over $\mathcal{T}_{1}, \mathcal{T}_{n-k}, \dots, \mathcal{T}_{n}$ for some integer $k$,
\begin{enumerate}
\item Initialisation:  $\tau \Smodels \mathcal{T}_{1}, \dots, \tau^{g(k+1)} \Smodels \mathcal{T}_{k+1}$

\item  $F$ provides an invariant, and this holds provably:
\begin{equation}
\forall n, \quad
F(\mathcal{T}_{1}, \mathcal{T}_{n-k},\dots, \mathcal{T}_{n}) \Smodels \mathcal{T}_{n+1}
%
\end{equation}
\item For all $n$, $\mathcal{T}_{n}$ contains a loop.
\end{enumerate}
\end{definition}

\begin{proposition}
If a tileset $\tau$ is provably robust then it admits a tiling.
\end{proposition}

\begin{proof}
	If we have $\vdash \phi$ for some formula $\phi$, then we have $\models \phi$. We did not what notion of proof $\vdash$ we use (this could be provability in ZF set theory or provability in first-order logic starting from Peano's axioms of arithmetic), but of course, we expect it to be sound. Consequently, if $\tau$ is provably robust it is semantically robust.
\end{proof}

\begin{proposition}
A periodic tileset is provably robust.
\end{proposition}

\begin{proof}
Let $\tau$ be a periodic tileset. Let $x$ be a tiling by $\tau$. Then it admits a period $(a,b)\in\Z^2\setminus\{(0,0)\}$ such that for all $(u,v)\in\Z^2$
\[
x(u,v) = x(a+u,b+v).
\]
Consider $\mathcal{T}$ the transducer associated with $\tau$. Then by periodicity of $x$ the transducer $\mathcal{T}^{|b|}$ contains a cyclic loop. By Proposition~\ref{proposition:circular_loop_periodic_loop} there exists $1\leq m\leq |b|$ such that $\mathcal{T}_1:=\mathcal{T}^{m\cdot|b|}$ contains a periodic loop. Consequently for every $n\in\N^*$ the transducer $\mathcal{T}_n:=\mathcal{T}^{n\cdot m\cdot|b|}$ contains a periodic loop. We are now ready to prove that $\tau$ is provably robust. Define $g(n):=n\cdot m\cdot|b|$, $k=0$ and $F$ the composition pattern $F(\mathcal{T},\mathcal{T}'):=\mathcal{T}\circ\mathcal{T}'$. Then all three conditions of provably robustness are satisfied:
\begin{enumerate}
\item $\mathcal{T}^{g(1)}=\mathcal{T}_1\Smodels \mathcal{T}_1$;
\item for all $n\in\N^*$, $F(\mathcal{T}_1,\mathcal{T}_n)=\mathcal{T}_1\circ\mathcal{T}_n=\mathcal{T}_{n+1}\Smodels\mathcal{T}_{n+1}$;
\item for all $n\in\N^*$, $\mathcal{T}_{n}$ contains a periodic loop.\qedhere
\end{enumerate}
\end{proof}

\subsection{On the role played by non-homogeneous linear recurrences}
\label{sec:nhlr}

We see that any tileset that admits a tiling necessarily has some invariant and vice-versa. The crucial question is whether this invariant is \emph{simple} or not and whether it is provable or not. By \emph{simple} we have in mind something similar to the invariant~\eqref{eq:thm:robinson:invariant:inductif} on page~\pageref{eq:thm:robinson:invariant:inductif} satisfied by Robinson's tileset. This invariant involves a meta-transducer with a constant number of edges (namely 7 edges $e_{-3}^{n}, \dots, e_{3}^{n}$), while in the general case, it may not be the case for  the invariant~\eqref{eq:recurrence} from Definition~\ref{defSemRobust}.

In the general case, as the height of $\mathcal{T}_{n+1}$ is  $g(n+1)$,  relation~\eqref{eq:recurrence} implies that $g(n+1)$ is necessarily obtained as the sum of the heights of the $\mathcal{T}_{i}$ involved in the composition pattern $F$. Consequently, $g$ must satisfy a (non-homogeneous) linear recurrence of the form
\[
g(n+1)=c_{0} g(n) +c_{1} g(n-1) +\dots + c_{k} g(n-k)+c
\]
for some non-negative integers $c_{0},\dots,c_{k}$ and $c$. As an example $g(n)=n$ from the proof of Proposition~\ref{prop:direct} satisfies $g(n+1)=g(n)+1$.

The previous remark, combined with the coming statements explains why tiling discussed in the literature is closely related to non-homogeneous linear recurrences: height $g(n)=2^{n}-1$ appears in Robinson's tiling. Similarly,  Fibonacci numbers appear in Jeandel-Rao's tiling \cite{jeandel2021aperiodic} discussed in Section \ref{section.JR}.

\subsection{Robinson's tileset is robust}
\label{subsection.examples_robust_tilesets}

\begin{proposition}\label{prop:Tn_Robi_contains_loop}
For all $n \ge 1$, $\mathcal{T}_{n}$ contains a loop.
\end{proposition}

\newcommand\IGNOREECRITURE[1]{}

\begin{remark}
For $n=1$, such a loop is

\begin{center}
\begin{tikzpicture}[scale=1,shorten >=1pt,>={Stealth[round]}]
	\node[state,rectangle,rounded corners=3pt] (V) {\nletterRobiV{-90}};
	\path[->] (V) edge[out=30,in=330,looseness=6, right] node{\patternLoopRobi{1}} (V);
\end{tikzpicture}
\end{center}

This loop corresponds to infinite horizontal tiling:

\begin{center}
\begin{tikzpicture}[scale=0.8]
\node at (-1.5,0.5) {\scalebox{1}{$\dots$}};
\robinsonemptybis{-1}{0}{0}{white}
\robinsoncornerbumpybis{0}{0}{0}{black!10}
\robinsoncornerbumpybis{3}{0}{90}{black!10}
\robinsonhorizontalbis{1}{0}{0}{white}
\robinsonemptybis{3}{0}{0}{white}
\robinsoncornerbumpybis{4}{0}{0}{black!10}
\robinsoncornerbumpybis{7}{0}{90}{black!10}
\robinsonhorizontalbis{5}{0}{0}{white}
\robinsonemptybis{7}{0}{0}{white}
\robinsoncornerbumpybis{8}{0}{0}{black!10}
\robinsoncornerbumpybis{11}{0}{90}{black!10}
\robinsonhorizontalbis{9}{0}{0}{white}
\robinsonemptybis{11}{0}{0}{white}
\node at (13,0.5) {\scalebox{1}{$\dots$}};

\end{tikzpicture}
\end{center}

\end{remark}

\begin{proof}
An attentive reader will be convinced that the following loop is contained in $e_{0}^{n} \cup e_{1}^{n} \cup e_{2}^{n} \cup e_{3}^{n}$ for every $n\in\N$.
\begin{center}
\begin{tikzpicture}[scale=1,shorten >=1pt,>={Stealth[round]}]
	\node[state,rectangle,rounded corners=3pt] (V) {$\pattern{\nletterRobiV{-90}}{n}$};
	\path[->] (V) edge[out=30,in=330,looseness=6, right] node{
	\tikz[scale=0.21]{\mymacrotilengeneric{0}{0}{0}{black!10}{3}{n}\atrightmacrotilengeneric{8}{0}{0}{black!10}{3}{n-1}\mymacrotilengeneric{16}{0}{90}{black!10}{3}{n}
	\macrorobinsonemptygeneric{24}{0}{0}{black!10}{4}{n}}}
	 (V);
\end{tikzpicture}
\end{center}

\end{proof}

\begin{proposition}\label{proposition:Robinson_provably_robust}
Robinson's tileset is provably robust.
\end{proposition}

\begin{proof}
We prove that the three items of Definition~\ref{defSemRobust} are satisfied: define $\mathcal{T}_{1}:=\mathcal{T}_{Robi}$ and $\mathcal{T}_{n}:=H_{n}$.  The first two items follow from Theorem \ref{thm:robinson:invariant:inductif}, considering the pattern $F(H_{n},\mathcal{T}_{1})= H_{n} \circ \mathcal{T}_{1} \circ H_{n}$. The last item follows from Proposition~\ref{prop:Tn_Robi_contains_loop}.
\end{proof}

\begin{corollary}
Robinson's tileset admits a tiling.
\end{corollary}

\subsection{Extension: Dealing with more general shapes}

Wang tiles are unit squares and can be arranged, provided local constraints are respected, to compose bigger squared patterns, rectangular patterns or patterns with more general shapes. Reasoning about non-rectangular patterns sometimes reveal easier, so we present here how the formalism of transducers can be adapted.

%
%
%
%
%

\begin{example} \label{ex:jr:moinsdeux}
Suppose we start from a Wang tileset and are able to build the following rectangular patterns.
%
%
%
%
%
\medskip

\renewcommand\NOCOLOR[1]{}
\renewcommand\BW[1]{#1}

\begin{tikzpicture}[xscale=0.4,yscale=0.25]
\grbeta{20}{-2}{0} 

\grgamma{16}{0}{0}: 
\grepsilon{3}{0}{0}

\grA{0}{0}{1}
\grB{9}{0}{1}
\grD{0}{-2}{1}
\grE{13}{-2}{1}
\end{tikzpicture}

By composing these patterns, and respected the local constraints given by the black decorations, patterns as pictured below can be obtained.

\medskip

\begin{tikzpicture}[xscale=0.4,yscale=0.25]
\grDdeux{0}{0}{0}
\end{tikzpicture}
\end{example}

\renewcommand\NOCOLOR[1]{#1}

We choose to define generalised shapes as supports of connected combinations of horizontal patterns. In particular, any connected combination of rectangular shapes similar to the one in Example~\ref{ex:jr:moinsdeux} falls under this definition.

\begin{definition}[Generalised shape]
A \emph{generalised shape} $S$ of height $h$ is a connected subset of $\Z \times [[0 \dots h-1]]$, such that for all $0 \le j < h$,
$S_{j}=\{i: (i,j) \in S\}$ is an non-empty interval $[[\min_{j} \dots \max_{j}]]$.
\end{definition}

The south (resp. north, left, right) frontier of a generalised shape is given by the interval $[[\min_{0}\dots\max_{0}]]$ (resp.  the interval $[[\min_{h-1}\dots\max_{h-1}]]$,  the sequence $\min_{0},\dots,\min_{h-1}$, the sequence $\max_{0},\dots,\max_{h-1}$). 

Such a sequence $\min_{0},\dots,\min_{h-1}$ (resp. $\max_{0},\dots,\max_{h-1}$) will be called a left (resp. right) frontier. A left frontier can be encoded, up to a translation, by the sequence of $h$ integers $d_{0}=\min_{0}$, $d_{1}=\min_{1}-\min_{0},d_{2}=\min_{2}-\min_{1}, \dots, d_{h-1}=\min_{h-1}-\min_{h-2}$. A right frontier can be encoded in a similar way.

%

A \emph{coloured frontier} is given by a left or right frontier $d_{0},\dots,d_{h-1}$ together with $h$ horizontal colours and $|d_{0}|+\dots+|d_{h-1}|$ vertical colours.

We can generalise our concept of transducers so that states are now coloured frontiers. There is an edge between two such states $q$ and $q'$  if there is a valid tiling of the shape whose left frontier is $q$ and the right frontier is $q'$. This edge is labelled by the word $w|w'$ corresponding to the vertical colours of the south and north frontiers of the shape. We can define a composition operation $\circ$ on transducers,  that corresponds to testing the vertical compatibility of edges.

\subsection{Extension: Tiling equivalence}

In Section~\ref{subsection.transducers} we defined meta-transducers not to get new or stronger results, but rather to simplify notations and proofs. With the same objective in mind we define a notion of equivalence between tilesets that are not necessarily Wang tilesets.

If $\tau$ is a tileset, we say that a tileset $\widetilde{\tau}$ is \emph{equivalent} to $\tau$ if (1) each tile in $\widetilde{\tau}$ is a valid pattern on $\tau$ (2) every tiling $x$ by $\tau$ can be decomposed into patterns from $\widetilde{\tau}$.

\begin{definition}
A tileset $\tau$ is \emph{tiling equivalent} to a tileset $\tau'$ if there exist $\widetilde{\tau}$ (resp. $\widetilde{\tau}'$) equivalent to $\tau$ (resp.  to $\tau'$) and a provable bijection $\Phi:\tau\to\tau'$ such that $x$ is a tiling by $\widetilde{\tau}$ iff $\Phi(x)$ is a tiling by $\widetilde{\tau}'$.
\end{definition}

\begin{example} \label{ex:jr}
The tileset $\tau_1$ made of the geometric tiles with decorations pictured below

\begin{tikzpicture}[xscale=0.45,yscale=0.25]
\grbeta{1}{0}{0}
\grgamma{7}{0}{0}
\grdelta{0}{-2}{0}
\grepsilon{9}{-2}{0}

\grA{0}{-4}{1}
\grB{3}{-4}{1}
\grC{10}{-4}{1}
\grD{0}{-6}{1}
\grE{12}{-6}{1}
\end{tikzpicture}

is tiling equivalent to the tileset $\tau_2$ pictured below, where not two north (resp. south) black decorations have the same width.

\renewcommand\NOCOLOR[1]{}
\renewcommand\BW[1]{#1}

\begin{tikzpicture}[xscale=0.45,yscale=0.25]
\grbeta{1}{0}{0}
\grgamma{7}{0}{0}
\grdelta{0}{-2}{0}
\grepsilon{9}{-2}{0}

\grA{0}{-4}{1}
\grB{3}{-4}{1}
\grC{10}{-4}{1}
\grD{0}{-6}{1}
\grE{12}{-6}{1}
\end{tikzpicture}

\renewcommand\NOCOLOR[1]{#1}
\renewcommand\BW[1]{}
Indeed, from a tiling from the first, by ignoring colours we get a tiling from the second. Conversely, a tiling from the second can be transformed into a tiling from the first, as colours can be recovered by only looking at the width of black decorations. 
\end{example}

\begin{proposition}\label{proposition.robust_aperiodicity_tiling_equivalence}
Provably robustness and aperiodicity are preserved by tiling equivalence.
\end{proposition}

\begin{proof}
For provably robustness this follows directly from the definitions, since the bijection involved in tiling equivalence is chosen provable. Assume that $x$ is a periodic tiling by $\tau$ which is tiling equivalent to $\tau'$ through sets of patterns $\widetilde{\tau}$, $\widetilde{\tau}'$ and a provable bijection $\Phi$. Then $\Phi(x)$ is a tiling by $\widetilde{\tau}'$ that is also periodic. So aperiodicity is also preserved by tiling equivalence.
\end{proof}

\section{Jeandel-Rao's tileset}
\label{section.JR}

Jeandel Rao's aperiodic tileset $\mathcal{T}$ is the Wang tileset considered in~\cite[Section 4]{jeandel2021aperiodic}, composed with the 11 tiles pictured below.
\begin{center}
\begin{tikzpicture}[scale = 0.8]
\wang{0}{0}{rouge}{rouge}{rouge}{vert}
\wang{1.5}{0}{bleu}{rouge}{bleu}{rouge}
\wang{3}{0}{rouge}{vert}{vert}{vert}
\wang{4.5}{0}{white}{bleu}{rouge}{bleu}
\wang{6}{0}{bleu}{bleu}{white}{bleu}
\wang{7.5}{0}{white}{white}{rouge}{white}
\wang{9}{0}{rouge}{vert}{bleu}{white}
\wang{10.5}{0}{blue}{white}{bleu}{rouge}
\wang{12}{0}{bleu}{rouge}{white}{rouge}
\wang{13.5}{0}{vert}{vert}{bleu}{rouge}
\wang{15}{0}{rouge}{white}{rouge}{vert}
\end{tikzpicture}
\end{center}

The two tilesets of Example~\ref{ex:jr} are tiling equivalent to the Jeandel-Rao's Wang tileset~\cite{jeandel2021aperiodic}. More pictures of relevant patterns made with this tileset can be found in Appendix \ref{sec:dessins:jr}.  

\begin{theorem}\label{th:jeandelrao}
Jeandel Rao's aperiodic tileset $\mathcal{T}$ is provably robust.
\end{theorem}

\begin{proof}
The discussion of \cite{jeandel2021aperiodic} proves that $\mathcal{T}$ is tiling equivalent to
a transducer made of the union of $T_{0}$, $T_{1}$ and $T_{2}$. Considering $g(0)=1$, $g(1)=2$ and $g(n+2)=g(n)+g(n+1)$ for all $n$, these transducers are defined by
\begin{itemize}
\item for $n$ even, as:

\begin{center}
\begin{tikzpicture}[baseline={(current bounding box.center)},scale=0.5]
\node[state] at (0,2) (a) {};
\node[state] at (6,2) (b) {};
\path[->]   (a) edge[bend left, above] node {
	$\alpha_{n}: 0^{g(n+2)-3} \mid 1^{g(n+2)-3}$
	} (b);
\path[->]   (b) edge[bend left, below] node {
	$\begin{array}{ll}
\beta_{n}: 1^{g(n+1)+3} & \mid(110) 0^{g(n+1)} \\
\gamma_{n}: 1^{g(n+3)+3} & \mid 0^{g(n+2)}(111) 0^{g(n+1)} \\
\delta_{n}: 1^{g(n+1)}(000) 1^{g(n+2)} & \mid 0^{g(n+3)+3} \\
\epsilon_{n}: 1^{g(n+1)}(100) & \mid 0^{g(n+1)+3} \\
\omega_{n}: 1^{g(n+3)}(100) 1^{g(n+1)} & \mid 0^{g(n+1)}(110) 0^{g(n+3)}
\end{array}$
	} (a);
\end{tikzpicture}
\end{center}
\item  for $n$ odd, as:

\begin{center}
\begin{tikzpicture}[baseline={(current bounding box.center)},scale=0.5]
\node[state] at (0,2) (a) {};
\node[state] at (6,2) (b) {};
\path[->]   (a) edge[bend left, above] node {
	$A_n: 1^{g(n+2)-3} \mid  0^{g(n+2)-3}$
	} (b);
\path[->]   (b) edge[bend left, below] node {
	$
\begin{array}{ll}
B_{n}: 0^{g(n+1)+3} & \mid(100) 1^{g(n+1)} \\
C_{n}: 0^{g(n+3)+3} & \mid 1^{g(n+2)}(000) 1^{g(n+1)} \\
D_{n}: 0^{g(n+1)}(111) 0^{g(n+2)} &\mid 1^{g(n+3)+3} \\
E_{n}: 0^{g(n+1)}(110) & \mid 1^{g(n+1)+3} \\
O_{n}: 0^{g(n+3)}(110) 0^{g(n+1)}&  \mid 1^{g(n+1)}(100) 1^{g(n+3)}
\end{array}$
	} (a);
\end{tikzpicture}
\end{center}
\end{itemize}

\begin{figure}[!ht]
\begin{center}
\begin{tikzpicture}[scale=0.3]
\grAtrois{0}{0}{0}
\end{tikzpicture}
\end{center}
\caption{\label{jr:acinq} The pattern that corresponds to the edge $A_{5}$  pictured above, composed with tiles from the tileset $\tau_1$ from Example~\ref{ex:jr}.}
\label{figure:A5}
\end{figure}
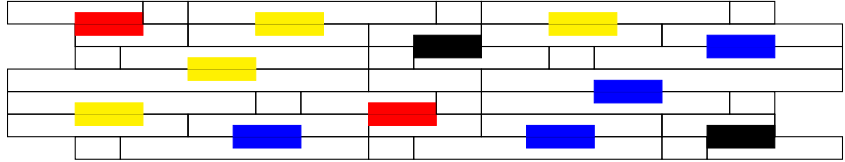

The proof of  \cite[Lemma 9]{jeandel2021aperiodic} can be rephrased as a proof that  for every $n$ even one has
\[
(A_{n+1} \cup C_{n+1}) \circ (\gamma_{n} \cup \alpha_{n} \cup \delta_{n})  \circ (D_{n+1} \cup A_{n+1}) \Smodels A_{n+3}.
\]
Similar expressions can be established for $B_{n+3}$, $C_{n+3}$, $D_{n+3}$, $E_{n+3}$, $O_{n+3}$, from the proof of same Lemma. 
For $n$ odd we can formulate similar relations, mainly by inverting the role of Greek and Latin letters. It follows that $T_{n+1} \circ T_n\circ T_{n+1} \Smodels T_{n+3}$. The fact that $T_{n}$ contains a loop is clear from its definition.
\end{proof}

\begin{remark}
The set of geometric tiles $\tau_2$ from Example~\ref{ex:jr} is tiling equivalent to Jeandel-Rao's tileset, which is known to be aperiodic. Hence by Proposition~\ref{proposition.robust_aperiodicity_tiling_equivalence} $\tau_2$ is also aperiodic.
\end{remark}


Let us focus on the size of the words that label the edges $\alpha_n,\beta_n,\gamma_n,\delta_n,\epsilon_n,\omega_n$ and $A_n,B_n,C_n,D_n,E_n,O_n$. A careful observation leads to the conclusion that all patterns produced must have a width that satisfies the recurrence relation
\[
g(n+2)=g(n+1)+g(n)
\]
which determines the Fibonacci numbers with $g(0)=0$ and $g(1)=1$. In general a tileset always satisfies some recurrence relation as~\eqref{eq:recurrence} from Definition~\ref{defSemRobust}. If additionally we ask for meta-transducers with a constant number of edges, as it is the case for Jeandel-Rao's tileset, then the same reasoning as above on the widths leads to an analogous conclusion: we get a sequence of arbitrarily large valid patterns whose widths satisfy some (non-homogenous) linear recurrence relation.


\section{Robust and non-robust Turing machines}
\label{sec:mt}

\subsection{Turing machines}

Following \cite{Sip97}, a (one-tape) we define a Turing machine as a 7-tuple $M=\left\langle Q, \Sigma, \Gamma, \delta, q_0, q_{accept},q_{reject} \right\rangle$ where $Q,\Sigma,\Gamma$ are finite non-empty sets, $B\in\Gamma$ is a blank symbol such that
\begin{itemize}
\item $Q$ is the set of states;
\item  $\Sigma$ is the set of input symbols,  with $B\notin\Sigma$.
\item  $\Gamma$ is the set of tape symbols, with $\Sigma \subseteq \Gamma$;
\item $\delta: Q \times \Gamma \rightarrow Q \times \Gamma \times\{\mathrm{L}, \mathrm{R}\}$ is the transition function, with $\mathrm{L}$ the left shift and $\mathrm{R}$ the right shift. If $\delta$ is not defined on some $(q,\gamma)\in Q\times\Gamma$ then the machine halts;
\item $q_0\in Q$ is the initial state;
\item $q_{accept} \in Q$ is the accept state;
\item $q_{reject} \in Q$ is the reject state, with $q_{reject} \neq q_{accept}$.
\end{itemize}

We assume our readers are familiar with Turing machines and their classical properties, we refer to~\cite{Sip97} for a complete presentation. The machine $M$ \emph{halts} on input $w$ if it eventually reaches state $q_{accept}$ or $q_{reject}$. A \emph{configuration} $C$ describes an instantaneous state of the machine: the state $q \in Q$, the tape content, and the head location.  We write $q=\internalstate(C)$ for the internal state of configuration $C$. We write  $\nothalt(C)$ for the formula $\internalstate(C) \neq q_{accept} \wedge \internalstate(C) \neq q_{reject} $.
The configuration $C[w]$ is the initial configuration associated with input $w$ ($w$ is written on the tape, $\internalstate(C)=q_0$ and the head is in front of the first letter of $w$). We write $C \vdash C'$ if configuration $C'$ is the immediate next configuration of configuration $C$.

\subsection{Robust Turing machine}

In this section and for the rest of the paper we fix some concept of proof: say, provability in first-order logic starting from Peano's axioms of arithmetic.

On some input $w$, a Turing machine either halts (in that case the sequence of successive configurations starting on $C[w]$ provides a proof of that fact) or it does not.  From G\"odel's incompleteness theorem, there is a difference between provable valid statements and valid statements. We thus distinguish the case where there exists a proof of non-termination from the case where there is no proof of it.
This leads to the introduction of the following concept:

\begin{definition}[Robust Turing machine]\label{definition.robust_TM}
	A Turing machine $M$ is \emph{robust} if for all inputs, either the machine $M$ halts, or there is proof that it does not halt.
\end{definition}

	In other words, a non-robust Turing machine is a Turing for which there is an input $w$ on which it does not halt, but for which there is no proof of that fact.

\begin{theorem}[A statement for Turing machines]\label{th:mt}
	The halting problem is decidable for robust Turing machines.
\end{theorem}

\begin{proof}
Given some Turing machine $M$ and some input $w$, run the two following procedures in parallel
\begin{enumerate}
\item simulate $M$ on $w$: if $M$ accepts $w$, then halt and accept; if $M$ rejects $w$, then halt and reject.
\item search for a proof $\pi$ of the fact that $M$ does not halt on $w$. If such a proof is found, then halt and reject.
\end{enumerate}
By the robustness hypothesis, this algorithm always halts and is correct.
\end{proof}

Definition~\ref{definition.robust_TM} has similarities with the concept of robustness for Turing machines, and more general dynamical systems, considered in \cite{asarin01perturbed}.  In~\cite{CSL24} the authors extended these concepts to complexity theory~\cite{CSL24}.  They also relate these concepts to $\delta$-decidability~\cite{gao2012delta}, or to statements such as~\cite{ratschan2023deciding}.   In all these references, the reasoning for establishing decidability can be interpreted as a variation on the previous proof, by playing on the concept of proof $\pi$ involved: the problem is proved c.e. and co-c.e., hence decidable. 

\begin{remark}
There is no "free lunch theorem": One can not decide if a Turing machine is robust. However, see the coming discussion (Example \ref{ex:nrtm} and Proposition \ref{prop:ne}) that can be related to the informal conjecture discussed in \cite{asarin01perturbed} about the fact that ``undecidability does not happen in practice'' (i.e. for, say, ``concrete programs of real life'') unless there have been devised on purpose to be related to unprovable statements, as in Example \ref{ex:nrtm}.
\end{remark}

\subsection{Uncompleteness and relative completeness}

The statement of Theorem~\ref{th:mt} might be considered as frustrating, as it does not provide an understanding of what proof of non-termination looks like.  We show in this section that this corresponds to finding an invariant, preserved by the machine's transitions.  We believe that the parallel between what is done in this section for Turing machines and our work on tilesets will help our readers to understand the essence of our results for tilesets.


 Turing machines are closely related to programs. To formally express the properties of programs, we can use a logic language $Assn$. We can use standard first-order logic for this purpose (with the signature of arithmetic). Any other choice of sufficiently expressive language, able to  talk about words, configurations,  and sequences, could have been appropriate.

From G\"odel's incompleteness theorem we get:

\begin{theorem}[{Uncompleteness, see e.g. \cite[Proposition 7.1]{winskel1993formal}}]
	There is no effective proof system for $Assn$ such that the theorems coincide with the valid assertions of $Assn$.
\end{theorem}

Asking whether a program terminates or not is equivalent to asking about its partial correctness.

\begin{proposition}[{Uncompleteness, see e.g. \cite[Proposition 7.2]{winskel1993formal}}]
	There is no effective proof system for partial correctness assertions such that its theorems are precisely valid partial correctness assertions.
\end{proposition}

However, we can devise \emph{relatively complete} proof systems. A famous example is Hoare's logic. In this logic, a partial correctness assertion has the form :
$
\{A\} c\{B\},
$
where $A$ and $B$ are in $Assn$, and $c$ is a program.
Such an assertion states that starting from any initial configuration $C$ satisfying $A$ if the system evolves according to the program -- or Turing machine -- $c$, then it will either diverge or reach a configuration $C'$ satisfying $B$.
Formulas $A$ and $B$ are allowed to depend on some parameters, i.e. some interpretation $I$ of some variables.

\begin{theorem}[{Relative completeness \cite[Theorem 7.5]{winskel1993formal}}] \label{hlrc} Hoare's logic is relatively complete.
\end{theorem}

Observe that the Hoare triplet $\{A\} c\{false\}$ holds iff the program $c$ diverges from any initial configuration $C$ satisfying $A$. This provides a relatively complete method for proving non-termination. A Turing machine can be seen as a program corresponding to a while loop, and so we can focus on the case of a while loop.

Using the same ideas as in the proof of Theorem \ref{hlrc}, we can derive the following Theorem. Given a configuration $C$, we write $C \models^{I} \phi$ for the fact that $\phi \in Assn$ holds in $C$. As in Hoare's logic, this is a semantic notion ($I$ stands for interpretation). 

\begin{theorem}[Relative Completeness] \label{mainth}
	A Turing machine $M$ provably does not halt (resp. does not halt on input $w$)
	iff there is some formula $A^{*} \in Assn$ such that:
	\begin{enumerate}
		\item Initialisation:  $C[w] \models^{I} A^{*}$:  any (resp. the)  initial configuration satisfies $A^{*}$
		\item $A^{*}$ is provably an invariant: 
		\[
		\frac{ C \models^{I} A^{*}  \quad C \vdash C' }
		{ C' \models^{I} A^{*} }
		\text{invariance of }A^{*}
		\]
		\item $A^{*}$ implies non-termination:
		$C \models^{I} A^{*} \Rightarrow \nothalt(C)$.
	\end{enumerate}
\end{theorem}

%
\subsection{Proof of Theorem \ref{mainth}}

Clearly, if there is such an $A^{*} \in Assn$, then the Turing machine $M$ is not terminating. Conversely, the point is that there always exists such an $A^{*}$.

\begin{remark}
Hoare's logic includes the following:
\begin{itemize}
\item
Rule for while loops:
$$
\frac{\{A \wedge b\} c\{A\}}{\{A\} \text { while } b \text { do } c\{A \wedge \neg b\}}
$$
\item Rule of consequence:
$$
\frac{\models^{I} \left(A \Rightarrow A^{\prime}\right) \quad\left\{A^{\prime}\right\} c\left\{B^{\prime}\right\} \quad \models^{I} \left(B^{\prime} \Rightarrow B\right)}{\{A\} c\{B\}}
$$
\end{itemize}
The rule for while loops corresponds to point 2) of the previous theorem, for the special case of triplets of the form $\{A\} c \{false\}$, when $c$ corresponds to a Turing machine.
\end{remark}

Namely, given a triplet $\{A\} c\{B\}$,  the idea is to reason about the weakest precondition: this is a predicate $Q=wp(c,B)$ such that for any precondition $A$, $\{A\} c \{B\}$ if and only if $A \Rightarrow Q$. This is the least restrictive requirement needed to guarantee that $B$ holds after $A$.

\begin{remark} The uniqueness of the weakest precondition follows easily from the definition:  If both $Q$ and $Q'$ are weakest preconditions, then by the definition $\{Q'\} c\{B\}$ so $Q' \Rightarrow Q$ and $\{Q'\} c\{B\}$ so $Q \Rightarrow Q'$ and hence $Q=Q'$.
\end{remark}

Then the key is that $Assn$ (first-order logic or arithmetic), or (any sufficiently expressive logic) is \emph{expressive}:  for every program $c$ and assertion $B \in Assn$ there is an assertion $A_{0} \in Assn$ corresponding to $wp(c,B)$.

\begin{theorem}[{\cite[Theorem 7.5]{winskel1993formal}}] \label{hlexp} $Assn$ is expressive.
\end{theorem}

As a consequence, $\{A\} c \{B\}$ holds iff only $A \Rightarrow A_{0}$ iff it can be established using the rule of consequence in Hoare's logic.  In the specific case of a triplet $\{A\} c \{false\}$ corresponding to the non-termination of Turing machine $C$, we get  the previous theorem.

\begin{remark} \label{rq:byhand}
	It is also possible to do an independent direct proof of the theorem. Consider $\NotHalt$ to be the set of configurations $C$ from which the execution of the Turing machine diverges. If we write $\mathcal{C}$ for the set of configurations, the trick is that for a configuration $C$ and an interpretation $I$, we have $C \in \NotHalt$ iff

	\begin{equation}
		\label{equnprefere}
		\begin{aligned}
			& \forall k \forall C_0, \ldots, C_k \in \mathcal{C} 
			 {\left[C=C_0 \&\right.} \\
			& \forall i(0 \leq i<k) .\left(C_i \models^I  \nothalt(C_{i}) \&\right. \\
			& \left.\qquad C_i \vdash C_{i+1}\right)  
			 \Rightarrow\left(C_k \models^I  \nothalt(C_{k}) \right)
		\end{aligned}
	\end{equation}

	This can be written as an equivalent formula of $Assn$, using $\beta$-predicate of G\"odel (see e.g. \cite[Lemma  7.4]{winskel1993formal}).  The obtained formula satisfies all the conditions 1), 2) and 3) of Theorem \ref{mainth}.
\end{remark}

In short, relatively complete means the difficulty is ``just'' being able to prove assertions such as
$A \Rightarrow w [ c, B ]$ (i.e. establishing $\models A \Rightarrow \NotHalt$) whenever $\models A \Rightarrow \NotHalt$, but not on the method of Theorem  \ref{mainth}.

And Theorem \ref{mainth}, means that a Turing machine is non-provably terminating iff there is an invariant that is preserved by transitions of the Turing machine and that holds provably.

\subsection{Examples of non-probably robust Turing machines}

By some argument of denumerability or by the fact that a computably enumerable set and co-computably enumerable set is recursive and that the halting problem of Turing machines is computable enumerable and non-recursive, we can deduce that there are some non-robust Turing machines.

It is possible to provide an explicit example.
The following example, inspired by \cite{kozen1997automata}, is based on two well-known facts from computability and logic\
\begin{enumerate}
	\item From Kleene's fixed point theorem, we can consider that a Turing machine can obtain its own code, using the terminology and its presentation in \cite{Sip97}.
	\item From logic that given the code $m$ of some Turing machine $M$, we can explicitly write some first-order arithmetic formula $\gamma_{m,w}$ that is true if $M$ accepts input $w$: this formula states, using the $\beta$-function from G\"odel, that there is some finite sequence of successive configurations of $M$, starting from the initial configuration corresponding to $w$, ending with some accepting configuration. This is basically the formula considered in Remark \ref{rq:byhand}. We write $\epsilon$ for the emptyword.
\end{enumerate}

\begin{example}[An example of non-robust Turing machine]\label{ex:nrtm}
	Let $M$ be the Turing machine that, on any input $w$ and using the recursion theorem, $M$ obtains its code $m$. Then it constructs the  arithmetic formula $\psi= \gamma_{m,\epsilon}$. Then, it enumerates all the provable formulas until it finds the formula $\psi$ in the iteration. If this loop eventually terminates, then it accepts (otherwise, it runs forever, of course).
\end{example}

We claim that $M$ is not terminating. The formula $\psi=\gamma_{m,\epsilon}$ is not provable: indeed, $\psi$ is true iff $m$ does not accept the empty word and if $M$ finds a proof of $\psi$, then $M$ accepts the empty word and hence formula $\psi$ is wrong. Unless the arithmetic is not coherent there is no way to prove some false formula, hence this case can not happen.  Hence, necessarily, the loop will run forever. Now, if $M$ does not find a proof of $\psi$, then $M$ does not accept the empty word by construction. In other words, $\psi$ is some valid formula that is not provable and $M$ is not terminating, but there is no proof of it in arithmetic.

\begin{proposition}[Non-robustness is necessarily related to some valid but non-provable formula] \label{prop:ne}
	We see that this example involves the existence of some valid sentence of arithmetic whose validity is non-provable. This is a necessary condition.
\end{proposition}

\begin{proof}
If a Turing machine $M$ is non-robust, it does not terminate but there is no proof of it on some input $w$.  We can always consider the arithmetic formula $\psi=\gamma(m,w)$ which states that the machine $M$ is not terminating. It is true, but we have no proof of $\phi$ by hypothesis. In other words,  we have some explicit examples of the formula, namely $\psi$ that is valid but not provable.  Hence, non-robustness necessarily involves some valid formula but non-provable formulas.
\end{proof}

\section{Kari's tilings}
\label{section.KC}

In this section, we briefly present a technique due to Kari to encode piecewise rational affine maps computations inside Wang tilings.  We encourage the readers to refer to~\cite{Kari2007} for details about the construction.

\subsection{Balanced representation of real numbers}

Let $i \in \mathbb{Z}$. We say that a bi-infinite sequence $\left(x_k\right)_{k \in \mathbb{Z}}$ of $i$'s and $(i+1)$'s \emph{represents} a real number $x \in$ $[i, i+1]$ if there exists a sequence of intervals $I_1, I_2, \cdots \subseteq \mathbb{Z}$ of increasing lengths $n_1<n_2<\ldots$ such that
\[
\lim _{k \rightarrow \infty} \frac{\sum_{j \in I_k} x_j}{n_k}=x,
\]
that is to say, the averages of $\left(x_k\right)_{k \in \mathbb{Z}}$ over the intervals converge to~$x$.
For a real number $x$, we write $\lfloor x\rfloor$ for the integer part of $x$, which is the largest integer which is less than or equal to $x$.

\begin{definition}[Balanced representation of real numbers]\label{def.balanced_representation}
Let $x\in \mathbb{R}$. For every $k \in \mathbb{Z}$ define
\[
B_k(x):=\lfloor k x\rfloor-\lfloor(k-1) x\rfloor .
\]
The bi-infinite sequence $\left(B_k(x)\right)_{k \in \mathbb{Z}}$ is called the \emph{balanced representation of $x$}.
\end{definition}

Clearly $B_k(x)\in \left\{ \lfloor x\rfloor,\lfloor x\rfloor+1\right\}$ and $\left(B_k(x)\right)_{k \in \mathbb{Z}}$ is a representation of $x$ in the sense defined above.  A key observation is that balanced representations of irrational $x\notin \mathbb{Q}$ are Sturmian sequences~\cite{Pytheas2002}, while for rational $x\in \mathbb{Q}$ the sequence is periodic.

\begin{remark}
Definition~\ref{def.balanced_representation} can be adapted to any $\vec{x}\in\R^d$: simply define $\lfloor \vec{x}\rfloor$ or $B_k(\vec{x})$ coordinate by coordinate.
\end{remark}

 \subsection{Rational piecewise affine maps}

A \emph{rational piecewise affine map} is given by finitely many pairs $\left(U_i, f_i\right)$ where
\begin{itemize}
\item  $U_i$ are disjoint unit cubes of $\mathbb{R}^d$ with integer corners,
\item  $f_i$ are affine maps with rational coefficients.
\end{itemize}

Each set $U_i$ is the domain where $f_i$ can be applied. The system $\left(U_i, f_i\right)_{i=1\dots n}$ determines a function
$f: D \longrightarrow \mathbb{R}^d$
whose domain is $D=\bigcup_i U_i$ and
$$f(\vec{x})=f_i(\vec{x}) \text { for all } \vec{x} \in U_i.$$

\subsection{Tileset associated with a rational piecewise affine map}

Kari's constructions are based on the idea that with rational (piecewise) affine map $f$ can be associated with some Wang tileset.

\begin{definition}[Tileset $\tau_{f_{i}}$ associated with $f_{i}$]
The tileset  $\tau_{f_{i}}$ corresponding to a rational affine map
$f_i(\vec{x})=M \vec{x}+\vec{b}$
and its domain cube $U_i$ consists of all tiles

\begin{center}
\hspace{-1.1cm}
\begin{tikzpicture}[scale=1]
\node at (0.8,1.3) [anchor=south]  {\small{$B_k(f_{i}(\vec{x}))$}};
\node at (0,0.7) [anchor=east]  {\small{$C_{f_i,k-1}(\vec{x})$}};
\draw  (0,0) rectangle (1.3,1.3);
\node at (0.8,0) [anchor=north]  {\small{$B_k(\vec{x})$}};
\node at (1.4,0.7) [anchor=west]  {\small{$C_{f_i,k}(\vec{x})$}};
\end{tikzpicture}
\end{center}

where $k \in \mathbb{Z}$, $\vec{x} \in U_i$, and $C_{f_i,k}(\vec{x})$ is a rational number depending on $f_i,\vec{x}$ and $k$ (details can be found in~\cite{Kari2007}).
\end{definition}

Every tile as pictured above locally computes $f_i$  with some error since we have
\[
f_i\left( B_k(\vec{x}) \right) + C_{f_i,k-1}(\vec{x}) = B_k\left(f_i\left( \vec{x} \right)\right) + C_{f_i,k}(\vec{x})
\]
Moreover, because $f_i$ is rational, there are only finitely many such tiles (even though there are infinitely many $k \in \mathbb{Z}$ and $\vec{x} \in U_i$).

\begin{lemma}[Main property of $\tau_{f_{i}}$]
For fixed $\vec{x} \in U_i$ the tiles for consecutive $k \in \mathbb{Z}$ match so that a horizontal row can be formed whose top and bottom labels bear the balanced representations of $\vec{x}$ and $f_i(\vec{x})$.
\end{lemma}

\begin{center}
\hspace{-1.1cm}
\begin{tikzpicture}[scale=1]
\node at (0,0.7) [anchor=east]  {\small{\dots}};
\node at (0.65,1.3) [anchor=south]  {\scalebox{0.9}{$B_k(f_{i}(\vec{x}))$}};
\node at (0.65,0) [anchor=north]  {\scalebox{0.9}{$B_k(\vec{x})$}};
\draw  (0,0) rectangle (1.3,1.3);
\node at (1.95,1.8) [anchor=south]  {\scalebox{0.9}{$B_{k+1}(f_{i}(\vec{x}))$}};
\node at (1.95,-0.5) [anchor=north]  {\scalebox{0.9}{$B_{k+1}(\vec{x})$}};
\draw  (1.3,0) rectangle (2.6,1.3);
\node at (3.25,1.3) [anchor=south]  {\scalebox{0.9}{$B_{k+2}(f_{i}(\vec{x}))$}};
\node at (3.25,0) [anchor=north]  {\scalebox{0.9}{$B_{k+2}(\vec{x})$}};
\draw  (2.6,0) rectangle (3.9,1.3);
\node at (4.55,1.8) [anchor=south]  {\scalebox{0.9}{$B_{k+1}(f_{i}(\vec{x}))$}};
\node at (4.55,-0.5) [anchor=north]  {\scalebox{0.9}{$B_{k+1}(\vec{x})$}};
\draw  (3.9,0) rectangle (5.2,1.3);
\begin{scope}[shift={(5.2,0)}]
\node at (0.65,1.3) [anchor=south]  {\scalebox{0.9}{$B_{k+4}(f_{i}(\vec{x}))$}};
\node at (0.65,0) [anchor=north]  {\scalebox{0.9}{$B_{k+4}(\vec{x})$}};
\draw  (0,0) rectangle (1.3,1.3);
\node at (1.4,0.7) [anchor=west]  {\small{$\dots$}};
\end{scope}
\end{tikzpicture}
\end{center}

Kari then considers the tileset $\tau_{f}$ obtained as the union of the $\tau_{f_{i}}$, with additional rules to ensure that only tiles from the same $\tau_{f_{i}}$ can appear on a given row.

\subsection{Tiling problem and iterations of rational piecewise map}

Let $(D,f)$ be a rational piecewise affine map. A real number $x\in D$ is an \emph{immortal point} for $f$ if for all $k \in \N^*$, $f^{[k]}(x) \in D$.  If such an $x$ exists, f is called \emph{immortal}. A non-empty set $I \subseteq D$ is a \emph{rational invariant box}  if $I$ is non-empty, $I\cap\mathbb{Q}\neq \emptyset$ and $f(I) \subset I$.  Every $x\in I$, where $I$ is a rational immortal box, is an immortal point for~$f$.

\begin{example}
In~\cite{kari1996small},  Kari considers the rational piecewise affine map:
$f(x)=
\left\{
\begin{array}{ll}
2 x, & \text { if } x \leq 1 \\
 \frac{2}{3} x, & \text { if } x>1
 \end{array}
 \right.$
which has $I=[1/2,2]$ as a rational invariant box.
\end{example}


\begin{lemma}\label{lemma:immortal_point_implies_rationals} \label{lemma:rationnal}
	Assume that a rational piecewise affine map $f$ has a (possibly irrational) immortal point $\vec{x} \in D$. Then for every $k\in\mathbb{N}^*$ there exists a rational $\vec{x}_{k} \in D$ such that $\vec{x}'_{k}=f^{[k]}(\vec{x}_{k}) \in D$ is rational.
\end{lemma}

\begin{proof}
When $k$ is fixed, finding such a rational point  $\vec{x}_{k} \in D$ corresponds to finding some rational solution to some disjunction of conjunctions of affine inequalities with rational coefficients: these constraints express the fact that $x_{k}$ and its successive images by the function $f$ remain in $D$, expressed in terms of the coefficients of piecewise affine function $f$.

We know that this
disjunction of conjunctions of affine inequalities with rational coefficients, is non-empty by hypothesis, as $\vec{x}$ is a (possibly irrational) point that satisfies all these constraints.

It is a well-known fact that a set that corresponds to some disjunction of conjunctions of affine inequalities with rational coefficients that is non-empty always admits at least a rational solution. For an explicit proof of this statement, see, for example, \cite[Lemma 2.4]{Maa94a} for a sketch of proof or \cite[Theorem 6]{Koi94a} for a full proof.

Then, as $x'_{k}=f^{[k]}(x_{k})$ and $f$ has only rational coefficients, $x'_{k}$ has rational coefficients, and the Lemma is proved.
\end{proof}

\begin{proposition}\label{propImmortality}
	Let $(D,f)$ be an immortal rational piecewise affine map. Then $\tau_{f}$ can tile the plane.
	Moreover if $f$ has a rational invariant box $I$ then $\tau_{f}$ is provably robust.
\end{proposition}

\begin{proof}
	The fact that $\tau_{f}$ can tile the plane is well-known~\cite{Kari2007} and follows from the construction: one can check that putting the tile $t\left(f_i,k,f^\ell\left(\vec{x}\right)\right)$ at position $(k,\ell)$ for every $(k,\ell)\in\mathbb{Z}^2$ works, where $f^\ell\left(\vec{x}\right) \in D_i$. Now if $f$ has a rational invariant box $I$ then the argument holds provably. It suffices to observe that $\tau_{f}^{k}$ contains a loop for every $k\in\mathbb{N}^*$ observing that a rational point has a periodic balanced representation.  Namely, 
 for every $k\in\N^*$ take $\vec{x}_k\in I$ rational as in Lemma \ref{lemma:immortal_point_implies_rationals}. Then $\vec{x}'_k:=f^k(\vec{x}_k)$ is also rational. Denote $p_k,p'_k$ the periods of the balanced representations of $\vec{x}_k,\vec{x}'_k$. Then $\tau_{f}^{k}$ contains the loop of length $m:=\lcm(p_k,p'_{k})$ labelled by $B_0(\vec{x}_k)\dots B_{m-1}(\vec{x}_k)$ and $B_0(\vec{x}'_{k})\dots B_{m-1}(\vec{x}'_{k})$.
\end{proof}
%


The immortality problem for rational piecewise affine maps is undecidable~\cite{BBKPC00}. The proof relies on the encoding of any Turing machine $\mathcal{M}$ inside a rational piecewise map $f_{\mathcal{M}}$,  such that $f_{\mathcal{M}}$ is immortal iff $\mathcal{M}$ does not halt. This extends classical ways to simulate a Turing machine using rational piecewise affine maps \cite{KCG94}. We can then consider the associated tiling $\tau_{f_{\mathcal{M}}}$. Putting everything together we have the following.

\begin{theorem}\label{th:robust:iff}
	Consider a Turing machine $\mathcal{M}$. Then $\mathcal{M}$ is provably robust and not terminating iff $\tau_{f_\mathcal{M}}$ is provably robust.
\end{theorem}

\section{Turing machines as tilesets through Robinson's construction}
\label{section.TM_Robinson}

Robinson described in \cite[section 7]{Robinson:1971:UNT}
a method to convert a Turing machine $\mathcal{M}$ into a Wang tileset $\tau_\mathcal{M}$ such that  $\mathcal{M}$ does not halt on the empty tape iff $\tau_\mathcal{M}$ tiles the plane. We call this \emph{Robinson's construction}.

\begin{theorem}\label{th:iff:robinsonun}
	If the Turing machine $\mathcal{M}$ is provably robust and not terminating,  then the tileset $\tau_\mathcal{M}$ is provably robust.
\end{theorem}

\begin{proof}
	Informally, the arguments of Robinson are based on the idea that Robinson's tileset is such that it constructs bigger and bigger squares and that they can be coloured in two colours of squares, such that one will be used to simulate computations of a Turing machine.

	Formally, reformulated using our framework, from the invariant \eqref{eq:thm:robinson:invariant:inductif}, we can consider odd and even $n$, to consider the invariants
	\begin{equation}\label{eq:un}
		H_{n} \circ \mathcal{T}_{Robi} \circ H_{n} \Smodels H'_{n}
	\end{equation}
	\begin{equation} \label{eq:deux}
		H'_{n} \circ \mathcal{T}_{Robi} \circ H'_{n} \Smodels H_{n}
	\end{equation}

	Then $H_{n}$ can be modified such that it embeds the computation of a Turing machine $M$ on $H_{n}$: this can be done exactly as in \cite{Robinson:1971:UNT}. This provides
	that the invariant~\ref{eq:un} holds for some $n$ iff the computation of $M$ on some fixed input does not terminate in a size related to the size of $H_{n}$. Observing that the size of $H_{n}$ is $\mathcal{O}\left({g(2n)}\right)$, the obtained tileset $\tau_\mathcal{M}$ will be provably robust iff $\mathcal{M}$ is not halting in the corresponding time.
	Indeed, the arguments of provability are preserved by the construction.
\end{proof}

\begin{theorem}\label{th:iff:robinsondeux}
	The converse is true: If $\tau_\mathcal{M}$ is provably robust, then $\mathcal{M}$ is provably robust and not terminating.
\end{theorem}

\begin{proof}
	If $\tau_\mathcal{M}$ is provably robust, then it means by construction that $\mathcal{M}$ is not halting and we have, from Robinson's constructions and arguments a proof of it. So $\mathcal{M}$ is provably robust.
\end{proof}

\section{Conclusions and discussions}
\label{sec:conclusion}

\subsection{Back to the domino problem}

\begin{theorem}\label{th:domino:decidable}
The domino problem is decidable for robust tilesets.
\end{theorem}

\begin{proof}
We prove the decidability of this problem by exhibiting two semi-algorithms taking as input a tileset $\tau$.
First note that since the set of tilings is a compact set, the fact that a tileset $\tau$ cannot tile the plane is equivalent to the existence of an integer $n$ such that $\tau^{n}$ is not compatible. A first semi-algorithm searches for such an integer $n$: construct successively the transducers $\mathcal{T}^n$, and look for a loop inside $\mathcal{T}^n$. If there exists an $n$ such that $\mathcal{T}^n$ has no loop, then the semi-algorithm stops. Otherwise it keeps running.
Now, the fact that a tileset $\tau$ can tile the plane is computably enumerable for a robust tileset: indeed, from the definition of provably robust, we just need to search for a proof of all the statements of the definition. If we find one, we are sure the plane can be tiled.

As the tileset is assumed to be robust, either the search for a $n$ in the first case or of a proof in the second case must terminate, and hence, in both cases, we can decide effectively which case holds.
\end{proof}

\subsection{Soundness and relative completeness}

The method we described to prove that a given tileset can tile the plane is based on proving invariants of the form of Equation  \eqref{eq:recurrence}. 
Consider a logic language $Assn$ sufficiently expressive, in particular able to talk about tiles, finite transducers, and sequences. We can use standard first-order logic for this purpose (with the signature of arithmetic) as before for that purpose.

\begin{theorem}[Soundness and relative completeness of the proof method]  \label{th:domino:rc} The method of finding an invariant of the form \ref{eq:recurrence} for some composition pattern $F$ to prove that a tileset can tile the plane is a sound and relatively complete method.
\end{theorem}

Once again, relatively complete means the difficulty is ``just'' on being able to prove assertions such as
$A \Rightarrow w [Equation~ \eqref{eq:recurrence} ]$ (i.e. establishing $\models A \Rightarrow Equation~\eqref{eq:recurrence}$) whenever $\models A \Rightarrow  Equation~ \eqref{eq:recurrence}$, but not on the method. Here, $A$ can be chosen as a formula of $Assn$.

Of course, it is not complete, because if it is applied to some tiling obtained from a non-robust Turing machine, then it will fail, exactly as a method like Hoare's logic fails for non-robust Turing machines.  Of course, we could repeat here all the (sometimes informal) discussions about robust/non-robust Turing machines, for tiling sets.

\subsection{Relations to experimental observations} \label{sec:experimental}
Jeandel-Rao's tileset $\mathcal{T}$ considered in Section~\ref{section.JR} arose from a systematic computer-assisted proof of tilesets by considering an increasing number of tiles, conducted in \cite{jeandel2021aperiodic}. The authors proved (using the help of about 23 CPU years of computing done on a cluster of parallel machines) that there is no aperiodic Wang set with 10 tiles or less. Another (also provably robust) tileset $\mathcal{T'}$ is considered in \cite[Section 7.2]{jeandel2021aperiodic}.  They mention that some other candidates with 11 tiles could also be aperiodic.

Our study provides theoretical explanations for the experimental facts mentioned in \cite[Section 8]{jeandel2021aperiodic}, when the authors discuss why they investigated the tileset $\mathcal{T}$, and not other candidates. They explain that ``\emph{it was very easy for a computer to produce the transducer for $\mathcal{T}^{k}$, even for large values of $k$ ($k \sim 1000$). In contrast, for almost all other tilesets, we were not able to reach even $k=30$}''. As demonstrated in the current article, the possibility of tiling the plane is closely related to the existence of an invariant of the form \eqref{eq:recurrence}. When such an invariant exists, then computing $\mathcal{T}^{k}$ becomes feasible. When it does not hold, then computing $\mathcal{T}^{k}$ is very costly, as the number of edges is exploding with $k$. So, their quest can be related to the experimental search of an invariant, 
and the existence of an invariant, seems to be closely related to the fact, that they mention that ``\emph{This suggested this tileset had some particular structure}''.  We somehow provide arguments that the involved ``structure'' is the existence of an invariant of the form \eqref{eq:recurrence} for some $F$.


\subsection{Perspectives and future work}

Our settings pave the way for important and fascinating developments. First, when a tileset is robust, we know it is decidable. But then it makes sense to discuss the complexity of the domino problem and not only its decidability. This can be done by quantifying the robustness in an approach similar to the one done in \cite{CSL24}.  Our concept of robustness is inspired by \cite{asarin01perturbed,CSL24}, but differs from their exact notion of robustness. Investigations, whether concepts in their spirit, based on a distance, also deserve investigations. Notice that \cite{asarin01perturbed} is, in turn, inspired by finding general arguments about various decidability proofs about hybrid and dynamical systems.

Furthermore, we think that similar statements can be obtained not only about the fact that the plane can be tiled but also about the fact that the plane can be tiled aperiodically. Furthermore, our study explains the common structures of the proofs and reasons why substitutive tilings appear in almost all known examples: this is necessarily the fact as soon as robust tilesets are considered and when the invariant involves a metatransducer with a constant number of edges. We believe this generalises to aperiodicity. 

We also believe that our reasoning opens the way to deriving logic similar to Hoare's logic suitable to talk about tilings and ways to discuss embedding of tilings into tilings, 
in the spirit of our presentation of Robin's construction by reasoning directly on invariants. 

%
%
%
%

\bibliographystyle{plainurl}

\bibliography{biblio, ../bournez,../perso}

%
%
%
%

\section*{Appendix}
\appendix

\section{More details on Example \ref{example:periodic_tileset}}
\label{more:example:periodic_tileset}

In this appendix we provide more details on the link between transducers and tilings by Wang tiles. As an illustration we go back to the tileset of Example~\ref{example:periodic_tileset} which is given by the four following Wang tiles.
\begin{center}
\begin{tikzpicture}[scale=1,,shorten >=1pt,>={Stealth[round]},transform shape]
\begin{scope}[scale=0.8]
	\wang{0}{0}{rouge}{bleu}{rouge}{rouge}
	\wang{1.5}{0}{vert}{vert}{rouge}{bleu}
	\wang{3}{0}{rouge}{rouge}{rouge}{vert}
	\wang{4.5}{0}{rouge}{vert}{vert}{rouge}
\end{scope}
\end{tikzpicture}
\end{center}

Remind that $\mathcal{T}^2$ the composition of $\mathcal{T}$ from Example~\ref{example:periodic_tileset} with itself is as pictured below.
\begin{center}
\begin{tikzpicture}[scale=1,shorten >=1pt,>={Stealth[round]}]
	\node[state, rectangle, rounded corners=3mm] (RV) {\lettreddT{rouge}{vert}};
    \node[state, rectangle, rounded corners=3mm] at (4,0) (BR) {\lettreddT{bleu}{rouge}};
    \node[state, rectangle, rounded corners=3mm] at (8,0) (VV) {\lettreddT{vert}{vert}};
    \node[state, rectangle, rounded corners=3mm] at (2,4) (VR) {\lettreddT{vert}{rouge}};
    \node[state, rectangle, rounded corners=3mm] at (6,4) (RB) {\lettreddT{rouge}{bleu}};
    \node[state, rectangle, rounded corners=3mm] at (2,-4) (VB) {\lettreddT{vert}{bleu}};
    \node[state, rectangle, rounded corners=3mm] at (6,-4) (RR) {\lettreddT{rouge}{rouge}};
    \path[->]   (VV) edge[right] node {\wangEdgeV{white}{rouge}{rouge}{vert}{rouge}{rouge}{white}{vert}} (RR)
	            		(RR) edge[above] node{\wangEdgeV{white}{vert}{vert}{rouge}{rouge}{bleu}{white}{rouge}} (VB)
		             (VB) edge[left] node{\wangEdgeV{white}{rouge}{rouge}{vert}{vert}{vert}{white}{bleu}} (RV)
		             (RV) edge[] node{\wangEdgeV{white}{bleu}{rouge}{rouge}{rouge}{rouge}{white}{vert}} (BR)
		             (RV) edge[above left] node{\wangEdgeV{white}{vert}{vert}{rouge}{rouge}{rouge}{white}{vert}} (VR)
		             (VR) edge[above] node{\wangEdgeV{white}{rouge}{rouge}{vert}{rouge}{bleu}{white}{rouge}} (RB)
		             (RB) edge[above right] node{\wangEdgeV{white}{vert}{rouge}{bleu}{rouge}{vert}{white}{rouge}} (VV)
		             (BR) edge[] node{\wangEdgeV{white}{vert}{rouge}{bleu}{rouge}{vert}{white}{rouge}} (VV);
\end{tikzpicture}
\end{center}

A more in-depth exploration leads to the following cycle in 
\[
\mathcal{T}^3=\mathcal{T}^2\circ\mathcal{T}
\]
\begin{center}
\begin{tikzpicture}[scale=1,shorten >=1pt,>={Stealth[round]}]
	\node[state, rectangle, rounded corners=3mm] (RVV) {\lettredddT{rouge}{vert}{vert}};
    \node[state, rectangle, rounded corners=3mm] at (3,2) (BRR) {\lettredddT{bleu}{rouge}{rouge}};
    \node[state, rectangle, rounded corners=3mm] at (7,2) (VVB) {\lettredddT{vert}{vert}{bleu}};
    \node[state, rectangle, rounded corners=3mm] at (7,-2) (RRV) {\lettredddT{rouge}{rouge}{vert}};
    \node[state, rectangle, rounded corners=3mm] at (3,-2) (VBR) {\lettredddT{vert}{bleu}{rouge}};
    \path[->]   (RVV) edge[above] node {\wangEdgeVVVun} (BRR)
	            		(BRR) edge[above] node{\wangEdgeVVVdeux} (VVB)
		             (VVB) edge[left] node{\wangEdgeVVVtrois} (RRV)
		             (RRV) edge[below] node{\wangEdgeVVVquatre} (VBR)
		             (VBR) edge[below] node{\wangEdgeVVVcinq} (RVV);
\end{tikzpicture}
\end{center}

Translate this cycle in terms of meta-transducers, and successively consider each vertex of the cycle as the initial vertex. We thus obtain:

\begin{center}
\begin{tikzpicture}[scale=0.75]
	\node[] at (-2.5,0) {\scalebox{1}{$\mathcal{T}^3 \Smodels~\mathcal{L}_1 := ~$}};
	\node[state, rectangle, rounded corners=3mm] (RVV) {\lettredddT{rouge}{vert}{vert}};
	\path[->] (RVV) edge[out=30,in=330,looseness=3, above right] node{\cyclicPun} (RVV);
\end{tikzpicture}
\end{center}
\begin{center}
\begin{tikzpicture}[scale=0.75,shorten >=1pt,>={Stealth[round]}]
	\node[] at (-2.5,0) {\scalebox{1}{$\mathcal{T}^3 \Smodels~\mathcal{L}_2 := ~$}};
	\node[state, rectangle, rounded corners=3mm] (BRR) {\lettredddT{bleu}{rouge}{rouge}};
	\path[->] (BRR) edge[out=30,in=330,looseness=3, above right] node{\cyclicPdeux} (BRR);
\end{tikzpicture}
\end{center}
\begin{center}
\begin{tikzpicture}[scale=0.75,shorten >=1pt,>={Stealth[round]}]
	\node[] at (-2.5,0) {\scalebox{1}{$\mathcal{T}^3 \Smodels~\mathcal{L}_3 := ~$}};
	\node[state, rectangle, rounded corners=3mm] (VVB) {\lettredddT{vert}{vert}{bleu}};
	\path[->] (VVB) edge[out=30,in=330,looseness=3, above right] node{\cyclicPtrois} (VVB);

\end{tikzpicture}
\end{center}
\begin{center}
\begin{tikzpicture}[scale=0.75,shorten >=1pt,>={Stealth[round]}]
	\node[] at (-2.5,0) {\scalebox{1}{$\mathcal{T}^3 \Smodels~\mathcal{L}_4 := ~$}};
	\node[state, rectangle, rounded corners=3mm] (RRV) {\lettredddT{rouge}{rouge}{vert}};
	\path[->] (RVV) edge[out=30,in=330,looseness=3, above right] node{\cyclicPquatre} (RVV);

\end{tikzpicture}
\end{center}
\begin{center}
\begin{tikzpicture}[scale=0.75,shorten >=1pt,>={Stealth[round]}]
	\node[] at (-2.5,0) {\scalebox{1}{$\mathcal{T}^3 \Smodels~\mathcal{L}_5 := ~$}};
	\node[state, rectangle, rounded corners=3mm] (VBR) {\lettredddT{vert}{bleu}{rouge}};
	\path[->] (VBR) edge[out=30,in=330,looseness=3, above right] node{\cyclicPcinq} (VBR);

\end{tikzpicture}
\end{center}

For instance the first loop $\mathcal{L}_1$ corresponds to the $5\times3$ pattern pictured below, which can be glued to itself to form an infinite horizontal strip of height $3$.

\begin{center}
\begin{tikzpicture}[scale=0.6]
\tikzmath{\x = 20;}
\node at (-7,1.5) {\dots};
\node at (7,1.5) {\dots};
\begin{scope}[shift={(-10,0)}]
\wang{4}{0}{rouge!\x}{rouge}{rouge}{vert!\x}
\wang{4}{1}{vert!\x}{vert}{rouge!\x}{bleu!\x}
\wang{4}{2}{rouge}{vert}{vert!\x}{rouge!\x}
\end{scope}
\begin{scope}[shift={(-5,0)}]
\wang{0}{0}{rouge!\x}{bleu!\x}{rouge}{rouge}
\wang{1}{0}{vert!\x}{vert!\x}{rouge}{bleu!\x}
\wang{2}{0}{rouge!\x}{rouge!\x}{rouge}{vert!\x}
\wang{3}{0}{rouge!\x}{vert!\x}{vert}{rouge!\x}
\wang{4}{0}{rouge!\x}{rouge}{rouge}{vert!\x}
\wang{0}{1}{rouge!\x}{rouge!\x}{rouge!\x}{vert}
\wang{1}{1}{rouge!\x}{vert!\x}{vert!\x}{rouge!\x}
\wang{2}{1}{rouge!\x}{rouge!\x}{rouge!\x}{vert!\x}
\wang{3}{1}{rouge!\x}{bleu!\x}{rouge!\x}{rouge!\x}
\wang{4}{1}{vert!\x}{vert}{rouge!\x}{bleu!\x}
\wang{0}{2}{rouge}{rouge!\x}{rouge!\x}{vert}
\wang{1}{2}{rouge}{bleu!\x}{rouge!\x}{rouge!\x}
\wang{2}{2}{vert}{vert!\x}{rouge!\x}{bleu!\x}
\wang{3}{2}{rouge}{rouge!\x}{rouge!\x}{vert!\x}
\wang{4}{2}{rouge}{vert}{vert!\x}{rouge!\x}
\end{scope}
\wang{0}{0}{rouge!\x}{bleu!\x}{rouge}{rouge}
\wang{1}{0}{vert!\x}{vert!\x}{rouge}{bleu!\x}
\wang{2}{0}{rouge!\x}{rouge!\x}{rouge}{vert!\x}
\wang{3}{0}{rouge!\x}{vert!\x}{vert}{rouge!\x}
\wang{4}{0}{rouge!\x}{rouge}{rouge}{vert!\x}
\wang{0}{1}{rouge!\x}{rouge!\x}{rouge!\x}{vert}
\wang{1}{1}{rouge!\x}{vert!\x}{vert!\x}{rouge!\x}
\wang{2}{1}{rouge!\x}{rouge!\x}{rouge!\x}{vert!\x}
\wang{3}{1}{rouge!\x}{bleu!\x}{rouge!\x}{rouge!\x}
\wang{4}{1}{vert!\x}{vert}{rouge!\x}{bleu!\x}
\wang{0}{2}{rouge}{rouge!\x}{rouge!\x}{vert}
\wang{1}{2}{rouge}{bleu!\x}{rouge!\x}{rouge!\x}
\wang{2}{2}{vert}{vert!\x}{rouge!\x}{bleu!\x}
\wang{3}{2}{rouge}{rouge!\x}{rouge!\x}{vert!\x}
\wang{4}{2}{rouge}{vert}{vert!\x}{rouge!\x}
\begin{scope}[shift={(5,0)}]
\wang{0}{0}{rouge!\x}{bleu!\x}{rouge}{rouge}
\wang{0}{1}{rouge!\x}{rouge!\x}{rouge!\x}{vert}
\wang{0}{2}{rouge}{rouge!\x}{rouge!\x}{vert}
\end{scope}
\end{tikzpicture}
\end{center}

None of the five loops $\mathcal{L}_1,\dots,\mathcal{L}_5$ in $\mathcal{T}^3$ is periodic, but we can combine them to get a periodic one in $\mathcal{T}^{15}$ as follows. Denote $(v_i,v_i,b_i,t_i)$ the loop $\mathcal{L}_i$ for every $i$ from $1$ to $5$. We observe that
\begin{align*}
b_1 &= t_4\\
b_2 &= t_5\\
b_3 &= t_1\\
b_4 &= t_2\\
b_5 &= t_3
\end{align*}
so that, using the composition rule of transducers, we get that
\[
\mathcal{T}^{15}  \Smodels (w,w,b_1,t_4)
\]
where
\[
w := \left(v_1,v_3,v_5,v_2,v_4\right),
\]
and since $b_1=t_4$ this loop on $w$ is periodic. This is formalized in Proposition~\ref{proposition:circular_loop_periodic_loop}.

%
%
%

\section{More details on Jeandel-Rao's tileset}
\label{sec:dessins:jr}

\newcommand\EPUR[1]{}

In this section, we provide more details on Jeandel Rao's aperiodic tileset $\mathcal{T}$ considered in \cite[Section 4]{jeandel2021aperiodic}.

The tileset is described in Example \ref{ex:jr}.

\def\sf{0.22}

Here is a graphical representation of $\alpha_{2},\beta_{2},\gamma_{2},\delta_{2}$ and $\epsilon_{2}$.

\def\n{2}

\begin{tikzpicture}[scale=\sf]
\EPURE{\node at (0,0) {0};}
\EPURE{\node at (10,0) {10};}
\gralpha{0}{0}{\n}
\grbeta{6}{0}{\n}
\grgamma{15}{0}{\n}
\grdelta{0}{-2}{\n}
\grepsilon{17}{-2}{\n}
\end{tikzpicture}

Here is a graphical representation of $A_{3},B_{3},C_{3},D_{3}$ and $E_{3}$.

\medskip

\def\n{0}

\begin{tikzpicture}[scale=\sf]
\EPURE{\node at (0,0) {0};}
\EPURE{\node at (10,0) {10};}
\grAdeux{0}{1}{0}

\EPUR{\grA{0}{-2}{3}}
\end{tikzpicture}

\medskip

\begin{tikzpicture}[scale=\sf]
\EPURE{\node at (0,0) {0};}
\EPURE{\node at (10,0) {10};}
\grBdeux{0}{0}{\n}

\EPUR{\grB{-3}{-2}{3}}
\end{tikzpicture}

\medskip

\begin{tikzpicture}[scale=\sf]
\EPURE{\node at (0,0) {0};}
\EPURE{\node at (10,0) {10};}
\grCdeux{0}{0}{\n}

\EPUR{\grC{-3}{-2}{3}}
\end{tikzpicture}

\medskip

\begin{tikzpicture}[scale=\sf]
\EPURE{\node at (0,0) {0};}
\EPURE{\node at (10,0) {10};}
\grDdeux{0}{0}{\n}

\EPUR{\grD{-3}{-2}{3}}
\end{tikzpicture}

\medskip

\begin{tikzpicture}[scale=\sf]
\EPURE{\node at (0,0) {0};}
\EPURE{\node at (10,0) {10};}
\grEdeux{0}{0}{\n}

\EPUR{\grE{-3}{-2}{3}}
\end{tikzpicture}

Here is a graphical representation of $\alpha_{4},\beta_{4},\gamma_{4},\delta_{4}$ and $\epsilon_{4}$.

\medskip

\def\n{1}

\begin{tikzpicture}[scale=\sf]
\EPURE{\node at (0,0) {0};}
\EPURE{\node at (10,0) {10};}
\gralphadeux{0}{0}{1}

\EPUR{\gralpha{0}{-2}{4}}
\end{tikzpicture}

\medskip

\begin{tikzpicture}[scale=\sf]
\EPURE{\node at (0,0) {0};}
\EPURE{\node at (10,0) {10};}
\grbetadeux{0}{0}{\n}

\EPUR{\grbeta{-3}{-2}{4}}
\end{tikzpicture}

\medskip

\begin{tikzpicture}[scale=\sf]
\EPURE{\node at (0,0) {0};}
\EPURE{\node at (10,0) {10};}
\grgammadeux{0}{0}{\n}

\end{tikzpicture}

\medskip

\begin{tikzpicture}[scale=\sf]
\EPURE{\node at (0,0) {0};}
\EPURE{\node at (10,0) {10};}
\grdeltadeux{0}{0}{\n}

\EPUR{\grdelta{-3}{-2}{4}}
\end{tikzpicture}

\medskip

\begin{tikzpicture}[scale=\sf]
\EPURE{\node at (0,0) {0};}
\EPURE{\node at (10,0) {10};}
\grepsilondeux{0}{0}{\n}

\EPUR{\grepsilon{-3}{-2}{4}}
\end{tikzpicture}

The graphical reprsentation of $A_{5}$ is given on Figure \ref{jr:acinq}.

\end{document}

\olivier{Pour moi, tout ce qui est après ce point, disparait.}

\newpage

\section{A mettre en forme (ou pas?}

\section{Turing machines and robustness}

\subsection{Turing machines as Wang tileset}

We explain here how to convert a Turing machine $\mathcal{M}$ into a Wang tileset $\tau_\mathcal{M}$ such that  $\mathcal{M}$ halts on the empty tape iff $\tau_\mathcal{M}$ tiles the plane.

We start by defining how we can encode an execution $\mathcal{M}$ into a tiling
of a Wang tileset.

The intuition is see the successive configurations as tiles, the horizontal
compatibility corresponds to what is written on the tape, the vertical compatibility
to the transition that has been made.

More formally, denoting by $\Sigma$ the extended alphabet including the starting
symbol $\$$ and the blank symbol $B$, the horizontal colours are :
$$ \{ ab | a, b \in \Sigma \cup Q, a = B \Rightarrow b = B, a\in Q \Rightarrow b \not\in Q \} $$
And the vertical colours are :
$$ (Q \cup \Sigma)^3 \cap \Sigma^*( Q + \epsilon) \Sigma^* $$

So a tile is :

\begin{tikzpicture}
	\wang{0}{0}{rouge}{blue}{white}{yellow}
\end{tikzpicture}
with : \lettrebT{rouge}:$\alpha' \beta' \gamma'$, \lettrebT{blue} : $\beta \gamma$,
\lettrebT{white} : $\alpha \beta \gamma$ and \lettrebT{yellow} : $\alpha \beta$.

\begin{itemize}
	\item $\delta(q, a) = q', a', \rightarrow$, on $\mathcal{M}$:
	\begin{tabular}{ccccccc}
		b & c & a' & q' & d & e & f \\
		\hline
		b & c & q & a & d & e & f \\
	\end{tabular}
	corresponding to: \lettrebT{rouge}:a'q'd, \lettrebT{blue} : ad, \lettrebT{white} : qad and \lettrebT{yellow} : qa.

	\item $\delta(q, a) = q', a', \leftarrow$, on $\mathcal{M}$ :
	\begin{tabular}{ccccccc}
		b & c & q' & d & a & e & f \\
		\hline
		b & c & d & q & a & e & f \\
	\end{tabular}
	corresponding to: \lettrebT{rouge}:q'da', \lettrebT{blue} : qa, \lettrebT{white} : dqa and \lettrebT{yellow} : dq.
\end{itemize}

\todo{define a \emph{product} transducer}

\begin{proposition}
If the Turing machine $\mathcal{M}$ is robust then the tileset $\tau_\mathcal{M}$ is semantically robust.
\end{proposition}

\begin{proof}
	Let us assume $\mathcal{M}$ is robust. Here, without loss of generality, we assume
	that if it reaches a terminating state, then it loops on the same final
	configuration. We define $\tau_\mathcal{M}$ and
	$\mathcal{T}_\mathcal{M}$ as the tileset and the associated tranducer defined
	above. Let $x$ be some input of $\mathcal{M}$. By definition, we must consider
	two cases:
	\begin{itemize}
		\item If $\mathcal{M}$ reaches a terminating state on $x$, then, we define
		$g : n \mapsto n$ and we have :
		\begin{itemize}
			\item $\mathcal{T}^{n}_\mathcal{M}$ can be effectively computed from $(\mathcal{T}_\mathcal{M},\mathcal{T}^{2}_\mathcal{M},\dots,\mathcal{T}^{n-1}_\mathcal{M})$, since it is just a transition in $\mathcal{M}$;
			\item $\mathcal{T}_\mathcal{M}$ is compatible;
			\item if $\mathcal{T}^{n}_\mathcal{M}$ is compatible then $\mathcal{T}^{n+1}_\mathcal{M}$ is compatible by definition
			of the transitions of $\mathcal{M}$.
		\end{itemize}
		Then $\tau_\mathcal{M}$ is semantically robust.
		\item If we have a proof that $\mathcal{M}$ does not reach terminating state,
		then $\tau_\mathcal{M}$ is semantically robust by the same arguments.
	\end{itemize}
\end{proof}

\begin{proposition}
If the Turing machine $\mathcal{M}$ is provably robust then the tileset $\tau_\mathcal{M}$ is provably robust.
\end{proposition}

\begin{proof}
\todo{write this}
\end{proof}

\todo{is this correctly written?}
\begin{theorem}
	A Turing machine $M$ is robust iff $\tau_M$ is inductive or $\tau_M$ does not tile $\N^2$.
\end{theorem}

\section{Robustness as a witness of decidability}
\label{section:robustness_Domino}

General idea: "Non robust tilesets are to the domino problem what non robust Turing machines are to the Halting problem."

An attempt of picture to summarize what happens:
    \begin{center}
    \includegraphics[width=0.9\textwidth]{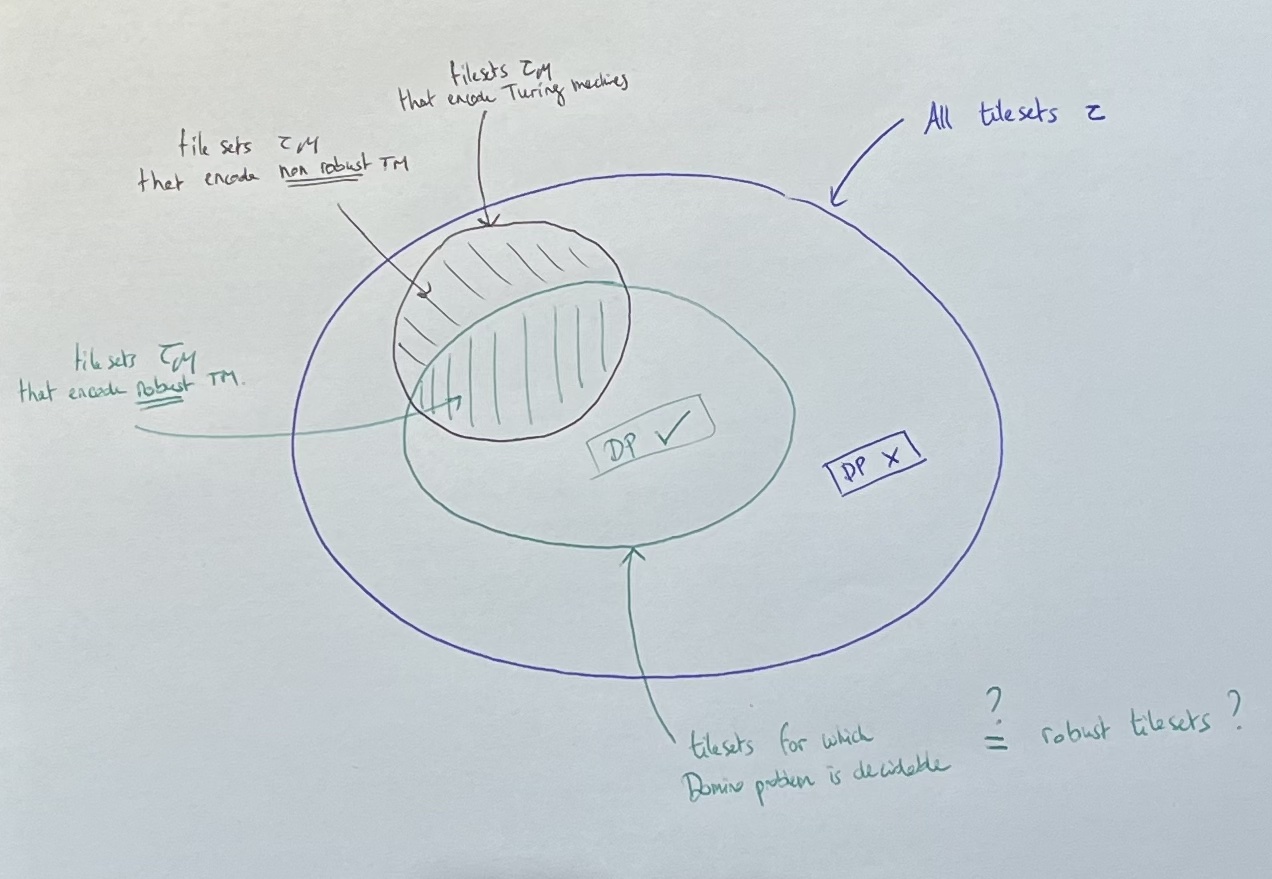}
    \end{center}

\begin{theorem}
	The domino problem is decidable for robust tilesets.
\end{theorem}

\begin{proof}
By compactness, the fact that a tileset $\tau$ cannot tile the plane is equivalent to find some integer $n$ such that $\tau^{n}$ is not compatible.

The fact that a tileset $\tau$ can tile the plane is computably enumerable for a robust tileset: indeed, from the definition of provably robust, we ust need to search for a proof of all the statements of the definition. If we find one, then we are sure that the plane can be tiled.

As the tileset is assumed to be robust, either the search for a $n$ in the first case, or of a proof in the second case, must terminate and hence, in both case, we can decided which case holds.
\end{proof}

\todo{is the following statement correct?}
\begin{theorem}
	If $\mathcal{D}$ is a set of tilesets with decidable domino problem, then~$\mathcal{D}$contains only robust tilesets.
\end{theorem}

\begin{proof}
\todo{write this?}
\end{proof}

\end{document}